%% file: eft_note.tex
\documentclass[a4paper]{article}
\pdfoutput=1
\usepackage[utf8]{inputenc}
\usepackage[T1]{fontenc}
\usepackage[numbers, elide, sort&compress]{natbib}
\usepackage{lmodern, amsmath, amssymb, xcolor, a4wide, graphicx, datetime, rotating, adjustbox, array}

\makeatletter\g@addto@macro\bfseries{\boldmath}\makeatother

\usepackage[colorlinks=true, citecolor=blue, urlcolor=blue, linkcolor=blue, breaklinks=true, pdfpagelabels=false]{hyperref}
\def\figureautorefname~#1\null{Fig.\,#1\null}

\def\equationautorefname~#1\null{Eq.\,(#1)\null}
\newcommand{\fullref}[2]{\hyperref[#2]{#1\,\ref*{#2}}}
\makeatletter
\newcommand{\mg}{\texttt{MG5\_aMC@NLO}}
\newcommand{\rcite}[1]{\hyper@@link[cite]{}{cite.#1}{Ref.\,\cite*{#1}}}
\makeatother
\newcommand*{\eqsref}[2]{\hyperref[#1]{Eqs.\,(\ref*{#1}--}\hyperref[#2]{\ref*{#2})}}

\makeatletter
\renewcommand{\@oddfoot}{\hfill\thepage}
\makeatother

\allowdisplaybreaks[4]

\makeatletter
\newcommand{\parenbar}{\mathpalette\p@renb@r}
\def\p@renb@r#1#2{%
	\vbox{%
		\ifx#1\scriptscriptstyle\dimen@.7em\dimen@ii.20em\else%
		\ifx#1\scriptstyle	\dimen@.8em\dimen@ii.25em\else%
					\dimen@.9em\dimen@ii.35em\fi\fi%
		\offinterlineskip%
		\ialign{%
			\hfill##\hfill\cr
			\vbox{\hrule width\dimen@ii height .35pt}\cr
			\noalign{\vskip-.35ex}%
			\hbox to\dimen@{$\mathchar300\hfil\mathchar301$}\cr
			\noalign{\vskip-.35ex}%
			$#1#2$\cr%
		}%
	}%
}%
\makeatother

\newcommand*{\tmp}[4]{\ensuremath{%
	{#4%
	\ifx\empty#3\empty\ifx\empty#1\empty\else^{#1}\fi\else^{#1(#3)}\fi%
	\ifx\empty#2\empty\else_{#2}\fi}%
}}
\newcommand*{\cc }[4][]{\tmp{#2}{#3}{#4}{#1{C}}}
\newcommand*{\ccc}[4][]{\tmp{#2}{#3}{#4}{#1{c}}}

\newcommand*{\qq }[4][]{\tmp{#2}{#3}{#4}{#1{O}}}

\let\Re\undefined
\let\Im\undefined
\DeclareMathOperator{\Re}{Re}
\DeclareMathOperator{\Im}{Im}
\newcommand{\ReIm}{{}_{\Re}^{[\Im]}\!}

\newcommand{\warsaw}[1]{\scriptsize\ensuremath{\equiv#1}}

\newcommand{\hc}[1]{{}^\ddagger #1}

\newcommand*{\sw}{s_W}
\newcommand*{\cw}{c_W}

\newcommand{\FDF}{(\varphi^\dagger i\!\!\overleftrightarrow{D}_\mu\varphi)}
\newcommand{\FDFI}{(\varphi^\dagger i\!\!\overleftrightarrow{D}^I_\mu\varphi)}

\newcommand*{\sd}{\sigma_{\mu\nu}}
\newcommand*{\su}{\sigma^{\mu\nu}}

\DeclareMathOperator{\diag}{diag}

\newcommand{\rien}[1]{}

\newcounter{affiliation}
\newcommand{\aff}[1]{$^{\ref{aff:#1}}$}
\newcommand{\twoaff}[2]{$^{\ref{aff:#1},\ref{aff:#2}}$}
\newcommand{\threeaff}[3]{$^{\ref{aff:#1},\ref{aff:#2},\ref{aff:#3}}$}
\newcommand{\naff}[2]{\refstepcounter{affiliation}\label{aff:#1}$^{\arabic{affiliation}}$\,#2\\}

\title{\vspace*{-37mm}%
{\normalsize\hfill CERN-LPCC-2018-01\\}\vspace*{37mm}
Interpreting top-quark LHC measurements\\in the standard-model effective field theory
}

\author{\hspace*{-.0\textwidth}\parbox{1.\textwidth}{\centering
J.\,A.\,Aguilar Saavedra,\aff{Granada}
C.\,Degrande,\aff{CERN}
G.\,Durieux,\aff{DESY}
\\
F.\,Maltoni,\aff{UCLouvain}
E.\,Vryonidou,\aff{CERN}
C.\,Zhang\aff{IHEP}
(editors),
\\
D.\,Barducci,\aff{SISSA_INFN}
I.\,Brivio,\aff{NBIA}
V.\,Cirigliano,\aff{LosAlamos}
W.\,Dekens,\twoaff{LosAlamos}{NewMexico}
J.\,de Vries,\aff{Nikhef}
C.\,Englert,\aff{Glasgow}
M.\,Fabbrichesi,\aff{INFN_Trieste}
C.\,Grojean,\twoaff{DESY}{Berlin}
U.\,Haisch,\twoaff{CERN}{Oxford}
Y.\,Jiang,\aff{NBIA}
J.\,Kamenik,\twoaff{JSI_Ljubljana}{U_Ljubljana}
M.\,Mangano,\aff{CERN}
D.\,Marzocca,\aff{INFN_Trieste}
E.\,Mereghetti,\aff{LosAlamos}
K.\,Mimasu,\aff{UCLouvain}
L.\,Moore,\aff{UCLouvain}
G.\,Perez,\aff{Weizmann}
T.\,Plehn,\aff{Heidelberg}
F.\,Riva,\aff{CERN}
M.\,Russell,\aff{Heidelberg}
J.\,Santiago,\aff{CAFPE_Granada}
M.\,Schulze,\aff{Berlin}
Y.\,Soreq,\aff{MIT}
A.\,Tonero,\aff{UNIFAL}
M.\,Trott,\aff{NBIA}
S.\,Westhoff,\aff{Heidelberg}
C.\,White,\aff{QueenMary}
A.\,Wulzer,\threeaff{CERN}{EPFL}{Padova_INFN}
J.\,Zupan.\aff{Cincinati}
\\[5mm]\small
\naff{Granada}{Departamento de Física Teórica y del Cosmos, U.\ de Granada, E-18071 Granada, Spain}
\naff{CERN}{CERN, Theoretical Physics Department, Geneva 23 CH-1211, Switzerland}
\naff{DESY}{DESY, Notkestraße 85, D-22607 Hamburg, Germany}
\naff{UCLouvain}{Centre for Cosmology, Particle Physics and Phenomenology (CP3), Université catholique de Louvain, B-1348 Louvain-la-Neuve, Belgium}
\naff{IHEP}{Institute of High Energy Physics, Chinese Academy of Sciences, Beijing 100049, China}
\naff{SISSA_INFN}{SISSA and INFN, Sezione di Trieste, via Bonomea 265, 34136 Trieste, Italy}
\naff{NBIA}{Niels Bohr International Academy and Discovery Center, Niels Bohr Institute, University of Copenhagen, DK-2100 Copenhagen, Denmark}
\naff{LosAlamos}{Theoretical Division, Los Alamos National Laboratory, Los Alamos, NM 87545, USA}
\naff{NewMexico}{New Mexico Consortium, Los Alamos Research Park, Los Alamos, NM 87544, USA}
\naff{Nikhef}{Nikhef, Theory Group, Science Park 105, 1098 XG, Amsterdam, The Netherlands}
\naff{Glasgow}{SUPA, School of Physics and Astronomy, University of Glasgow, Glasgow G12 8QQ, UK}
\naff{INFN_Trieste}{INFN, Sezione di Trieste, Via Valerio 2, 34127 Trieste, Italy}
\naff{Berlin}{Institut für Physik, Humboldt-Universität zu Berlin, D-12489 Berlin, Germany}
\naff{Oxford}{Rudolf Peierls Centre for Theoretical Physics, University of Oxford, OX1 3NP Oxford, UK}
\naff{JSI_Ljubljana}{Jožef Stefan Institute, Jamova 39, 1000 Ljubljana, Slovenia}
\naff{U_Ljubljana}{Faculty of Mathematics and Physics, University of Ljubljana, Jadranska 19, 1000 Ljubljana, Slovenia}
\naff{Weizmann}{Department of Particle Physics and Astrophysics, Weizmann Institute of Science,\\Rehovot 7610001, Israel}
\naff{Heidelberg}{Institut für Theoretische Physik, Universität Heidelberg, Germany}
\naff{CAFPE_Granada}{CAFPE and Departamento de Física Teórica y del Cosmos, U.\ de Granada, E-18071 Granada, Spain}
\naff{MIT}{Center for Theoretical Physics, Massachusetts Institute of Technology, Cambridge, MA 02139, USA}
\naff{UNIFAL}{UNIFAL-MG, Rodovia José Aurélio Vilela 11999, 37715-400 Poços de Caldas, MG, Brazil}
\naff{QueenMary}{Centre for Research in String Theory, School of Physics and Astronomy, Queen Mary University of London, 327 Mile End Road, London E1 4NS, UK}
\naff{EPFL}{Institut de Théorie des Phénomènes Physiques, EPFL, Lausanne, Switzerland}
\naff{Padova_INFN}{Dipartimento di Fisica e Astronomia, Universitá di Padova and INFN Padova, Italy}
\naff{Cincinati}{Department of Physics, University of Cincinnati, Cincinnati, Ohio 45221,USA}
}}
\date{}

\begin{document}
\thispagestyle{empty}
\maketitle
\begin{abstract}
This note proposes common standards and prescriptions for the effective-field-theory interpretation of top-quark measurements at the LHC.
\end{abstract}

\newpage
\setcounter{tocdepth}{1}
\tableofcontents

\section{Introduction}

Summarising efforts undertaken under the auspices of the LHC TOP Working Group, this theory note aims at establishing basic common standards for the interpretation of top-quark measurements at the LHC in the standard-model effective field theory (SMEFT).

Guiding principles are first stated (\autoref{sec:principles}). In a nutshell, we rely on the Warsaw basis of gauge-invariant dimension-six operators, focus on the ones which give rise to top-quark interactions (\autoref{sec:operators}) and limit ourselves to the tree level. Three different assumptions about beyond-the-standard-model (BSM) flavour structures are considered to prioritize studies among the otherwise overwhelming number of four-fermion operators (\autoref{sec:flavour}). Top-quark flavour-changing neutral currents (FCNCs) are examined separately (\autoref{sec:fcnc}). For convenience, degrees of freedom are defined from combinations of Warsaw-basis operator coefficients (\hyperref[app:dof]{Appendices\,\ref{app:dof}} and \ref{app:less_restrictive_flavour}). They are aligned with the directions of the effective-field-theory (EFT) parameter space which appear in given processes, in interferences with standard-model (SM) amplitudes, and in top-quark interactions with some of the gauge boson mass eigenstates. Naming and normalization conventions are fixed. Indicative constraints on these degrees of freedom are compiled in \autoref{sec:dof_limits}. Model implementations are provided for tree-level Monte Carlo simulation (\autoref{sec:ufo}). Finally, a simple example of analysis strategy is outlined to illustrate how the challenges of a global EFT treatment could be addressed and to identify the experimental outputs that would be useful in this case (\autoref{sec:strategy}). It rests on the common knowledge established over the years through many EFT studies in the top-quark, Higgs, electroweak or flavour sectors. The literature quoted in this note is by no means comprehensive or even representative of the fields of top-quark physics and EFTs.

\section{Guiding principles}
\label{sec:principles}
\begin{enumerate}\itemsep0pt \topsep0pt \parsep0pt
	\item The so-called Warsaw basis of dimension-six operators~\cite{Grzadkowski:2010es} is adopted. See also Ref.\,\cite{AguilarSaavedra:2008zc, Zhang:2010dr} for early discussions of top-quark related operators.

	\item Our discussion exclusively concerns processes involving at least a top quark. Only operators involving such a particle are considered. Other operators affecting the considered processes are assumed to be well constrained by measurements in processes that do not involve top quarks. This assumption may not always be justified and explicit checks should be performed. It was for instance shown that jet production sufficiently tightly constrain modifications to the triple gluon vertex~\cite{Krauss:2016ely}.

	\item Three different assumptions about BSM flavour structures are considered ---mostly based on symmetries--- to effectively reduce the overwhelming number of four-quark operators~\cite{AguilarSaavedra:2010zi}. A baseline scenario is defined. Less and more restrictive assumptions are also considered. Top-quark FCNC operators are treated separately.
	      
		Minimal flavour violation in the quark sector is used in the baseline scenario, under the assumption of a unit Cabibbo-Kobayashi-Maskawa (CKM) matrix and finite Yukawa couplings only for the top and bottom quarks, i.e.\ we impose a $U(2)_q\times U(2)_u \times U(2)_d$ flavour symmetry among the first two generations. In the lepton sector, flavour diagonality, i.e.\ $[U(1)_{l+e}]^3$, is chosen as baseline.
	      
	\item For convenience, we identify the linear combinations of Warsaw-basis operators that appear in interferences with SM amplitudes and in interactions with physical fields after electroweak symmetry breaking. This may reduce the number of relevant parameters and unconstrained combinations in a given measurement. Independent combinations are defined and referred to as \emph{degrees of freedom} from now on. Normalizations and notations are fixed. We recommend to provide experimental and theoretical results in terms of those parameters in the future. We refrain from stating which degrees of freedom are relevant for which process as such a statement is observable dependent. Different factorization schemes (four- or five-flavour) and approximations (e.g., about the CKM matrix, light fermion masses and Yukawa couplings, or subleading electroweak contributions) would impact numerical results. We recommend to determine systematically the dependence of each observable of interest on the listed degrees of freedom. Some illustrative dependences are provided in \hyperref[sec:dependences]{Appendix\,\ref{sec:dependences}} for total rates. More sophisticated observables may be sensitive to additional degrees of freedom or have enhanced dependences on particular combinations.

	\item As a starting point, we rely on a tree-level description, working at zeroth order in the loop expansion. All tree-level contributions are considered on an equal footing. Hierarchies between them are only derived from the available experimental constraints. The dependence of a specific observable on a given EFT parameter is possibly omitted only if other measurements constrain that parameter much below the level of sensitivity of this observable. As a result, the inclusion of its dependence should affect neither the pre-existing constraints on that parameter, nor the resulting constraints on others, in a combination of this new measurement with existing ones. This approach is very phenomenological and agnostic about specific theories. In practice, as further and further constraints are collected and combined, a picture will progressively emerge of what specific measurement is particularly relevant to constrain a given direction in parameter space.
	      
		The higher-order dependences of observables on SM couplings will be considered in a second step. They could either induce corrections to existing tree-level contributions, or generate a dependence on new EFT parameters. A discussion of next-to-leading order QCD effects on EFT predictions is planned. A variety of results is already available in the literature \cite{Zhang:2014rja, Degrande:2014tta, Rontsch:2014cca, Franzosi:2015osa, Rontsch:2015una, Zhang:2016omx, Bylund:2016phk, Maltoni:2016yxb}.
\end{enumerate}

\section{Operator definitions}
\label{sec:operators}
The definitions of the operators that will be relevant to our discussion ---those which contain a top quark for a suitable flavour assignment--- are collected here. The associated degrees of freedom, following the flavour assumptions detailed in \autoref{sec:flavour}, will be defined in \autoref{app:dof} and used in the rest of this note. The notation employed in this section is that of Ref.~\cite{Grzadkowski:2010es} with flavour indices labelled by $ijkl$; left-handed fermion doublets denoted by $q$, $l$; right-handed fermion singlets by $u$, $d$, $e$; the Higgs doublet by $\varphi$; the antisymmetric $SU(2)$ tensor by $\varepsilon\equiv i\tau^2$; $\tilde{\varphi}=\varepsilon\varphi^*$; $\FDF\equiv \varphi^\dagger(iD_\mu \varphi) - (iD_\mu\varphi^\dagger) \varphi$; $\FDFI\equiv \varphi^\dagger\tau^I(iD_\mu \varphi) - (iD_\mu\varphi^\dagger) \tau^I\varphi$ where $\tau^I$ are the Pauli matrices; $T^A\equiv \lambda^A/2$ where $\lambda^A$ are Gell-Mann matrices.
\pagebreak[3]
\begin{align}
\noalign{Four-quark operators:}
	\qq{1}{qq}{ijkl}
	&= (\bar q_i \gamma^\mu q_j)(\bar q_k\gamma_\mu q_l)
	\label{eq:LLLL_1}
	,\\
	\qq{3}{qq}{ijkl}
	&= (\bar q_i \gamma^\mu \tau^I q_j)(\bar q_k\gamma_\mu \tau^I q_l)
	\label{eq:LLLL_2}
	,\\
	\qq{1}{qu}{ijkl}
	&= (\bar q_i \gamma^\mu q_j)(\bar u_k\gamma_\mu u_l)
	,\\
	\qq{8}{qu}{ijkl}
	&= (\bar q_i \gamma^\mu T^A q_j)(\bar u_k\gamma_\mu T^A u_l)
	,\\
	\qq{1}{qd}{ijkl}
	&= (\bar q_i \gamma^\mu q_j)(\bar d_k\gamma_\mu d_l)
	,\\
	\qq{8}{qd}{ijkl}
	&= (\bar q_i \gamma^\mu T^A q_j)(\bar d_k\gamma_\mu T^A d_l)
	,\\
	\qq{}{uu}{ijkl}
	&=(\bar u_i\gamma^\mu u_j)(\bar u_k\gamma_\mu u_l)
	,\\
	\qq{1}{ud}{ijkl}
	&=(\bar u_i\gamma^\mu u_j)(\bar d_k\gamma_\mu d_l)
	,\\
	\qq{8}{ud}{ijkl}
	&=(\bar u_i\gamma^\mu T^A u_j)(\bar d_k\gamma_\mu T^A d_l)
	,\\
	\hc{\qq{1}{quqd}{ijkl}}
	&=(\bar q_i u_j)\:\varepsilon\;
	  (\bar q_k d_l)
	,\\
	\hc{\qq{8}{quqd}{ijkl}}
	&=(\bar q_iT^A u_j)\;\varepsilon\;
	  (\bar q_kT^A d_l)
	,\\
\noalign{Two-quark operators:}
	\hc{\qq{}{u\varphi}{ij}}
	&=\bar{q}_i u_j\tilde\varphi\: (\varphi^{\dagger}\varphi)
	,\\
	\qq{1}{\varphi q}{ij}
	&=\FDF (\bar{q}_i\gamma^\mu q_j)
	,\\
	\qq{3}{\varphi q}{ij}
	&=\FDFI (\bar{q}_i\gamma^\mu\tau^I q_j)
	,\\
	\qq{}{\varphi u}{ij}
	&=\FDF (\bar{u}_i\gamma^\mu u_j)
	,\\
	\hc{\qq{}{\varphi ud}{ij}}
	&=(\tilde\varphi^\dagger iD_\mu\varphi)
	  (\bar{u}_i\gamma^\mu d_j)
	,\\
	\hc{\qq{}{uW}{ij}}
	&=(\bar{q}_i\sigma^{\mu\nu}\tau^Iu_j)\:\tilde{\varphi}W_{\mu\nu}^I
	,\\
	\hc{\qq{}{dW}{ij}}
	&=(\bar{q}_i\sigma^{\mu\nu}\tau^Id_j)\:{\varphi} W_{\mu\nu}^I
	,\\
	\hc{\qq{}{uB}{ij}}
	&=(\bar{q}_i\sigma^{\mu\nu} u_j)\quad\:\tilde{\varphi}B_{\mu\nu}
	,\\
	\hc{\qq{}{uG}{ij}}
	&=(\bar{q}_i\sigma^{\mu\nu}T^Au_j)\:\tilde{\varphi}G_{\mu\nu}^A
	,\\
\noalign{Two-quark-two-lepton operators:}
	\qq{1}{lq}{ijkl}
	&=(\bar l_i\gamma^\mu l_j)
	  (\bar q_k\gamma^\mu q_l)
	,\\
	\qq{3}{lq}{ijkl}
	&=(\bar l_i\gamma^\mu \tau^I l_j)
	  (\bar q_k\gamma^\mu \tau^I q_l)
	,\\
	\qq{}{lu}{ijkl}
	&=(\bar l_i\gamma^\mu l_j)
	  (\bar u_k\gamma^\mu u_l)
	,\\
	\qq{}{eq}{ijkl}
	&=(\bar e_i\gamma^\mu e_j)
	  (\bar q_k\gamma^\mu q_l)
	,\\
	\qq{}{eu}{ijkl}
	&=(\bar e_i\gamma^\mu e_j)
	  (\bar u_k\gamma^\mu u_l)
	,\\
	\hc{\qq{1}{lequ}{ijkl}}
	&=(\bar l_i e_j)\;\varepsilon\;
	  (\bar q_k u_l)
	,\\
	\hc{\qq{3}{lequ}{ijkl}}
	&=(\bar l_i \sigma^{\mu\nu} e_j)\;\varepsilon\;
	  (\bar q_k \sigma_{\mu\nu} u_l)
	,\\
	\hc{\qq{}{ledq}{ijkl}}
	&=(\bar l_i e_j)
	  (\bar d_k q_l)
	,\\
\noalign{Baryon- and lepton-number-violating operators:\footnote{In the latest version of Ref.\,\cite{Grzadkowski:2010es}, $\qq{1,3}{qqq}{}$ are merged into one single operator with $SU(2)_L$ indices mixed between the two fermion bilinears. The two conventions are technically speaking equivalent~\cite{Abbott:1980zj}.}}
	\hc{\qq{}{duq}{ijkl}}
	&=(\overline{d^c}_{i\alpha}			u_{j\beta})
	  (\overline{q^c}_{k\gamma}	\varepsilon	l_l)
	  \;\epsilon^{\alpha\beta\gamma}
	,\\
	\hc{\qq{}{qqu}{ijkl}}
	&=(\overline{q^c}_{i\alpha}	\varepsilon	q_{j\beta})
	  (\overline{u^c}_{k\gamma}	e_l)
	  \;\epsilon^{\alpha\beta\gamma}
	,\\
	\hc{\qq{1}{qqq}{ijkl}}
	&=(\overline{q^c}_{i\alpha}	\varepsilon	q_{j\beta})
	  (\overline{q^c}_{k\gamma}	\varepsilon	l_l)
	  \;\epsilon^{\alpha\beta\gamma}
	,\\
	\hc{\qq{3}{qqq}{ijkl}}
	&=(\overline{q^c}_{i\alpha}	\tau^I\varepsilon	q_{j\beta})
	  (\overline{q^c}_{k\gamma}	\tau^I\varepsilon	l_l)
	  \;\epsilon^{\alpha\beta\gamma}
	,\\
	\hc{\qq{}{duu}{ijkl}}
	&=(\overline{d^c}_{i\alpha}	u_{j\beta})
	  (\overline{u^c}_{k\gamma}	e_l)
	  \;\epsilon^{\alpha\beta\gamma}
	,
\end{align}
Non-Hermitian operators are marked with a double dagger (only in the above list): $\hc{\qq{}{}{}}$. For Hermitian operators which involve vector Lorentz bilinears, complex conjugation is equivalent to the transposition of generation indices: $\qq{}{}{ij}{}^* = \qq{}{}{ji}$ and by extension, for four-fermion operators, $\qq{}{}{ijkl}{}^* = \qq{}{}{jilk}$.
The corresponding effective Lagrangian then takes the form:
\begin{equation}
\mathcal{L} =
 \sum_a \left( \frac{\cc{}{a}{}}{\Lambda^2}  \hc{\qq{}{a}{}} +\text{h.c.}\right)
+\sum_b \frac{\cc{}{b}{}}{\Lambda^2} \qq{}{b}{}\,,
\label{eq:eft_lagrangian}
\end{equation}
where the conjugates of the Hermitian operators, labelled here with an index $b$, are not added. Conventionally and unless otherwise specified, the arbitrary scale $\Lambda$ will be set to $1\,$TeV. Equivalently, one could consider that numerical values quoted are in units of TeV$^{-2}$ for the dimensionful coefficients $\tilde{C}_i\equiv C_i/\Lambda^2$. It is understood that the implicit sum over flavour indices only includes independent combinations, i.e., the symmetry in flavour space of the operator coefficients is taken into account.

\section{Flavour assumptions}
\label{sec:flavour}
As mentioned in the introduction, prioritizing the study of certain flavour structures is required among the otherwise overwhelming number of four-quark operators.
In the lepton sector, it seems manageable to only assume flavour diagonality,  i.e. $[U(1)_{l+e}]^3$ which includes a separate $U(1)_l\times U(1)_e$ diagonal subgroup for each of the three lepton generations. This leaves independent parameters for each lepton-antilepton pair of a given generation. This assumption adopted as baseline could, for instance, easily be further restricted to $[U(1)_{l} \times U(1)_{e}]^3$, $U(3)_{l+e}$ or even $U(3)_l\times U(3)_e$. As in the quark sector (see below), the third generation of leptons could also be put aside and $U(2)$ symmetries assumed among the first two generations.
We discuss different flavour assumptions for the quark sector in the next three sections, starting with a baseline symmetry (\autoref{sec:41}) subsequently made less and more restrictive (\hyperref[sec:flavour-less-restrictive]{Sections\,\ref{sec:flavour-less-restrictive}} and \ref{sec:flavour-more-restrictive}).

\subsection{Baseline \texorpdfstring{$U(2)_q\times U(2)_u\times U(2)_d$}{U(2)q x U(2)u x U(2)d} scenario}\label{sec:41}

As a baseline flavour scenario in the quark sector and motivated ---as detailed below--- by the minimal flavour violation (MFV) ansatz~\cite{Chivukula:1987py, Hall:1990ac, DAmbrosio:2002vsn}, we impose a $U(2)_q\times U(2)_u\times U(2)_d$ symmetry among the first two generations. The MFV expansion of quark bilinear coefficients is the following:
\begin{flalign}
	\bar q_i q_j&: a_1 \mathbb{I}+a_2 Y_u Y_u^\dagger+a_3 Y_d Y_d^\dagger +\cdots
	\\
        \bar u_i u_j&: b_1 \mathbb{I}+b_2 Y_u^\dagger Y_u+\dots
	\\
	\bar d_i d_j&: c_1 \mathbb{I} + c_2 Y_d^\dagger Y_d +\dots
	\\
	\bar u_i d_j&: d_1 Y_u^\dagger Y_d+\dots
	\\
	\bar q_i u_j&: e_1 Y_u + e_2 Y_u Y_u^\dagger Y_u+e_3 Y_d Y_d^\dagger Y_u+\dots
	\\
	\bar q_i d_j&: f_1 Y_d + f_2 Y_d Y_d^\dagger Y_d+f_3 Y_u Y_u^\dagger Y_d+\dots
\end{flalign}
where $a_1$, $b_1$, etc.\ are order-one coefficients (see Ref.\,\cite{Kagan:2009bn} for the resummation  of large contributions in such series). As a first approximation, we assume a unit CKM matrix, and retain only the top and bottom Yukawa couplings, so that $Y_u=\diag(0,0,y_t)$ and $Y_d=\diag(0,0,y_b)$. When considering low-energy constraints and restoring the full CKM matrix, one would need to worry about whether UV flavour structures are aligned with the up-, down-quark sectors, or in between those limits~\cite{Gedalia:2010zs}. Denoting the left-handed quark doublet and right-handed quark singlets of the third generation as $Q$, $t$, and $b$,
\begin{align*}
        \bar q_i q_i,\ \bar u_i u_i,\ \bar d_i d_i\quad
	&\mbox{bilinears are allowed in the first two generations,}
	\\
        \bar Q Q,\ \bar t t,\ \bar b b,\ 
        \bar t b,\ \bar Q t,\ \bar Q b\quad
	&\mbox{bilinears are allowed in the third generation,}
\end{align*}
under the above assumptions.
The coefficients of the first-generation bilinears do not depend on the
$i\in\{1,2\}$ index which is thus implicitly summed over. Fierz transformations may be required on four-fermion operators to bring such quark-antiquark pairs in the same Lorentz bilinear.
Equivalently, a $U(2)_q\times U(2)_u\times U(2)_d$ symmetry is assumed between the first two quark generations and no restriction is imposed on the third-generation bilinears. This assumption simplifies four-fermion operators but does not affect third-generation two-fermion ones.
Compared to flavour diagonality, i.e. $[U(1)_{q+u+d}]^3$, which would just force quarks and antiquarks to appear in same-flavour pairs, $U(2)_q\times U(2)_u\times U(2)_d$ effectively imposes the following additional requirements:
\begin{enumerate}
\item the right-handed charged currents of the first generations ($\bar u d$, $\bar du$) are forbidden, \label{item:light-r-cc}
\item the chirality-flipping bilinears of the first generations ($\bar qu$, $\bar qd$) are forbidden, \label{item:light-ch-flip}
\item the coefficients of the bilinears of the first and second generations are forced to be identical.
\end{enumerate}
The $U(2)_q\times U(2)_u\times U(2)_d$ flavour symmetry assumption is used by default in this note where not otherwise specified. The following numbers of degrees of freedom are produced for the operators of each category of field content:
\begin{center}
\begin{tabular}[t]{r@{\quad}l}
four heavy quarks & $11+2$ CPV\\
two light and two heavy quarks & $14$\\
two heavy quarks and bosons & $9+6$ CPV\\
two heavy quarks and two leptons & $(8+3\text{ CPV}) \times 3$ lepton flavours
\end{tabular}
\end{center}
where we counted separately CP-conserving and CP-violating (CPV) parameters. They are constructed explicitly in \autoref{app:dof} and listed in \autoref{tab:limits} together with their definitions in terms of Warsaw-basis operator coefficients.

Finally, a more restrictive variant of this $U(2)_q\times U(2)_u\times U(2)_d$ scenario would retain only the four-fermion operators and exclude the operators with two heavy quarks and bosons. This would be justified when heavy bosons only couple to the SM fermions, so that the low-energy effects are dominated by the tree-level exchanges of heavy mediators between fermionic currents.

\subsection{Less restrictive \texorpdfstring{$U(2)_{q+u+d}$}{U(2)q+u+d} scenario}
\label{sec:flavour-less-restrictive}
In order to allow for the light-quark bilinears listed in \autoref{item:light-r-cc} and \autoref{item:light-ch-flip} above, one can limit the flavour symmetry imposed to $U(2)_{q+u+d}$ only, the diagonal subgroup of $U(2)_q\times U(2)_u\times U(2)_d$. The additional $10+10$~CPV degrees of freedom that then appear for operators containing two light and two heavy quarks are discussed in \autoref{app:less_restrictive_flavour}.

\subsection{More restrictive \emph{top-philic} scenario}
\label{sec:flavour-more-restrictive}
A more restrictive \emph{top-philic} scenario is not obtained by imposing a specific flavour symmetry but rather by assuming that new physics couples dominantly to the left-handed doublet and right-handed up-type quark singlet of the third generation as well as to bosons. All possible operators with this field content are thus constructed. Purely bosonic operators which lead to flavour-universal effects are discarded. A projection onto the Warsaw basis is subsequently performed, notably by employing the equations of motion to trade operators with more derivatives for operators with more fields. In this process, the CKM matrix is again approximated by a unit matrix and all Yukawa couplings but the top and bottom ones are neglected. Only a limited number of independent Lorentz and colour structures are generated. In terms of the degrees of freedom of the baseline scenario, the ones generated in this case are:
\begin{align}
&\hspace{-7mm}
	\ccc{[I]}{t\varphi}{}	,\qquad
	\ccc{-}{\varphi q}{}	,\qquad
	\ccc{3}{\varphi q}{}	,\qquad
	\ccc{}{\varphi t}{}	,\qquad
	\ccc{[I]}{tW}{}		,\qquad
	\ccc{[I]}{tB}{}		,\qquad
	\ccc{[I]}{tG}{}		,\qquad
	\\
&\hspace{-7mm}
	\ccc{[I]}{\varphi tb}{}	\quad\text{and}\quad
	\ccc{[I]}{bW}{}		\qquad\text{appear proportional to $y_b$}
	\\
&\hspace{-7mm}
	\ccc{1}{QQ}{}	,\qquad
	\ccc{8}{QQ}{}	,\qquad
	\ccc{1}{Qt}{}	,\qquad
	\ccc{8}{Qt}{}	,\qquad
	\ccc{1}{tt}{}	,\\
\ccc{}{QDW}{}&
	= \ccc{3,1}{Qq}{}
	= \ccc{3}{Ql}{\ell},\\
\ccc{}{QDB}{}&
	= 6\ccc{1,1}{Qq}{}
	= \frac{3}{2}\ccc{1}{Qu}{}
	= -3\ccc{1}{Qd}{}
	= -3\ccc{1}{Qb}{}
	= -2\ccc{1}{Ql}{\ell}
	= - \ccc{}{Qe}{\ell},\\
\ccc{}{tDB}{}&
	= 6\ccc{1}{tq}{}
	= \frac{3}{2}\ccc{1}{tu}{}
	= -3\ccc{1}{td}{}
	= -3\ccc{1}{tb}{}
	= -2\ccc{}{tl}{\ell}
	= - \ccc{}{te}{\ell},\\
\ccc{}{QDG}{}&
	= \ccc{8}{Qq}{}
	= \ccc{8}{Qu}{}
	= \ccc{8}{Qd}{}
	= \ccc{8}{Qb}{},\\[2mm]
\ccc{}{tDG}{}&
	= \ccc{8}{tq}{}
	= \ccc{8}{tu}{}
	= \ccc{8}{td}{}
	= \ccc{8}{tb}{}.
\end{align}
All other degrees of freedom then vanish. Counting the degrees of freedom in this scenario, distinguishing operator categories by their field content, one obtains:
\begin{center}
\begin{tabular}[t]{r@{\quad}l}\\
two heavy quarks and bosons & $9+6$ CPV\\
four heavy quarks & $5$\\
four fermions & $5$.
\end{tabular}
\end{center}

\section{Example of EFT analysis strategy}
\label{sec:strategy}

An example of analysis strategy is sketched in this section to illustrate how the challenges of global EFT interpretations could be addressed and to highlight what would, in that case, be useful experimental outputs. It relies on fiducial observables defined at, and unfolded to, the particle level.\footnote{A higher level, like the parton one, could be envisioned but would enhance the model-dependence of the unfolding. If advantageous, the \emph{forward folding} of particle-level predictions to the detector level could be considered instead.\label{foot:unfold-level}} We stress that other strategies could be equally suitable, more practical to implement in some cases or lead to better sensitivities, by avoiding unfolding. As our aim is illustrative only and as it is in any case difficult to make prescriptions applicable to any type of experimental analysis, we refrain from considering other possibilities in detail. We expect, however, many of the points raised and recommendations made to apply more generally to other analysis strategies. A wealth of EFT analyses has for instance been carried out in the Higgs and electroweak sectors at the LHC and treats issues related to those discussed here (see e.g.\ Refs.\,\cite{Butter:2016cvz, Murphy:2017omb, DiVita:2017eyz, Falkowski:2016cxu, Brivio:2016fzo, Englert:2015hrx}).

The approach presented here is meant to be simultaneously practical and useful on a long-term basis. Importantly, it allows for relatively precise reinterpretations without full detector simulation. It could thus easily benefit from future improvements in the accuracy of our EFT predictions and adapt to the evolving picture of global EFT constraints or to changes in underlying assumptions. It should allow to derive robust global constraints and be applicable in a wide variety of situations but could however not lead to the tightest constraints on individual operators. Our rationale is that global constraints have more value than individual ones. A global analysis covers systematically the theory space in direct vicinity of the SM and is able to identify new physics through correlated deviations in precise measurements. A tentative \emph{recipe} could be the following:

\begin{enumerate}
\item Define observables at the particle level in a fiducial phase-space volume close enough to the detector level so as to make the unfolding model independent.
	\begin{enumerate}
	\item Several bins of a differential distribution would qualify as examples of observables. A total rate would qualify too, as well as ratios~\cite{Plehn:2015cta, Schulze:2016qas}, asymmetries, etc.
	\item One could also consider observables that are based on multivariate analysis (MVA) techniques. The binned MVA output then yields suitable observables (or, e.g., one bin in several MVA outputs, each maximizing sensitivity to different operator coefficients). The trained classifier(s) should then be provided, for instance as \texttt{C++} code taking as input a particle-level event sample or kinematic variables.
	\item Statistically optimal observables~\cite{Atwood:1991ka, Davier:1992nw, Diehl:1993br}, similar to matrix element methods (see also Ref.\,\cite{Brehmer:2016nyr, Brehmer:2017lrt}), could be very useful from both theoretical and experimental points of view. Their definitions rely on firm theoretical bases, and encode our physical understanding instead of requiring a resource-intensive and opaque training. They moreover constitute a discrete set exactly sufficient to maximally exploit the available kinematic information on which the effect of higher-order corrections (see e.g.\ Ref.\,\cite{Gritsan:2016hjl}) or systematic uncertainties can be transparently studied. Applications in experimental analyses include anomalous triple gauge coupling studies at LEP~\cite{Abbiendi:2003mk, Achard:2004ji, Schael:2004tq, Abdallah:2010zj} and, recently, CP properties of the Higgs boson in di-tau final state~\cite{Aad:2016nal}.
	\end{enumerate}

\item Unfold the measurement of these observables, as well as the estimates for the various SM contributions, to the particle level.
      \label{item:composition}
	\begin{enumerate}
	\item For a suitably defined fiducial region, the particle and detector levels should be sufficiently close to each other for the unfolding to be performed under the SM hypothesis only. Full simulation at various EFT parameter points would then be avoided altogether. The validity of this approximation should however be checked explicitly. \label{item:fiducial-goodness}
	\item The SM contributions would include both the \emph{signal} and \emph{backgrounds} of a SM measurement. It is important to detail the background composition as these processes could have some EFT dependence which may be neglected at first but which could be desirable to account for at some later point in time. \label{item:sm_contrib}
	\end{enumerate}

\item For the various measurements, provide the statistical and systematics likelihoods, error source breakdown, their correlations, and whether they follow a flat or Gaussian distribution.
      
\item[] Notice that the information above should be sufficient for anybody able to generate an EFT (or any NP) sample at the particle level to set constraints. In order to obtain global EFT constraints, one could further proceed as follows.

\item \label{item:dim-6-expansion} Compute, numerically or analytically, for each observable $O^k$, the linear $S_i$ and quadratic $S_{ij}$ contributions of dimension-six operators, in addition to the SM contributions $B_l$ already mentioned in \autoref{item:sm_contrib}. Schematically, the expansion in powers of dimension-six operator coefficients can the be written as:
      \begin{equation}
      O^k = B_l^k 
      	+ \frac{C_i}{\Lambda^2} S_i^k 
      	+ \frac{C_i C_j}{\Lambda^4} S_{ij}^k
      	+ \cdots
      \label{eq:observable}
      \end{equation}
      Quadratic and higher powers of dimension-six operator coefficients can be generated at tree level when several operator insertions are possible in one amplitude. The expansion of normalized distributions, ratios, asymmetries, or of the total width appearing in top-quark propagators also generates such higher powers. In some cases, the interference between SM amplitudes and EFT ones could be suppressed (see for instance helicity selection rules discussed in Ref.\,\cite{Azatov:2016sqh} and references therein) or even vanishingly small (in the case of FCNCs for instance). The dominant contribution could then arise at the quadratic level. Operators of dimension larger than six, in particular that of dimension-eight which were not considered in this note, also start contributing at order $1/\Lambda^4$ in the EFT expansion.
	
	\begin{enumerate}
	\item At tree level, with \mg, if a different coupling order \texttt{EFT$_\texttt{i}$} is given to each of the relevant $N$ operator coefficients $C_i$, the computation of $S_i^k$ and $S_{ij}^k$ can most easily be done by generating at most $N+N(N+1)/2$ samples (see \autoref{sec:ufo} for more details, and for the \texttt{DIM6\_i\^{}2==1} syntax of \mg~\cite{Alwall:2014hca} in particular). If the approximation in \autoref{item:fiducial-goodness} holds, these samples do not need to be passed through detector simulation.

	\item As explained in \autoref{sec:principles}, it is in principle desirable to include all operators that contribute at some given loop order, say first at the tree level. Exceptions could be made for operators that are much better constrained than the others, at the time at which the limits are extracted. This comparison of the relative strength of different measurements could however evolve with time and the goodness of this approximation be re-evaluated.
	\end{enumerate}

\item Within some statistical framework, use the measured $O^k$, the estimated $B^k_l$ with their statistical and systematic uncertainties, and the $S_i^k$, $S_{ij}^k$ to derive global constraints on the $C_i$ operator coefficients.
      
	\begin{enumerate}
	\item It is instructive to also quote individual constraints, set by considering one operator at a time. A comparison between global and individual constraints gives some indication about the magnitude of approximate degeneracies between EFT parameters.
	
	\item For easy combination with other measurements, provide the full covariance matrices of uncertainties on the $C_i$ determination in the Gaussian approximation. The eigenvectors of its inverse corresponding to vanishing eigenvalues indicate which directions of the EFT parameter space remain unconstrained by the measured observables. Other eigenvectors and eigenvalues provide the constraints in other directions.
	
	\item \label{item:linear-and-quadratic} Repeat this procedure twice, with and without including the $S_{ij}^k$ quadratic EFT contributions. The comparison between those two sets of results can explicitly establish where quadratic dimension-six EFT contributions are subleading compared to linear ones. Remarkably, when the linear dependence dominates, the constraints obtained can easily be translated from one basis of dimension-six operators to the other and are therefore of greater generality.
	
	\end{enumerate}

\item \label{item:validity-Ecut} For the purpose of this note, let us define our EFT description to be valid in the regime where it is dominated by dimension-six operator contributions, at the linear level or at some finite higher order. Determining the relative magnitude of the coefficients of operators of various dimensions requires a specific model or power counting. Moreover, the magnitudes of operator-coefficient contributions to a specific observable also depend on $E$, the characteristic energy scale of the process examined. In a production process at the LHC, the reconstructed partonic centre-of-mass energy is often, theoretically, such a suitable quantity. Other proxies may also be considered like $H_T$, the scalar sum of all transverse energies. Their correlation with the physically relevant variables should then be studied, as done for instance in Ref.\,\cite{Falkowski:2016cxu} (see Fig.\,2 and related discussion). In decays, this characteristic scale $E$ is the mass of the decaying particle.
      
      When it is practically feasible, displaying the variation of the limit as a function of an upper cut on $E$ at $E_\text{cut}$ (see \autoref{fig:lim-Ecut}) allows for the valid interpretation of EFT results in a larger class of models~\cite{Brehmer:2015rna, Contino:2016jqw}: featuring a new-physics scale lower than the nominal $E$, or leading to significant contributions from operators of dimension larger than six when no $E_\text{cut}$ is applied.
      
      Interpreting EFT results in specific models proceeds by expressing each $C_i/\Lambda^2$ in terms of the model couplings and mass scales, through a so-called \emph{matching} procedure. The EFT description of a resonance of mass $M$ would be dominated by the operators of lowest dimension when $M\gg E_\text{cut}$. Stricter constraints may apply for instance if the resonance has a large width.

\begin{figure}
\begin{minipage}[b]{.5\textwidth}\centering
	\includegraphics[scale=1]{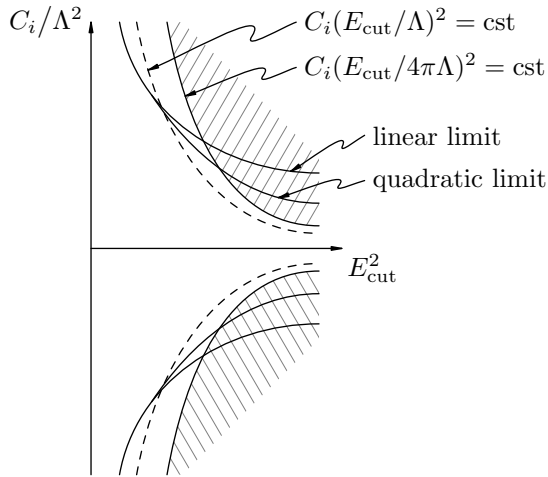}
\end{minipage}\hfill
\begin{minipage}[b]{.49\textwidth}
\caption{Illustration of the limit set on an EFT parameter as function of a cut on the characteristic energy scale of the process considered (see \autoref{item:validity-Ecut}). Qualitatively, one expects the limits to be progressively degraded as $E_\text{cut}$ is pushed towards lower and lower values. At high cut values, beyond the energy directly accessible in the process considered, a plateau should be reached. The regions excluded when the dimension-six EFT is truncated to linear and quadratic orders are delimited by solid lines (see \autoref{item:linear-and-quadratic}). The hatched regions indicate where the dimension-six EFT loses perturbativity (see \autoref{item:perturbativity}). In practice, curves will not be symmetric with respect to $C_i/\Lambda^2=0$.}
\label{fig:lim-Ecut}
\end{minipage}%
\end{figure}

\item \label{item:perturbativity} Besides validity, the quantum perturbativity of the dimension-six EFT can be examined using the same picture. This perturbativity is a necessary condition for the validity (defined as in the previous item) and can be established without reference to a specific model or power counting.
	\begin{enumerate}

	\item This quantum perturbativity of the dimension-six EFT is examined through the sum of contributions involving more and more operator insertions in diagrams with higher and higher numbers of loops. The convergence of such a series (involving one additional operator insertion per loop) requires that $C_i (E_\text{cut}/4\pi\Lambda)^2$ be smaller than a constant which would roughly be of order one if the normalization of $C_i$ is \emph{natural} to the observable considered. The exact condition would need to be determined on a case-by-case basis. It defines a hyperbola in the $(C_i/\Lambda^2,E_\text{cut}^2)$ plane. At tree level, all allowed powers of $C_i/\Lambda^2$ can moreover become relevant as soon as $C_i (E_\text{cut}/\Lambda)^2$ approaches a constant, possibly of order one. If contributions from higher dimensional operators at the same order in $1/\Lambda$ had similar magnitudes, they would also become relevant at that point.
	      
	\item If it exists, the point at which the perturbativity hyperbola crosses the limit curves establishes a minimal $E_\text{cut}$ that has to be imposed for the limit computed perturbatively to make sense. For a sufficiently tight constraint, the limit curves would only cross the perturbativity hyperbola for a cut value beyond the maximal energy directly accessible in the process considered.
	      
	\item If it exists, the point at which the $C_i(E_\text{cut}/\Lambda)^2=\text{\it constant}$ hyperbola crosses the limit curve provides a sense of where higher-order terms in the tree-level expansion in powers of dimension-six operator coefficients could become relevant. In particular, one may expect the linear and quadratic limits to start diverging around that point.
	      
	\item When discussing the relevance of higher-order contributions, definite statements are difficult to make, as is always the case when trying to estimate the terms not computed in a truncated series. As mentioned in \autoref{item:dim-6-expansion}, it is worth bearing in mind the first term could for instance have an accidental suppression as it generally arises from the interference of a SM amplitude with an EFT one.
	\end{enumerate}

\end{enumerate}

\section{Summary and outlook}
In this note, we established the first bases of a framework for the EFT interpretation of top-quark measurements at the LHC. Experimental collaborations and theorists are encouraged to use this common language in future publications to facilitate comparisons and combinations. Having mentioned the caveats associated with those choices, we limited our scope to dimension-six operators of the Warsaw basis which contain a top quark, prioritized the study of possible flavour structures with consistent flavour scenarios, defined the degrees of freedom relevant in each case, compiled existing indicative constraints on these parameters, discussed the use of simulation tools to extract their dependence, and tabulated several benchmark results. An example of analysis strategy was also sketched to illustrate one possible way in which the challenges of a global EFT analysis could be addressed. The experimental outputs that are in that case desirable were highlighted. It is expected that the statements made in the discussion of this example could be transposed to concrete analyses and adapted to different strategies.

Having laid those bases, we did not address more involved issues like next-to-leading order corrections (especially in QCD), theoretical uncertainties, or the treatment of unstable tops. These important topics may deserve further discussion in the LHC TOP Working Group. It is left to the communities of both theorists and experimentalists to study the observables that are best suited to constrain particular direction of the EFT parameter space and to examine the complementarity between them as well as between different processes. Combining the constraints from various sources or determining which dependences are relevant in each observable are tasks that should also be addressed to make progress in the program outlined here.
Ultimately, it will be important to combine top-quark, Higgs and electroweak measurements within the SMEFT. Studies targeting these two other sectors are discussed in the corresponding LHC HXS and EW Working Groups. Several methodological points raised in this note, such as those discussed in \autoref{sec:strategy}, have also been addressed in these contexts, see for example Ref.\,~\cite{deFlorian:2016spz}.

Finally, as stressed several times in this note, the global EFT picture will necessarily evolve with time as new measurements and more accurate predictions become available. Some of the assumptions made here may therefore need to be revised in the future as a finer picture is obtained.

\section*{Acknowledgements}
We would like to warmly thank the experimentalists of the ATLAS and CMS Collaborations having provided us with extensive feedback on the content of this note, the participants to the LHC TOP Working Group meetings in which it was discussed, and the conveners of the Working Group for their support.

\appendix

\begin{table}
\renewcommand{\arraystretch}{1.2}%
\vspace{-3cm}
	\begin{tabular}{@{}l@{}ll}
	\multicolumn{2}{@{}l@{}}{Four-heavy ($11+2\text{ CPV}$ d.o.f.)}
		& Indicative direct limits
	\\[0mm]\hline\noalign{\vskip1mm}
	$\ccc{1}{QQ}{}$
		&\warsaw{2\cc{1}{qq}{3333}-\frac{2}{3}\cc{3}{qq}{3333}}
	\\  
	$\ccc{8}{QQ}{}$
		&\warsaw{8\cc{3}{qq}{3333}}
	\\
	!$\ccc{+}{QQ}{}$
		&\warsaw{\cc{1}{qq}{3333}+\cc{3}{qq}{3333}}
		& $[-2.92, 2.80]$ ($E_{cut}= 3\,$ TeV) \cite{Zhang:2017mls}
	\\
	$\ccc{1}{Qt}{}$
		&\warsaw{\cc{1}{qu}{3333}}
		& $[-4.97,4.90]$ ($E_{cut}= 3\,$TeV) \cite{Zhang:2017mls}
	\\
	$\ccc{8}{Qt}{}$
		&\warsaw{\cc{8}{qu}{3333}}
		& $[-10.3,9.33]$ ($E_{cut}= 3\,$TeV) \cite{Zhang:2017mls}
	\\
	$\ccc{1}{Qb}{}$
		&\warsaw{\cc{1}{qd}{3333}}
	\\
	$\ccc{8}{Qb}{}$
		&\warsaw{\cc{8}{qd}{3333}}
	\\
	$\ccc{1}{tt}{}$
		&\warsaw{\cc{}{uu}{3333}}
		& $[-2.92,2.80]$ ($E_{cut}= 3\,$TeV) \cite{Zhang:2017mls}
	\\
	$\ccc{1}{tb}{}$
		&\warsaw{\cc{1}{ud}{3333}}
	\\
	$\ccc{8}{tb}{}$
		&\warsaw{\cc{8}{ud}{3333}}
	\\
	$\ccc{1[I]}{QtQb}{}$
		&\warsaw{\ReIm\{\cc{1}{quqd}{3333}\}}
	\\
	$\ccc{8[I]}{QtQb}{}$
		&\warsaw{\ReIm\{\cc{8}{quqd}{3333}\}}
	\\[0mm]
	\multicolumn{2}{@{}l}{Two-light-two-heavy (14 d.o.f.)}
	\\[0mm]\hline\noalign{\vskip1mm}
	$\ccc{3,1}{Qq}{}$
		&\warsaw{\cc{3}{qq}{ii33} +\frac{1}{6}(\cc{1}{qq}{i33i} -\cc{3}{qq}{i33i})}
		& $[-0.66,1.24]$ \cite{Buckley:2015lku}, $[-3.11,3.10]$ \cite{Zhang:2017mls}
	\\$\ccc{3,8}{Qq}{}$
		&\warsaw{\cc{1}{qq}{i33i}-\cc{3}{qq}{i33i}}
		& $[-6.06,6.73]$ \cite{Zhang:2017mls}
	\\$\ccc{1,1}{Qq}{}$
		&\warsaw{\cc{1}{qq}{ii33} +\frac{1}{6}\cc{1}{qq}{i33i} +\frac{1}{2}\cc{3}{qq}{i33i}}
		& $[-3.13,3.15]$ \cite{Zhang:2017mls}
	\\$\ccc{1,8}{Qq}{}$
		&\warsaw{\cc{1}{qq}{i33i}+3\cc{3}{qq}{i33i}}
		& $[-6.92,4.93]$ \cite{Zhang:2017mls}
	\\$\ccc{1}{Qu}{}$
		&\warsaw{\cc{1}{qu}{33ii}}
		& $[-3.31,3.44]$ \cite{Zhang:2017mls}
	\\$\ccc{8}{Qu}{}$
		&\warsaw{\cc{8}{qu}{33ii}}
		& $[-8.13,4.05]$ \cite{Zhang:2017mls}
	\\$\ccc{1}{Qd}{}$
		&\warsaw{\cc{1}{qd}{33ii}}
		& $[-4.98,5.02]$ \cite{Zhang:2017mls}
	\\$\ccc{8}{Qd}{}$
		&\warsaw{\cc{8}{qd}{33ii}}
		& $[-11.7,9.39]$ \cite{Zhang:2017mls}
	\\$\ccc{1}{tq}{}$
		&\warsaw{\cc{1}{qu}{ii33}}
		& $[-2.84,2.84]$ \cite{Zhang:2017mls}
	\\$\ccc{8}{tq}{}$
		&\warsaw{\cc{8}{qu}{ii33}}
		& $[-6.80,3.49]$ \cite{Zhang:2017mls}
	\\$\ccc{1}{tu}{}$
		&\warsaw{\cc{}{uu}{ii33}+\frac{1}{3} \cc{}{uu}{i33i}}
		& $[-3.62,3.57]$ \cite{Zhang:2017mls}
	\\$\ccc{8}{tu}{}$
		&\warsaw{2 \cc{}{uu}{i33i}}
		& $[-8.05,4.75]$ \cite{Zhang:2017mls}
	\\$\ccc{1}{td}{}$
		&\warsaw{\cc{1}{ud}{33ii}}
		& $[-4.95,5.04]$ \cite{Zhang:2017mls} 
	\\$\ccc{8}{td}{}$
		&\warsaw{\cc{8}{ud}{33ii}}
		& $[-11.8,9.31]$ \cite{Zhang:2017mls} 
	\\[1mm]
	\multicolumn{2}{@{}l@{}}{Two-heavy ($9 + 6$ CPV d.o.f.)}
	\\[0mm]\hline\noalign{\vskip1mm}
	$\ccc{[I]}{t\varphi}{}$
		&\warsaw{\ReIm\{\cc{}{u\varphi}{33}\}}
	\\$\ccc{-}{\varphi q}{}$
		&\warsaw{\cc{1}{\varphi q}{33}-\cc{3}{\varphi q}{33}}
		& $\ccc{1}{\varphi q}{}$
		  $[-3.1,3.1]$ \cite{Buckley:2015lku},
		  $[-8.3,8.6]$ \cite{Englert:2017dev}
	\\$\ccc{3}{\varphi Q}{}$
		&\warsaw{\cc{3}{\varphi q}{33}}
		& $[-4.1,2.0]$ \cite{Buckley:2015lku},
		  $[-8.6,8.3]$ \cite{Englert:2017dev}
	\\$\ccc{}{\varphi t}{}$
		&\warsaw{\cc{}{\varphi u}{33}}
		& $[-9.7,8.3]$ \cite{Buckley:2015lku},
		  $[-9.1,9.1]$ \cite{Englert:2017dev}
	\\$\ccc{[I]}{\varphi tb}{}$
		&\warsaw{\ReIm\{\cc{}{\varphi ud}{33}\}}
	\\$\ccc{[I]}{tW}{}$
		&\warsaw{\ReIm\{\cc{}{uW}{33}\}}
		&  $\ccc{}{tW}{}$
		  $[-4.0,3.5]$ \cite{Buckley:2015lku},
		  $[-4.1,4.1]$ \cite{Englert:2017dev}
	\\$\ccc{[I]}{tZ}{}$
		&\warsaw{\ReIm\{-\sw\cc{}{uB}{33}+\cw\cc{}{uW}{33}\}}
		& $\ccc{}{tB}{}$
		  $[-6.9,4.6]$ \cite{Buckley:2015lku},
		  $[-7.6,7.6]$ \cite{Englert:2017dev}
	\\$\ccc{[I]}{bW}{}$
		&\warsaw{\ReIm\{\cc{}{dW}{33}\}}
	\\$\ccc{[I]}{tG}{}$
		&\warsaw{\ReIm\{\cc{}{uG}{33}\}}
		& $c_{tG}$  $[-1.32,1.24]$ \cite{Buckley:2015lku}
	\\[0mm]
	\multicolumn{3}{@{}l@{}}{Two-heavy-two-lepton ($8+3$ CPV d.o.f. $\times 3$ lepton flavours)}
	\\\hline\noalign{\vskip1mm}
	$\ccc{3}{Ql}{\ell}$
		&\warsaw{\cc{3}{lq}{\ell\ell33}}
	\\$\ccc{-}{Ql}{\ell}$
		&\warsaw{\cc{1}{lq}{\ell\ell33}-\cc{3}{lq}{\ell\ell33}}
	\\$\ccc{}{Qe}{\ell}$
		&\warsaw{\cc{}{eq}{\ell\ell33}}
	\\$\ccc{}{tl}{\ell}$
		&\warsaw{\cc{}{lu}{\ell\ell33}}
	\\$\ccc{}{te}{\ell}$
		&\warsaw{\cc{}{eu}{\ell\ell33}}
	\\$\ccc{S[I]}{t}{\ell}$
		&\warsaw{\ReIm\{\cc{1}{lequ}{\ell\ell33}\}}
	\\$\ccc{T[I]}{t}{\ell}$
		&\warsaw{\ReIm\{\cc{3}{lequ}{\ell\ell33}\}}
	\\$\ccc{S[I]}{b}{\ell}$
		&\warsaw{\ReIm\{\cc{}{ledq}{\ell\ell33}\}}
	\\\hline
\end{tabular}
\caption{Indicative limits on top-quark operator coefficients for $\Lambda=1$ TeV. For details on the fit procedure, information on the input data and set of operators over which the results are marginalised please consult the corresponding references (see also Ref.~\cite{Buckley:2015nca}). Coefficients marked with a `!' are not independent of the ones previously defined.}
\label{tab:limits}
\end{table}

\section{Indicative constraints}
\label{sec:dof_limits}
As a reference, we collect here various limits set by theoretical studies on the relevant degrees of freedom. We note that there is presently no study considering all the operators listed above and therefore no marginalised constraints are available. Typically, the fits considering one operator at a time (or marginalising over a smaller subset of operators) provide more stringent constraints. The limits listed should therefore only serve as a guidance for potential sensitivity studies or fits. Direct limits arising from the top-quark measurements are given in \autoref{tab:limits}, where available. Indirect limits from low-energy observables also exist, some of which can be very strong yet more model dependent. Limits from $B$ decays, dilepton production, electroweak precision observables, as well as electric dipole moments and CP asymmetries are discussed in the following three subsections. 

%
%
\input{bdecays.tex}

\subsection{Indirect constraints from electroweak precision observables}
Electroweak precision observables (EWPOs) provide additional constraints on
the top-quark operators.  These measurements include the $Z$-pole data from LEP
and SLC, the fermion pair production and $W$ pair production from LEP2, $W$ mass
and width measurements from LEP and Tevatron, and other low-energy measurements
such as DIS and atomic parity violation. These measurements do not directly
involve resonant top quarks. The constraints come only from the top-quark
loop-induced contributions in the two-point functions of the electroweak gauge
bosons, which are modified by top-quark operators.  Unlike in the SM, these
contributions are in general UV divergent, and therefore they need to be interpreted
with care.  On the one hand, we aim to get as much information as possible,
while on the other hand, we would like to base our approach only on minimum set
of assumptions, so that the resulting constraints are meaningful for a wide
range of BSM scenarios.  The approach adopted in Refs.~\cite{Greiner:2011tt,
Zhang:2012cd} is an example of a reasonable balance between these two aspects.

Consider the following subset of the two-heavy degrees of freedom
\begin{flalign}
	\ccc{-}{\varphi Q}{},
	\quad \ccc{3}{\varphi Q}{},
	\quad \ccc{}{\varphi t}{},
	\quad \ccc{}{tW}{},
	\quad \ccc{}{tB}{},
	\quad \ccc{}{bW}{}
\end{flalign}
that are relevant in the EWPOs.  One specific contribution, 
\begin{flalign}
	\ccc{-}{\varphi Q}{}+2\ccc{3}{\varphi Q}{}\,,
\end{flalign}
corresponds to $Z\to b\bar b$ and can be constrained at the tree-level.
All the other degrees of freedom enter only through the modification of the self-energies of $W$, $Z$ and $\gamma$ at the loop level.  These contributions consist of two parts:
\begin{description}
	\item[(a)] the UV poles and the corresponding logarithmic terms,
	\item[(b)] the remaining finite terms.
\end{description}
The UV poles in the first part need to be cancelled by the renormalization of the following two non-top operators, to get physical results:
\begin{flalign}
	&\qq{}{\varphi WB}{}=\left( \varphi^\dagger\tau^I\varphi \right)
	W_{\mu\nu}^I B^{\mu\nu}
	\\
	&\qq{}{\varphi D}{}=(\varphi^\dagger D^\mu \varphi)^*(\varphi^\dagger D_\mu \varphi)
\end{flalign}
In other words, the top-quark operators mix into these two
operators.\footnote{This statement is basis dependent.  In this study we use
	the ``HISZ'' basis~\cite{Hagiwara:1993ck}, which is convenient for
	oblique physics effects, but we keep the definitions of $\qq{}{\varphi
	WB}{}$ and $\qq{}{\varphi D}{}$ to be consistent with the Warsaw basis.
	The physics result is basis-independent.  In the Warsaw basis, some of
	the counter terms for the oblique parameters will be provided by other
	operators involving fermion fields.}
After the renormalization, contribution (a) gives a linear $q^2$-dependent
part in all the self-energy functions $\Pi_{VV}(q^2)$. These contributions can
be identified as a change of the $S$ and $T$ parameters.  It is well known that
the EWPOs places stringent bounds on the $S$ and $T$ parameters, and so
naively one could apply these bounds to constrain the top-quark operators.
However, given that $\qq{}{\varphi WB}{}$ and $\qq{}{\varphi D}{}$ are
included to obtain physical results, they need to be incorporated in the
calculations of $S$ and $T$.  One finds, for example for $\ccc{}{tW}{}$,
\begin{flalign}
	\hat S = \frac{c_{\varphi WB}(\mu)v^2}{\Lambda^2}\frac{c_W}{s_W}
	-N_c\frac{g\ccc{}{tW}{}}{4\pi^2}\frac{\sqrt{2}vm_t}{4\Lambda^2}
	\frac{5}{3}\ln\frac{m_t^2}{\mu^2}
	\label{eq:sparam}
\end{flalign}
which prevents us from directly setting bounds on $\ccc{}{tW}{}$, as the value of $c_{\varphi WB}(\mu)$ is unknown. Here $\mu$ is the renormalization scale. Indicative bounds may be obtained by assuming ``no accidental cancellation'' between the two terms, allowing us to set $c_{\varphi WB}(\mu)=0$. However, since the mixing effect exits, the results strongly depend on the scale $\mu$ at which this assumption is made.\footnote{The most constraining bounds obtained in this way corresponds to setting $\mu=\Lambda$ in \autoref{eq:sparam} and assuming $c_{\varphi WB}(\Lambda)=0$. This is often called the renormalization-group-induced bounds in the literature. They are however strongly model-dependent as assumptions about the matching scale are required, and are not consistent with the global EFT picture as a bottom-up approach. Therefore we do not discuss these bounds here and refer the interested readers to Refs.~\cite{Brod:2014hsa, deBlas:2015aea, Feruglio:2016gvd, Feruglio:2017rjo}.}

Nevertheless, useful information can be obtained from contribution (b). By definition the $S$ and $T$ parameters assume a linear $q^2$ dependence for all two-point functions $\Pi_{VV}(q^2)$.  This is not the case when loop-contributions are the dominant ones, in particular the finite terms in (b) are in general not linear $q^2$ functions. On the other hand, the EWPOs contain more information than the $S$ and $T$ parameters.  To extract this information, one needs to abandon the oblique parameter formalism and perform a global fit for all measurements, including the top-loop contributions in all theory predictions. $c_{\varphi WB}(\mu)$ and $c_{\varphi D}(\mu)$ should be included, to obtain physical predictions. In the resulting $\chi^2$, one then marginalizes over the $c_{\varphi WB}(\mu)$ and $c_{\varphi D}(\mu)$ coefficients. The remaining constraints become weaker, but they are more reliable as they do not depend on any specific assumptions on these two coefficients.  One can also check that the constraints obtained in this way are independent of the renormalization scale $\mu$, to confirm that these results are physical.

The details of this analysis can be found in Ref.~\cite{Zhang:2012cd}. We summarize the results on the relevant degrees of freedom, as indicative constraints in \autoref{tab:ewpo}, assuming only one top operator is considered at a time.\footnote{Ref.\,\cite{Efrati:2015eaa} also constrained both $\ccc{-}{\varphi q}{}+2\ccc{3}{\varphi Q}{}$ and $\ccc{-}{\varphi q}{}$ by combining $e^+e^-\to Z\to b\bar{b}$ measurements with single top-quark $t$-channel production at hadron colliders.}

\begin{table}
\renewcommand{\arraystretch}{1.2}%
	\begin{tabular}{l@{\;}l@{\hspace{2cm}}c@{\hspace{2cm}}}
	\multicolumn{2}{l}{Two-heavy}
		&
	\\[0mm]\hline\noalign{\vskip1mm}

	$\ccc{-}{\varphi q}{}+2\ccc{3}{\varphi Q}{}$
		&\warsaw{\cc{1}{\varphi q}{33}+\cc{3}{\varphi q}{33}}
		& $0.016\pm0.021$
	\\$\ccc{-}{\varphi q}{}$
		&\warsaw{\cc{1}{\varphi q}{33}-\cc{3}{\varphi q}{33}}
		& $2.0\pm2.7$
	\\$\ccc{}{\varphi t}{}$
		&\warsaw{\cc{}{\varphi u}{33}}
		& $1.8\pm1.9$
	\\$\ccc{}{tW}{}$
		&\warsaw{\Re\{\cc{}{uW}{33}\}}
		& $-0.4\pm1.2$
	\\$\ccc{}{tB}{}$
		&\warsaw{\Re\{\cc{}{uB}{33}\}}
		& $4.8\pm5.3$
	\\$\ccc{}{bW}{}$
		&\warsaw{\Re\{\cc{}{dW}{33}\}}
		& $11\pm13$
	\\[1mm]\hline
\end{tabular}
\caption{Indicative limits on top-quark operators arising from precision electroweak observables, with $\Lambda=1\,$TeV.}
\label{tab:ewpo}
\end{table}

As a final remark, apart from $O_{\varphi WB}$ and $O_{\varphi D}$, one could certainly include more non-top operators and perform a more global EW fit. The results will be more model-independent, at the cost of weakening the limits one would get after marginalizing over the non-top operators. The reason for only incorporating the $O_{\varphi WB}$ and $O_{\varphi D}$ operators here, is that they are the only ones to which the top-quark operators mix, and therefore the minimum set of operators one has to incorporate in order to extract exactly the part (b) contributions, which are $\mu$ independent and therefore more physical. This however implies that the resulting limits are based on the assumption that the dominant BSM effects are captured by top-quark operators.

%
%
\input{low-energy.tex}

\section{UFO models}
\label{sec:ufo}
The \texttt{dim6top} implementation of the degrees of freedom introduced in this note, as well the \texttt{SMEFTsim} implementation of Warsaw-basis operators will be described in this appendix. Such UFO~\cite{Degrande:2011ua} models can be used for Monte Carlo simulation.

\subsection{The \texttt{dim6top} implementation}
The flavour-, $B$- and $L$-conserving parameters implemented in the \texttt{dim6top} UFO model (available at \url{https://feynrules.irmp.ucl.ac.be/wiki/dim6top}) are listed in \autoref{tab:ufo_parameters}. The FCNC degrees of freedom defined in \autoref{sec:fcnc} and implemented in this model are listed in \autoref{tab:dim6top_fcnc}.

\begin{table}[p]
\vspace*{-3cm}%
\begin{minipage}[b]{.5\textwidth}
\scalebox{.8}{\begin{tabular}{@{}l@{}c@{}l@{}}
	\multicolumn{2}{@{}l}{Four-heavy}
	\\\hline\noalign{\vskip 1mm}
	$\ccc{1}{QQ}{}$ &
		{\tt cQQ1} \\
	$\ccc{8}{QQ}{}$ &
		{\tt cQQ8} \\
	$\ccc{1}{Qt}{}$ &
		{\tt cQt1} \\
	$\ccc{8}{Qt}{}$ &
		{\tt cQt8} \\
	$\ccc{1}{Qb}{}$ &
		{\tt cQb1} \\
	$\ccc{8}{Qb}{}$ &
		{\tt cQb8} \\
	$\ccc{1}{tt}{}$ &
		{\tt ctt1}\\
	$\ccc{1}{tb}{}$ &
		{\tt ctb1}\\
	$\ccc{8}{tb}{}$ &
		{\tt ctb8}\\
	$\ccc{1[I]}{QtQb}{}$ &
		{\tt cQtQb1[I]}
			& ({\tt I} stands for imaginary part)\\
	$\ccc{8[I]}{QtQb}{}$ &
		{\tt cQtQb8[I]}
\\[2mm]
\multicolumn{2}{@{}l}{Two-heavy-two-light}
\\\hline\noalign{\vskip 1mm}
	$\ccc{3,1}{Qq}{}$ &
	{\tt cQq13} & $U(2)_q\times U(2)_u\times U(2)_d$ assumed\\
	$\ccc{3,8}{Qq}{}$ &
	{\tt cQq83} & \\
	$\ccc{1,1}{Qq}{}$ &
	{\tt cQq11}\\
	$\ccc{1,8}{Qq}{}$ &
	{\tt cQq81}\\
	$\ccc{1}{Qu}{}$ &
	{\tt cQu1}\\
	$\ccc{8}{Qu}{}$ &
	{\tt cQu8}\\
	$\ccc{1}{Qd}{}$ &
	{\tt cQd1}\\
	$\ccc{8}{Qd}{}$ &
	{\tt cQd8}\\
	$\ccc{1}{tq}{}$ &
	{\tt ctq1}\\
	$\ccc{8}{tq}{}$ &
	{\tt ctq8}\\
	$\ccc{1}{tu}{}$ &
	{\tt ctu1}\\
	$\ccc{8}{tu}{}$ &
	{\tt ctu8}\\
	$\ccc{1}{td}{}$ &
	{\tt ctd1}\\
	$\ccc{8}{td}{}$ &
	{\tt ctd8}
\\[2mm]
\multicolumn{2}{@{}l}{Two-heavy}
\\\hline\noalign{\vskip 1mm}
	$\ccc{[I]}{t\varphi}{}$&
	{\tt ctp,ctpI} &\\
	$\ccc{-}{\varphi Q}{} $& 
	{\tt cpQM}\\
	$\ccc{3}{\varphi Q}{} $&
	{\tt cpQ3}\\
	$\ccc{}{\varphi t}{} $&
	{\tt cpt}\\
	$\ccc{}{\varphi b}{} $&
	{\tt cpb} & (implemented but involves no top)\\
	$\ccc{[I]}{\varphi tb}{} $&
	{\tt cptb,cptbI}\\
	$\ccc{[I]}{tW}{} $& 
	{\tt ctW,ctWI}\\
	$\ccc{[I]}{tZ}{} $&
	{\tt ctZ,ctZI}\\
	$\ccc{[I]}{bW}{} $&
	{\tt cbW,cbWI} &\\
	$\ccc{[I]}{tG}{} $&
	{\tt ctG,ctGI}
\\[2mm]
\multicolumn{2}{@{}l}{Two-heavy-two-lepton}
\\\hline\noalign{\vskip 1mm}
	$\ccc{3}{Ql}{\ell} $& 
	{\tt cQl3(l)} &
	assuming lepton flavour diagonality:\\
	$\ccc{-}{Ql}{\ell} $& 
	{\tt cQlM(l)} &
	 \quad $\texttt{(l)}\in\{1,2,3\}$\\
	$\ccc{}{Qe}{\ell} $&
	{\tt cQe}\\
	$\ccc{}{tl}{\ell} $&
	{\tt ctl(l)}\\
	$\ccc{}{te}{\ell} $&
	{\tt cte(l)}\\
	$\ccc{S[I]}{t}{\ell} $&
	{\tt ctlS[I](l)} & \\
        $\ccc{T[I]}{t}{\ell} $&
	{\tt ctlT[I](l)}\\
	$\ccc{S[I]}{b}{\ell} $&
        {\tt cblS[I](l)}
\\[2mm]
\multicolumn{3}{@{}l}{Two-heavy-two-light, preserving only $U(2)_{q+u+d}$}
\\\hline\noalign{\vskip 1mm}
	$\ccc{1[I]}{tQqu}{}$	& \texttt{ctQqu1[I]}	\\
	$\ccc{8[I]}{tQqu}{}$	& \texttt{ctQqu8[I]}	\\
	$\ccc{1[I]}{bQqd}{}$	& \texttt{cbQqd1[I]}	\\
	$\ccc{8[I]}{bQqd}{}$	& \texttt{cbQqd8[I]}	\\
	$\ccc{1[I]}{Qtqd}{}$	& \texttt{cQtqd1[I]}	\\
	$\ccc{8[I]}{Qtqd}{}$	& \texttt{cQtqd8[I]}	\\
	$\ccc{1[I]}{Qbqu}{}$	& \texttt{cQbqu1[I]}	\\
	$\ccc{8[I]}{Qbqu}{}$	& \texttt{cQbqu8[I]}	\\
	$\ccc{1[I]}{btud}{}$	& \texttt{cbtud1[I]}	\\
	$\ccc{8[I]}{btud}{}$	& \texttt{cbtud8[I]}	\\\hline
\end{tabular}}
\caption{\texttt{dim6top} UFO model parameter names for the flavour-, $B$- and $L$-conserving degrees of freedom defined in \hyperref[app:dof]{Appendices\,\ref{app:dof}} and \ref{app:less_restrictive_flavour}.}
\label{tab:ufo_parameters}
\end{minipage}\quad
\begin{minipage}[b]{.5\textwidth}
\scalebox{.75}{\begin{tabular}{lrl@{}}
	\multicolumn{2}{@{}l}{One-light-three-heavy}
	\\\hline\noalign{\vskip1mm}
	\ccc{1[I]}{qq}{333a}
		& \texttt{cqq11[I]x333(a)}	
		& $\texttt{(a)}\in\{1,2\}$: light quark gen.
	\\
	\ccc{3[I]}{qq}{333a}
		& \texttt{cqq13[I]x333(a)}
		& \hspace{.5cm}{\tt I}: imaginary part\\
	\ccc{ [I]}{uu}{333a}
		& \texttt{cuu1[I]x333(a)}
		&\\
	\ccc{1[I]}{qu}{333a}
		& \texttt{cqu1[I]x333(a)}
		&\\
	\ccc{8[I]}{qu}{333a}
		& \texttt{cqu8[I]x333(a)}
		&\\
	\ccc{1[I]}{qu}{3a33}
		& \texttt{cqu1[I]x3(a)33}
		&\\
	\ccc{8[I]}{qu}{3a33}
		& \texttt{cqu8[I]x3(a)33}
		&\\
	\ccc{1[I]}{qd}{333a}
		& \texttt{cqd1[I]x333(a)}
		&\\
	\ccc{8[I]}{qd}{333a}
		& \texttt{cqu8[I]x333(a)}
		&\\
	\ccc{1[I]}{qd}{3a33}
		& \texttt{cqd1[I]x3(a)33}
		&\\
	\ccc{8[I]}{qd}{3a33}
		& \texttt{cqd8[I]x3(a)33}
		&\\
	\ccc{1[I]}{ud}{333a}
		& \texttt{cud1[I]x333(a)}
		&\\
	\ccc{8[I]}{ud}{333a}
		& \texttt{cud8[I]x333(a)}
		&\\
	\ccc{1[I]}{ud}{3a33}
		& \texttt{cud1[I]x3(a)33}
		&\\
	\ccc{8[I]}{ud}{3a33}
		& \texttt{cud8[I]x3(a)33}
		&\\
	\ccc{1[I]}{quqd}{333a}
		& \texttt{cquqd1[I]x333(a)}
		&\\
	\ccc{1[I]}{quqd}{33a3}
		& \texttt{cquqd1[I]x33(a)3}
		&\\
	\ccc{1[I]}{quqd}{3a33}
		& \texttt{cquqd1[I]x3(a)33}
		&\\
	\ccc{1[I]}{quqd}{a333}
		& \texttt{cquqd1[I]x(a)333}
		&\\
	\ccc{8[I]}{quqd}{333a}
		& \texttt{cquqd8[I]x333(a)}
		&\\
	\ccc{8[I]}{quqd}{33a3}
		& \texttt{cquqd8[I]x33(a)3}
		&\\
	\ccc{8[I]}{quqd}{3a33}
		& \texttt{cquqd8[I]x3(a)33}
		&\\
	\ccc{8[I]}{quqd}{a333}
		& \texttt{cquqd8[I]x(a)333}
\\[2mm]
\multicolumn{2}{@{}l}{Three-light-one-heavy}
\\\hline\noalign{\vskip1mm}
	\ccc{1,1[I]}{qq}{3a}
		& \texttt{cqq11[I]x3(a)ii}
		&\\
	\ccc{3,1[I]}{qq}{3a}
		& \texttt{cqq13[I]x3(a)ii}
		&\\
	\ccc{1,8[I]}{qq}{3a}
		& \texttt{cqq81[I]x3(a)ii}
		&\\
	\ccc{3,8[I]}{qq}{3a}
		& \texttt{cqq83[I]x3(a)ii}
		&\\
	\ccc{1[I]}{uu}{3a}
		& \texttt{cuu1[I]x3(a)ii}
		&\\
	\ccc{8[I]}{uu}{3a}
		& \texttt{cuu8[I]x3(a)ii}
		&\\
	\ccc{1[I]}{ud}{3a}
		& \texttt{cud1[I]x3(a)ii}
		&\\
	\ccc{8[I]}{ud}{3a}
		& \texttt{cud8[I]x3(a)ii}
		&\\
	\ccc{1[I]}{qu}{3a}
		& \texttt{cqu1[I]x3(a)ii}
		&\\
	\ccc{8[I]}{qu}{3a}
		& \texttt{cqu8[I]x3(a)ii}
		&\\
	\ccc{1[I]}{qu}{a3}
		& \texttt{cqu1[I]xii3(a)}
		&\\
	\ccc{8[I]}{qu}{a3}
		& \texttt{cqu8[I]xii3(a)}
		&\\
	\ccc{1[I]}{qd}{3a}
		& \texttt{cqd1[I]x3(a)ii}
		&\\
	\ccc{8[I]}{qd}{3a}
		& \texttt{cqd8[I]x3(a)ii}
		&
\\[2mm]
\multicolumn{2}{@{}l}{One-light-one-heavy}
\\\hline\noalign{\vskip 1mm}
	\ccc{[I]}{t\varphi}{3a}
		& \texttt{ctp[I]x3(a)}
		& \\
	\ccc{[I]}{t\varphi}{a3}
		& \texttt{ctp[I]x(a)3}
		& \\
	\ccc{-[I]}{\varphi q}{3+a}
		& \texttt{cpQM[I]x3(a)}
		& \\
	\ccc{3[I]}{\varphi q}{3+a}
		& \texttt{cpQ3[I]x3(a)}
		& \\
	\ccc{[I]}{\varphi u}{3+a}
		& \texttt{cpt[I]x3(a)}
		& \\
	\ccc{[I]}{\varphi ud}{3a}
		& \texttt{cptb[I]x3(a)}
		& \\
	\ccc{[I]}{\varphi ud}{a3}
		& \texttt{cptb[I]x(a)3}
		& \\
	\ccc{[I]}{uW}{3a}
		& \texttt{ctW[I]x3(a)}
		& \\
	\ccc{[I]}{uW}{a3}
		& \texttt{ctW[I]x(a)3}
		& \\
	\ccc{[I]}{uZ}{3a}
		& \texttt{ctZ[I]x3(a)}
		& \\
	\ccc{[I]}{uZ}{a3}
		& \texttt{ctZ[I]x(a)3}
		& \\
	\ccc{[I]}{dW}{3a}
		& \texttt{cdW[I]x3(a)}
		& \\
	\ccc{[I]}{dW}{a3}
		& \texttt{cdW[I]x(a)3}
		& \\
	\ccc{[I]}{uG}{3a}
		& \texttt{cdG[I]x3(a)}
		& \\
	\ccc{[I]}{uG}{a3}
		& \texttt{cdG[I]x(a)3}
		&
\\[2mm]
\multicolumn{2}{@{}l}{One-light-one-heavy-two-lepton}
\\\hline\noalign{\vskip1mm}
	\ccc{3}{lq}{\ell,3+a}
		& \texttt{cQl3[I]x(l)x3(a)}
		& $\texttt{(l)}\in\{1,2,3\}$ lepton gen.\\
	\ccc{-}{lq}{\ell,3+a}
		& \texttt{cQlM[I]x(l)x3(a)}
		& \\
	\ccc{}{eq}{\ell,3+a}
		& \texttt{cQe[I]x(l)x3(a)}
		& \\
	\ccc{}{lu}{\ell,3+a}
		& \texttt{ctl[I]x(l)x3(a)}
		& \\
	\ccc{}{eu}{\ell,3+a}
		& \texttt{cte[I]x(l)x3(a)}
		& \\
	\ccc{S}{lequ}{\ell,3a}
		& \texttt{ctlS[I]x(l)x3(a)}
		& \\
	\ccc{S}{lequ}{\ell,a3}
		& \texttt{ctlS[I]x(l)x(a)3}
		& \\
	\ccc{T}{lequ}{\ell,3a}
		& \texttt{ctlT[I]x(l)x3(a)}
		& \\
	\ccc{T}{lequ}{\ell,a3}
		& \texttt{ctlT[I]x(l)x(a)3}
		& \\
	\ccc{S}{ledq}{\ell,3a}
		& \texttt{cblS[I]x(l)x3(a)}
		& (implemented but involve\\
	\ccc{S}{ledq}{\ell,a3}
		& \texttt{cblS[I]x(l)x(a)3}
		& \quad no top FCNC)
\\\hline
\end{tabular}}
\caption{\texttt{dim6top} UFO model parameter names for the FCNC degrees of freedom introduced in \autoref{sec:fcnc}.}
\label{tab:dim6top_fcnc}
\end{minipage}
\end{table}

\paragraph{General comments}
\begin{list}{--}{\topsep 3pt \itemsep 0pt \parsep 3pt}
	\item The implementation is a tree-level one.
	
	\item $\Lambda$ is conventionally fixed to $1\,$TeV. Equivalently, EFT input parameters can be thought of as being the dimensionful $\tilde{c}_i\equiv c_i/\Lambda^2$ expressed in units of TeV$^{-2}$.
        \item The CKM matrix is assumed to be a unit matrix.
	\item The masses of $u,d,s,c,e,\mu$ fermions are set to zero by default.
	\item The unitary gauge is used and Goldstone bosons are removed.
	\item Are only added to the Lagrangian, the Hermitian conjugates of the non-Hermitian operators, and independent flavour assignments.
	
	\item Two versions of the model are made available. In the first one, \texttt{dim6top\_LO\_UFO}, all the flavour-, $B$- and $L$-conserving parameters are assigned the same \emph{coupling order}: \texttt{DIM6=1}. Similarly, all the flavour-changing parameters are assigned one single \emph{coupling order}: \texttt{FCNC=1}.
	
	In a second version, \texttt{dim6top\_LO\_UFO\_each\_coupling\_order}, an individual \emph{coupling order} is additionally assigned to each EFT parameter: the \texttt{cQQ1} and \texttt{cqq11x3331} parameters are for instance assigned \texttt{DIM6\_cQQ1} and \texttt{FCNC\_cqq11x3331} \emph{coupling orders}.
	
	This allows for the selection of individual degrees-of-freedom interferences in \mg, using a syntax such as
	\\\texttt{> generate p p > t t\~{} FCNC=0 DIM6\^{}2==1 DIM6\_ctZ\^{}2==1}
	\\\texttt{> generate p p > t t\~{} FCNC=0 DIM6\^{}2==2 DIM6\_ctZ\^{}2==1 DIM6\_ctW\^{}2==1}
	\\which would for instance respectively yield the interference between SM amplitudes and that in which \texttt{ctZ} is inserted once, and between amplitudes in which \texttt{ctZ} and \texttt{ctW} are inserted once. Specifying the order of the squared amplitude is however not supported yet when decay chains are specified.
	
	A positive \texttt{QED=n} \emph{coupling order} has also been assigned to EFT parameters corresponding to operators involving \texttt{n} Higgs doublet fields in the unbroken phase. Given that the Higgs vacuum expectation value has \texttt{QED$=-$1}, this avoids technical problems related to the appearance of interactions with net negative \texttt{QED} \emph{coupling order}.
	
\end{list}

\paragraph{Syntax}
\begin{itemize}
	\item In \mg, import the model by
	\\\texttt{> import model dim6top\_LO\_UFO}, or
	\\\texttt{> import model dim6top\_LO\_UFO\_each\_coupling\_order}

	\item By default, the bottom quark is massive and the four-flavour scheme is used. Loading the model with a restriction card where \texttt{MB=0} would automatically switch to the five flavour scheme. One can otherwise redefine manually:
	\\\texttt{> define p = p b b\~}
	\\\texttt{> define j = p}
	\\and set \texttt{MB=0} in the \texttt{param\_card}. For consistency, one may then also set \texttt{ymb=0}.

	\item Processes can be generated through commands like
	\\\texttt{> generate p p > t t\~{} FCNC=0 DIM6=1}
	which allows for one (or no) insertion of a parameter of \texttt{DIM6} \emph{coupling order} per amplitude.

	\item To focus on the leading QCD amplitudes, one can also restrict their maximal allowed \texttt{QED} order using for instance:
	\\\texttt{> generate p p > t t\~{} ~~FCNC=0 DIM6=1 QED=0}
	\\\texttt{> generate p p > t t\~{} Z FCNC=0 DIM6=1 QED=1}
\end{itemize}


\subsection{The \texttt{SMEFTsim} implementation}
{\input{smeftsim.tex}}

\subsection{Benchmark results}
\label{sec:dependences}
The linear and quadratic EFT dependences of some total rates are displayed in \hyperref[tab:lin_pair]{Tables~\ref{tab:lin_pair}}-\ref{tab:quad_tjh}. These are meant to provide benchmark results against which simulation procedures could be checked. They are also indicative of which degree of freedom is relevant for which process.

These numbers are obtained using \mg\ (v2.6.0 or 2.6.1)~\cite{Alwall:2014hca} for the $13\,$TeV LHC, with \texttt{nn23lo1} as PDF set. All fermion masses and Yukawa couplings but the top ones (\texttt{MT=172}) are taken vanishing. The five-flavour scheme, default running renormalization and factorization scales are used, and simple $p_T>20\,$GeV, $|\eta|<2.5$, $\Delta R>0.4$ generation cuts are imposed on ($b$-)jets, charged leptons and photons.

All tree-level contributions are included irrespectively of their \texttt{QED} order. Additional dependences may also be generated: with finite fermion masses and Yukawas (especially the bottom ones), when considering other observables (e.g., sensitive to CPV contributions), beyond the tree level (e.g., at NLO in QCD or when accounting for loop-induced Higgs couplings), etc. Cuts may also affect the hierarchies of sensitivities.

Obtained with \texttt{dim6top}, these benchmark results have been cross checked with {\tt SMEFTsim\_A} version 2.0, in the flavour general setup and with the $\{\hat\alpha_\text{ew}, \hat{m}_Z,\hat{G}_F\}$ input scheme (equally valid results can be obtained with set B and/or switching to a $\{\hat{m}_{W}, \hat{m}_Z,\hat{G}_F\}$ input scheme):
\begin{list}{--}{\topsep 3pt \itemsep 0pt \parsep 3pt}
	
	\item A set of dedicated restriction cards is provided with the \texttt{SMEFTsim\_A\_general\_alphaScheme} model to reproduce these numerical results. All of them set to zero the masses and Yukawa couplings of all fermions except the bottom and top quarks. They also approximate the CKM matrix by a unit matrix and fix the input parameters to values consistent with those adopted in \texttt{dim6top}. In particular they modify
	\begin{center}
	\begin{tabular}{llll}
	{\tt MB = ymb = 4.7},& {\tt MT = 172.0},& {\tt MH = 125.0}, & {\tt MW = 79.824360}  \\
	{\tt aEW = 7.818608e-03},& {\tt aS = 0.1184}
	\end{tabular}
	\end{center}
compared to the default settings.
In addition:
	\begin{description}
	\item[\tt restrict\_SMlimit\_top:] sets to zero all the Wilson coefficients. This restriction is available for all the flavour setups.
	\item[\tt restrict\_TopEFT:] sets to 1 the absolute values of all the Wilson coefficients that are relevant for top physics and to 0 the remaining ones. This restriction is available for all the flavour setups, and the list of coefficients retained in each case is given in \autoref{tab:parameters_topEFT_rst}.
	\item[\tt restrict\_(ci)\_top:] with {\tt (ci)} one of the degrees of freedom of \autoref{tab:ufo_parameters} in the notation of \texttt{dim6top}: 
	turns on the combination of Wilson coefficients equivalent to setting {\tt ci = 1}, while vanishing the others.
	\end{description}
 
	\item As \texttt{dim6top} does not include loop-level interactions of the Higgs ($hgg$, $h\gamma\gamma$, or $hZ\gamma$), a meaningful comparison with \texttt{SMEFTsim} would require setting \texttt{SMHLOOP=0}.

	\item \texttt{SMEFTsim} utilizes the opposite convention, compared to \texttt{dim6top}, for the sign of covariant derivatives (schematically: $D_\mu = \partial_\mu + i A_\mu)$. At the linear level, some interferences with SM amplitudes are affected, in particular those proportional to \texttt{ctW[I]} and \texttt{ctZ[I]}.
	
	\item Due to different choices concerning the inclusion of non-independent flavour contractions (only independent flavour assignments are included in \texttt{dim6top} while all are in \texttt{SMEFTsim}), factor-of-two differences arise when considering the coefficients {\tt cQq83, cQq81, cQq13, cQq11, ctu1, ctu8}, that stem from operators with four identical fields: $Q_{qq}^{1,3}$, $O_{uu}$.

	\item Due to the different parametrization of complex coefficients (in (Abs, Arg) rather than (Re, Im)), the evaluation of $S_i^k$ for purely imaginary coefficients can be less accurate in \texttt{SMEFTsim}.

	\item In \texttt{SMEFTsim}, due to the internal implementation of Warsaw basis operators instead of the degrees of freedom specific to top-quark physics, some small interferences may need to be obtained from cancellations between large contributions and thus suffer from larger numerical uncertainties. This concerns for instance the evaluation of the vanishing linear dependence on the colour octet {\tt cQq83} in single top production. Phenomenological consequences may be limited.
\end{list}

\section{Flavour-, \texorpdfstring{$B$}{B}- and \texorpdfstring{$L$}{L}-conserving degrees of freedom}
\label{app:dof}

Let us define the EFT degrees of freedom, operator category by operator category. We recommend to quote results in terms of these degrees of freedom. Their definitions, in terms of Warsaw-basis operator coefficients are also summarized in \autoref{tab:limits}.

\subsection{Four-heavy operators}
We consider first four-quark operators. All vector ($\bar LL\bar LL$, $\bar LL\bar RR$, $\bar RR\bar RR$) operators have real $\cc{}{}{3333}=\cc{}{}{3333}^*$ coefficients, unlike the scalar ($\bar LR\bar LR$) ones. For $\bar LL\bar LL$ operators, one can define the colour singlet and octets that would or not interfere with the QCD amplitudes:\footnote{Note that we wrote matrix equations involving operators as $O=M^To$. The Lagrangian $C^TO$ terms are then equated to $c^To$ to establish the definition of the coefficients of the $o$ operators as $c\equiv M\,C$.}
\begin{equation}
	\begin{pmatrix}
		\qq{1}{qq}{3333}	\\
		\qq{3}{qq}{3333}
	\end{pmatrix}
	=
	\begin{pmatrix}
		1	& -1/3	\\
		0	&  4
	\end{pmatrix}^{\!\!T}
	\left(\begin{array}{@{(}l@{\,}r@{\,}r@{)\:(}l@{\,}r@{\,}r@{)}}
	 \bar Q		\gamma_\mu& & Q	
	&\bar Q		\gamma^\mu& & Q
	\\
	 \bar Q		\gamma_\mu& T^A& Q	
	&\bar Q		\gamma^\mu& T^A& Q
	\end{array}\right),
\end{equation}
where $Q$ represents the third-generation left-handed quark doublet, while $t$ and $b$ will represent the right-handed quark singlets in the unbroken electroweak phase. In the electroweak broken phase, one further obtains:
\begin{equation}
	\begin{pmatrix}
	\qq{1}{qq}{3333}\\
	\qq{3}{qq}{3333}
	\end{pmatrix}
	=
	\begin{pmatrix}
	1	& 1	\\
	2	& -2/3	\\
	0	& 8	\\
	1	& 1		
	\end{pmatrix}^{\!\!T}
	\begin{pmatrix}
	(\bar t\gamma^\mu P_L t)(\bar t\gamma_\mu P_L t)\\
	(\bar t\gamma^\mu P_L t)(\bar b\gamma_\mu P_L b)\\
	(\bar t\gamma^\mu T^A P_L t)(\bar b\gamma_\mu T^A P_L b)\\
	(\bar b\gamma^\mu P_L b)(\bar b\gamma_\mu P_L b)
	\end{pmatrix}
\end{equation}
which is useful to identify the colour singlet and octet combinations of two top and two $b$ quarks, four tops and four $b$'s. We use the combinations leading to $\bar tt\bar bb$ interactions as independent degrees of freedom. No further manipulation is performed on other operators featuring four heavy quarks. One thus defines:\begin{equation}
	\begin{aligned}
	\ccc{1}{QQ}{} &\equiv 2\cc{1}{qq}{3333}-\frac{2}{3}\cc{3}{qq}{3333}, \\
	\ccc{8}{QQ}{} &\equiv 8\cc{3}{qq}{3333},	\\
	\end{aligned}
	\qquad
	\begin{aligned}
	\ccc{1}{Qt}{} &\equiv  \cc{1}{qu}{3333},	\\
	\ccc{8}{Qt}{} &\equiv  \cc{8}{qu}{3333},	\\
	\ccc{1}{Qb}{} &\equiv  \cc{1}{qd}{3333},	\\
	\ccc{8}{Qb}{} &\equiv  \cc{8}{qd}{3333},	\\
	\end{aligned}
	\qquad
	\begin{aligned}
	\ccc{1}{tt}{} &\equiv  \cc{1}{uu}{3333},	\\
	\end{aligned}
	\qquad
	\begin{aligned}
	\ccc{1}{tb}{} &\equiv  \cc{1}{ud}{3333},	\\
	\ccc{8}{tb}{} &\equiv  \cc{8}{ud}{3333},
	\end{aligned}
\end{equation}
and
\begin{equation}
	\ccc{1[I]}{QtQb}{}	\equiv	\ReIm\{\cc{1}{quqd}{3333}\},	\qquad
	\ccc{8[I]}{QtQb}{}	\equiv	\ReIm\{\cc{8}{quqd}{3333}\}.
\end{equation}
In total, there are thus $11+2\text{ CPV}$ degrees of freedom for four-heavy operators. One can also define the combination $\ccc{+}{QQ}{}\equiv \cc{1}{qq}{3333}+\cc{8}{qq}{3333} = (3\, \ccc{1}{QQ}{} + \ccc{8}{QQ}{})/6$ which appears in front of operators involving four left-handed top and bottom quarks but is not independent of the ones defined above.

\subsection{Two-light-two-heavy operators}
Scalar operators involving a light-quark current are not allowed by our baseline flavour assumption (see \autoref{app:less_restrictive_flavour} where this assumption is relaxed). The coefficients of the vector $\bar LL\bar LL$ operators $\qq{3,1}{qq}{}$ have the following symmetries under permutations of their generation indices:
\begin{equation*}
		\cc{}{qq}{ijkl} = \cc{}{qq}{klij},
	\qquad	\cc{}{qq}{jilk} = \cc{}{qq}{ijkl}^*,
	\qquad (\text{and thus } \cc{}{qq}{lkji} = \cc{}{qq}{ijkl}^*),
\end{equation*}
which namely implies that $\cc{}{qq}{iijj}$ and $\cc{}{qq}{ijji}$ elements are real.
From the $\cc{}{qq}{ijkl}$ and $\cc{}{qq}{klij}$ combinations, only one is thus
retained in the sum over flavour indices in the EFT
Lagrangian of \autoref{eq:eft_lagrangian}. For each of these two operators,
there are thus two independent assignments of third-generation indices to a
quark-antiquark pairs that are compatible with our $U(2)_q\times U(2)_u\times U(2)_d$ baseline flavour symmetry:
\begin{flalign}
	\cc{}{qq}{ii33}=\cc{}{qq}{ii33}^*
	=
	\cc{}{qq}{33ii}=\cc{}{qq}{33ii}^*
	,\qquad 
	\cc{}{qq}{i33i}=\cc{}{qq}{i33i}^*
	=
	\cc{}{qq}{3ii3}=\cc{}{qq}{3ii3}^*
	.
\end{flalign}
To understand better the structure of their interferences with SM amplitudes it is useful to decompose them, using Fierz identities, onto quadrilinears featuring a heavy- and a light-quark bilinear. One obtains:
\begin{equation}
	\begin{pmatrix}
		\qq{1}{qq}{3aii}	\\
		\qq{1}{qq}{3iia}	\\
		\qq{3}{qq}{3aii}	\\
		\qq{3}{qq}{3iia}	\\
	\end{pmatrix}
	=
	\begin{pmatrix}
		1	& 1/6	& 0	&  1/2	\\
		0	& 1/6	& 1	& -1/6	\\
		0	& 1	& 0	&  3	\\
		0	& 1	& 0	& -1
	\end{pmatrix}^{\!\!T}
	\left(\begin{array}{@{(}l@{\,}r@{\,}r@{)\:(}l@{\,}r@{\,}r@{)}}
	 \bar Q		\gamma_\mu& & Q	
	&\bar q_i	\gamma^\mu& & q_i
	\\
	 \bar Q		\gamma_\mu& &\tau^I Q
	&\bar q_i	\gamma^\mu& &\tau^I q_i
	\\
	 \bar Q		\gamma_\mu& T^A& Q	
	&\bar q_i	\gamma^\mu& T^A& q_i
	\\
	 \bar Q		\gamma_\mu& T^A&\tau^I Q	
	&\bar q_i	\gamma^\mu& T^A&\tau^I q_i
	\end{array}\right)
	,
\end{equation}
and similarly, for the $\bar RR\bar RR$ operator $\qq{}{uu}{}$ :
\begin{equation}
	\begin{pmatrix}
	\qq{}{uu}{ii33}\\
	\qq{}{uu}{i33i}
	\end{pmatrix}
	=
	\begin{pmatrix}
	1	& 1/3\\
	0	& 2
	\end{pmatrix}^{\!\!T}
	\left(\begin{array}{@{(}l@{\,}r@{\,}r@{)\:(}l@{\,}r@{\,}r@{)}}
	 \bar t		\gamma_\mu& & t	
	&\bar u_i	\gamma^\mu& & u_i
	\\
	 \bar t		\gamma_\mu& T^A& t	
	&\bar u_i	\gamma^\mu& T^A& u_i
	\end{array}\right)
	.
\end{equation}
There are four four-quark operators of $\bar LL\bar RR$ type in the Warsaw basis: $\qq{1}{qu}{ijkl}$, $\qq{8}{qu}{ijkl}$, $\qq{1}{qd}{ijkl}$, $\qq{8}{qd}{ijkl}$ which form six degrees of freedom when accounting for the two $\cc{}{qu}{ii33}=\cc{}{qu}{ii33}^*$ and $\cc{}{qu}{33ii}=\cc{}{qu}{33ii}^*$ flavour assignments. Altogether, one thus defines the following degrees of freedom for two-light-two-heavy operators:
\begin{gather}
\begin{aligned}
	\ccc{1,1}{Qq}{}&\equiv
		\cc{1}{qq}{ii33}
		+\frac{1}{6}\cc{1}{qq}{i33i}
		+\frac{1}{2}\cc{3}{qq}{i33i},
	\\
	\ccc{3,1}{Qq}{}&\equiv
		\cc{3}{qq}{ii33}
		+\frac{1}{6}(\cc{1}{qq}{i33i}-\cc{3}{qq}{i33i}),
	\\
	\ccc{1,8}{Qq}{}&\equiv
		\cc{1}{qq}{i33i}+3\cc{3}{qq}{i33i},
	\\
	\ccc{3,8}{Qq}{}&\equiv
		\cc{1}{qq}{i33i}-\cc{3}{qq}{i33i},
\end{aligned}
\qquad
\begin{aligned}
	\ccc{1}{tu}{} &\equiv \cc{}{uu}{ii33} +\frac{1}{3} \cc{}{uu}{i33i},	\\
	\ccc{8}{tu}{} &\equiv 2 \cc{}{uu}{i33i},	\\
	\ccc{1}{td}{} &\equiv \cc{1}{ud}{33ii},		\\
	\ccc{8}{td}{} &\equiv \cc{8}{ud}{33ii},		\\
\end{aligned}
\qquad
\begin{aligned}
	\ccc{1}{tq}{}	&\equiv\cc{1}{qu}{ii33},		\\
	\ccc{1}{Qu}{}	&\equiv\cc{1}{qu}{33ii},		\\
	\ccc{1}{Qd}{}	&\equiv\cc{1}{qd}{33ii},		\\
	\ccc{8}{tq}{}	&\equiv\cc{8}{qu}{ii33},		\\
	\ccc{8}{Qu}{}	&\equiv\cc{8}{qu}{33ii},		\\
	\ccc{8}{Qd}{}	&\equiv\cc{8}{qd}{33ii},		\\
\end{aligned}
\end{gather}
where $i$ is either $1$ or $2$. In total, there are thus $14$ degrees of freedom for two-light-two-heavy operators.

\subsection{Two-heavy operators}
The operator involving two quarks and boson which possibly contain a top quark where listed in \autoref{sec:operators}. There are a few choices to be made in the definitions of the associated degrees of freedom, in view of the electroweak broken phase decomposition of $\qq{}{\varphi q}{}$
\begin{equation}
	\begin{pmatrix}
	\qq{1}{\varphi q}{33}\\
	\qq{3}{\varphi q}{33}
	\end{pmatrix}
	=
	\begin{pmatrix}
	-1	& 1	\\
	1	& 1	\\
	0	& 1	\\
	0	& 1
	\end{pmatrix}^{\!\!T}
	\left(\begin{array}{@{}c*{3}{@{\,}c}@{}}
	\frac{+e}{2\sw\cw}	&(\bar t\gamma^\mu P_L t)	&Z_\mu	&(v+h)^2\\
	\frac{-e}{2\sw\cw}	&(\bar b\gamma^\mu P_L b)	&Z_\mu	&(v+h)^2\\
	\frac{e}{\sw\sqrt{2}}	&(\bar t\gamma^\mu P_L b)	&W^+_\mu&(v+h)^2\\
	\frac{e}{\sw\sqrt{2}}	&(\bar b\gamma^\mu P_L t)	&W^-_\mu&(v+h)^2
	\end{array}\right),
\end{equation}
and of electroweak dipole operators:
\begin{equation}
	\begin{pmatrix}
	\qq{}{uB}{33}\\
	\qq{}{uW}{33}
	\end{pmatrix}
	=
	\begin{pmatrix}
	\cw	& \sw	\\
	-\sw	& \cw	\\
	0	& 2	
	\end{pmatrix}^{\!\!T}
	\left(\begin{array}{@{}c*{2}{@{\,}c}@{}}
		(\bar t\sigma^{\mu\nu} P_R t)	&A_{\mu\nu}	&(v+h)\\
		(\bar t\sigma^{\mu\nu} P_R t)	&Z_{\mu\nu}	&(v+h)\\
		(\bar b\sigma^{\mu\nu} P_R t)	&W^-_{\mu\nu}	&(v+h)
	\end{array}\right)
	,
\end{equation}
where $\sw$ and $\cw$ are the sine and cosine of the weak mixing angle (in the unitary gauge). Among the possible definitions, we choose to use as degrees of freedom the combinations that involve charged $W$, $Z$ bosons and tops in the broken phase. In this sector, one thus defines:
\begin{gather}
	\begin{aligned}
	\ccc{[I]}{t\varphi}{}	&\equiv \ReIm\{\cc{}{u\varphi}{33}\},	\\
	\end{aligned}
	\qquad
	\begin{aligned}
	\ccc{-}{\varphi Q}{}	&\equiv \cc{1}{\varphi q}{33}-\cc{3}{\varphi q}{33},	\\
	\ccc{3}{\varphi Q}{}	&\equiv \cc{3}{\varphi q}{33},	\\
	\ccc{}{\varphi t}{}	&\equiv \cc{}{\varphi u}{33},	\\
	\ccc{[I]}{\varphi tb}{}	&\equiv \ReIm\{\cc{}{\varphi ud}{33}\},	\\
	\end{aligned}
	\qquad
	\begin{aligned}
	\ccc{[I]}{tW}{}		&\equiv \ReIm\{\cc{}{uW}{33}\} ,	\\
	\ccc{[I]}{tZ}{}		&\equiv \ReIm\{-\sw\cc{}{uB}{33}+\cw\cc{}{uW}{33}\},\\
	\ccc{[I]}{bW}{}		&\equiv \ReIm\{\cc{}{dW}{33}\} ,	\\
	\ccc{[I]}{tG}{}		&\equiv \ReIm\{\cc{}{uG}{33}\} .
	\end{aligned}
\end{gather}
On the other hand, the combination of $\qq{1,3}{\varphi q}{33}$ operators that modifies the SM coupling of the $b$ quark to the $Z$ is $\ccc{+}{\varphi Q}{} \equiv \cc{3}{\varphi q}{33}+\cc{1}{\varphi q}{33} = \ccc{-}{\varphi Q}{}+2\ccc{3}{\varphi Q}{}$ and the combination appearing electromagnetic dipole of the top is $\ccc{[I]}{tA}{} \equiv \ReIm\{\cw \cc{}{uB}{33}+\sw \cc{}{uW}{33}\} = (\ccc{[I]}{tW}{}-\cw \ccc{[I]}{tZ}{})/\sw$. In total, there are thus $9+6\text{ CPV}$ degrees of freedom for two-heavy operators.

\subsection{Two-heavy-two-lepton operators}
Let us know address the definition of the degrees of freedom associated to the operators involving two quarks and two leptons. We decompose the $\qq{1,3}{lq}{}$ operators in the broken phase,
\begin{equation}
	\begin{pmatrix}
	\qq{1}{lq}{\ell\ell33}\\
	\qq{3}{lq}{\ell\ell33}
	\end{pmatrix}
	=
	\begin{pmatrix}
	1	& 1	\\
	1	& -1	\\
	1	& -1	\\
	1	& 1	\\
	0	& 2	\\
	0	& 2	
	\end{pmatrix}^{\!\!T}
	\begin{pmatrix}
	(\bar\nu_\ell	\gamma^\mu P_L\nu_\ell)(\bar t\gamma_\mu P_Lt)\\
	(\bar\nu_\ell	\gamma^\mu P_L\nu_\ell)(\bar b\gamma_\mu P_Lb)\\
	(\bar\ell	\gamma^\mu P_L\ell    )(\bar t\gamma_\mu P_Lt)\\
	(\bar\ell	\gamma^\mu P_L\ell    )(\bar b\gamma_\mu P_Lb)\\
	(\bar\nu_\ell	\gamma^\mu P_L\ell    )(\bar b\gamma_\mu P_Lt)\\
	(\bar\ell	\gamma^\mu P_L\nu_\ell)(\bar t\gamma_\mu P_Lb)\\
	\end{pmatrix},
\end{equation}
and select as degrees of freedom the combinations that give rise to a top-quark interaction with a pair of charged leptons, and to charged currents. One therefore defines:
\begin{equation}
	\begin{aligned}
	\ccc{-}{Ql}{\ell}	&\equiv \cc{1}{lq}{\ell\ell33}-\cc{3}{lq}{\ell\ell33},\\
	\ccc{3}{Ql}{\ell}	&\equiv \cc{3}{lq}{\ell\ell33}	,\\
	\end{aligned}
	\qquad
	\begin{aligned}
	\ccc{}{tl}{\ell}	&\equiv \cc{}{lu}{\ell\ell33} 	,\\
	\ccc{}{Qe}{\ell}	&\equiv \cc{}{eq}{\ell\ell33}	,\\
	\end{aligned}
	\qquad
	\begin{aligned}
	\ccc{}{te}{\ell}	&\equiv \cc{}{eu}{\ell\ell33}	,\\
	\end{aligned}
	\qquad
	\begin{aligned}
	\ccc{S[I]}{t}{\ell}	&\equiv \ReIm\{\cc{1}{lequ}{\ell\ell33}\}	,\\
	\ccc{T[I]}{t}{\ell}	&\equiv \ReIm\{\cc{3}{lequ}{\ell\ell33}\}	,\\
	\ccc{S[I]}{b}{\ell}	&\equiv \ReIm\{\cc{3}{ledq}{\ell\ell33}\}	.
	\end{aligned}
\end{equation}
These constitute $8+3\text{ CPV}$ degrees of freedom, for each of the three generations of leptons.

\section{Less restrictive flavour symmetry}
\label{app:less_restrictive_flavour}
We examine here the consequences of relaxing the flavour symmetry among the first two generations of quarks from $U(2)_q\times U(2)_u\times U(2)_d$ to $U(2)_{q+u+d}$. New contributions are only generated in the two-light--two-heavy category of four-quark operators as flavour-diagonal chirality-flipping light quark-antiquark pairs as well as right-handed charged currents within the first generations are now allowed.

\subsection{Four-quark \texorpdfstring{$\bar LL\bar RR$}{LLRR} operators}
For this category of operators, the $(i33i)$ and $(3ii3)$ flavour assignments become allowed:
\begin{flalign}
	\cc{1}{qu}{i33i}=\cc{1}{qu}{3ii3}^*
	,
	\qquad
	\cc{8}{qu}{i33i}=\cc{8}{qu}{3ii3}^*
	,\\[1mm]
	\cc{1}{qd}{i33i}=\cc{1}{qd}{3ii3}^*
	,
	\qquad
	\cc{8}{qd}{i33i}=\cc{8}{qd}{3ii3}^*
	.
\end{flalign}
Using Fierz identities, one then obtains:
\begin{equation}
	\begin{pmatrix}
	\qq{1}{qu}{i33i}\\
	\qq{8}{qu}{i33i}
	\end{pmatrix}
	=
	\begin{pmatrix}
	-2/3	&-8/9\\
	-4	&2/3
	\end{pmatrix}^{\!\!T}
	\left(\begin{array}{@{(}l@{}c@{}c@{}l@{)\:(}l@{}c@{}c@{}l@{)}}
	 \bar t		&	&	&Q
	&\bar q_i	&	&	&u_i
	\\
	 \bar t		&	&T^A	&Q
	&\bar q_i	&	&T^A	&u_i
	\end{array}\right)
\end{equation}
and similarly for $\qq{}{qd}{}$ operators. The corresponding degrees of freedom
\begin{equation}
\begin{aligned}
	\ccc{1[I]}{tQqu}{}
	&=\ReIm\{
		-\frac{2}{3} \cc{1}{qu}{i33i} -\frac{8}{9} \cc{8}{qu}{i33i}
	\},\\
	\ccc{1[I]}{bQqd}{}
	&=\ReIm\{
		-\frac{2}{3} \cc{1}{qd}{i33i} -\frac{8}{9} \cc{8}{qd}{i33i}
	\},
\end{aligned}
\qquad
\begin{aligned}
	\ccc{8[I]}{tQqu}{}
	&=\ReIm\{
		-4\cc{1}{qu}{i33i} +\frac{2}{3} \cc{8}{qu}{i33i}
	\},\\
	\ccc{8[I]}{bQqd}{}
	&=\ReIm\{
		-4\cc{1}{qd}{i33i} +\frac{2}{3} \cc{8}{qd}{i33i}
	\},
\end{aligned}
\end{equation}
are complex.

\subsection{Four-quark \texorpdfstring{$\bar LR\bar LR$}{LRLR} operators} The $\qq{}{quqd}{}$ operators belong to that category. Proceeding to the Fierz decomposition:
\begin{equation}
	\begin{pmatrix}
	\qq{1}{quqd}{i33i}	\\
	\qq{1}{quqd}{3ii3}	\\[2mm]
	\qq{8}{quqd}{i33i}	\\
	\qq{8}{quqd}{3ii3}	
	\end{pmatrix}
	=
	\begin{pmatrix}
	-1/6	&0	&-2/9	&0\\
	-1	&0	&1/6	&0\\
	-1/24	&0	&-1/18	&0\\
	-1/4	&0	&1/24	&0\\[2mm]
	0	&-1/6	&0	&-2/9\\
	0	&-1	&0	&1/6\\
	0	&-1/24	&0	&-1/18\\
	0	&-1/4	&0	&1/24
	\end{pmatrix}^{\!\!T}
	\left(\begin{array}{@{(}l@{}c@{}c@{}l@{)\:\varepsilon\:(}l@{}c@{}c@{}l@{)}}
	 \bar Q		&	&	&t
	&\bar q_i	&	&	&d_i
	\\
	 \bar Q		&	&T^A	&t
	&\bar q_i	&	&T^A	&d_i
	\\
	 \bar Q		&\sd	&	&t
	&\bar q_i	&\su	&	&d_i
	\\
	 \bar Q		&\sd	&T^A	&t
	&\bar q_i	&\su	&T^A	&d_i
	\\[2mm]
	 \bar Q		&	&	&b
	&\bar q_i	&	&	&u_i
	\\
	 \bar Q		&	&T^A	&b
	&\bar q_i	&	&T^A	&u_i
	\\
	 \bar Q		&\sd	&	&b
	&\bar q_i	&\su	&	&u_i
	\\
	 \bar Q		&\sd	&T^A	&b
	&\bar q_i	&\su	&T^A	&u_i
	\end{array}\right),
\end{equation}
one can for instance select the scalar operator coefficients as independent degrees of freedom:
\begin{equation}
	\begin{aligned}
	\ccc{1[I]}{Qtqd}{}
	&=\ReIm\{
		-\frac{1}{6} \cc{1}{quqd}{i33i} -\frac{2}{9} \cc{8}{quqd}{i33i}
	\},\\
	\ccc{1[I]}{Qbqu}{}
	&=\ReIm\{
		-\frac{1}{6} \cc{1}{quqd}{3ii3} -\frac{2}{9} \cc{8}{quqd}{3ii3}
	\},
	\end{aligned}
	\qquad
	\begin{aligned}
	\ccc{8[I]}{Qtqd}{}
	&=\ReIm\{
		-\cc{1}{quqd}{i33i} +\frac{1}{6} \cc{8}{quqd}{i33i}
	\},\\
	\ccc{8[I]}{Qbqu}{}
	&=\ReIm\{
		-\cc{1}{quqd}{3ii3} +\frac{1}{6} \cc{8}{quqd}{3ii3}
	\},
	\end{aligned}
\end{equation}
which are complex.

\subsection{Four-quark \texorpdfstring{$\bar RR\bar RR$}{RRRR} operators}
In this category, new structures only arise for $\qq{}{ud}{}$ operators. The Fierz decomposition gives:
\begin{equation}
	\begin{pmatrix}
	\qq{1}{ud}{i33i}	\\
	\qq{8}{ud}{i33i}	
	\end{pmatrix}
	=
	\begin{pmatrix}
	1/3	&4/9\\
	2	&-1/3
	\end{pmatrix}^{\!\!T}
	\left(\begin{array}{@{(}l@{}c@{}c@{}l@{)\:(}l@{}c@{}c@{}l@{)}}
	 \bar b		&\gamma_\mu	&	&t
	&\bar u_i	&\gamma^\mu	&	&d_i
	\\
	 \bar b		&\gamma_\mu	&T^A	&t
	&\bar u_i	&\gamma^\mu	&T^A	&d_i
	\end{array}\right),
\end{equation}
and two additional complex degrees of freedom can be defined:
\begin{equation}
\ccc{1[I]}{btud}{}
=\ReIm\{
	\frac{1}{3} \cc{1}{ud}{i33i} +\frac{4}{9} \cc{8}{ud}{i33i}
\},
\qquad
\ccc{8[I]}{btud}{}
=\ReIm\{
	2\cc{1}{ud}{i33i} -\frac{1}{3} \cc{8}{ud}{i33i}
\}.
\end{equation}

\section{FCNC degrees of freedom}
\label{sec:fcnc}
We now somewhat relax the benchmark flavour symmetry imposed to consider the important case of top-quark FCNC interactions. We therefore assume the $U(2)_q\times U(2)_u\times U(2)_d$ is broken by a small parameter connecting generations $3$ and $a\in\{1,2\}$. In each operator, a quark-antiquark pair mixing these two generations arises at the linear order in this breaking parameter. We restrict ourselves to this order and require the other quark bilinear in a four-quark operator to still satisfy the $U(2)_q\times U(2)_u\times U(2)_d$ symmetry, possibly after a Fierz transformation has been applied. We now construct the degrees of freedom satisfying this condition, operator type by operator type.

\subsection{One-light-one-heavy operators}
The FCNC operators in this category are trivially obtained from the flavour conserving ones. Note Hermitian couplings $\cc{}{}{ji}^*=\cc{ij}{}{}$ have complex off-diagonal components. The degrees of freedom are the following:
\begin{equation}
\begin{aligned}
\ccc{[I]}{t\varphi}{3a} &\equiv \ReIm\{\cc{}{u\varphi}{3a}\},\\
\ccc{[I]}{t\varphi}{a3} &\equiv \ReIm\{\cc{}{u\varphi}{a3}\},
\end{aligned}
\qquad
\begin{aligned}
\ccc{-[I]}{\varphi q}{3+a}  &\equiv \ReIm\{\cc{1}{\varphi q}{3a}-\cc{3}{\varphi q}{3a}\},\\
\ccc{3[I]}{\varphi q}{3+a}  &\equiv \ReIm\{\cc{3}{\varphi q}{3a}\},\\
\ccc{[I]}{\varphi u}{3+a}  &\equiv \ReIm\{\cc{}{\varphi u}{3a}\},\\
\ccc{[I]}{\varphi ud}{3a} &\equiv \ReIm\{\cc{}{\varphi ud}{3a}\},\\
\ccc{[I]}{\varphi ud}{a3} &\equiv \ReIm\{\cc{}{\varphi ud}{a3}\},\\
\end{aligned}
\qquad
\begin{aligned}
\ccc{[I]}{uW}{3a} &\equiv \ReIm\{\cc{}{uW}{3a}\},\\
\ccc{[I]}{uW}{a3} &\equiv \ReIm\{\cc{}{uW}{a3}\},\\
\ccc{[I]}{uZ}{3a} &\equiv \ReIm\{-\sw\cc{}{uB}{3a}+\cw\cc{}{uW}{3a}\},\\
\ccc{[I]}{uZ}{a3} &\equiv \ReIm\{-\sw\cc{}{uB}{a3}+\cw\cc{}{uW}{a3}\},\\
\ccc{[I]}{dW}{3a} &\equiv \ReIm\{\cc{}{dW}{3a}\},\\
\ccc{[I]}{dW}{a3} &\equiv \ReIm\{\cc{}{dW}{a3}\},\\
\ccc{[I]}{uG}{3a} &\equiv \ReIm\{\cc{}{uG}{3a}\},\\
\ccc{[I]}{uG}{a3} &\equiv \ReIm\{\cc{}{uG}{a3}\}.\\
\end{aligned}
\end{equation}
Altogether, these are $15$ CP-conserving degrees of freedom and $15$ CP-violating ones.

\subsection{One-light-one-heavy-two-lepton operators}
There is no difficulty either to list the FCNC degrees of freedom for operators containing two quarks and two leptons:
\begin{equation}
\begin{aligned}
\ccc{3}{lq}{\ell,3+a} &\equiv \ReIm\{\cc{3}{lq}{\ell\ell 3a}\},\\
\ccc{-}{lq}{\ell,3+a} &\equiv \ReIm\{\cc{-}{lq}{\ell\ell 3a}\},\\
\ccc{}{eq}{\ell,3+a} &\equiv \ReIm\{\cc{}{eq}{\ell\ell 3a}\},\\
\ccc{}{lu}{\ell,3+a} &\equiv \ReIm\{\cc{}{lu}{\ell\ell 3a}\},\\
\ccc{}{eu}{\ell,3+a} &\equiv \ReIm\{\cc{}{eu}{\ell\ell 3a}\},\\
\end{aligned}
\qquad
\begin{aligned}
\ccc{S}{lequ}{\ell,3a} &\equiv \ReIm\{\cc{1}{lequ}{\ell\ell 3a}\},\\
\ccc{S}{lequ}{\ell,a3} &\equiv \ReIm\{\cc{1}{lequ}{\ell\ell a3}\},\\
\ccc{T}{lequ}{\ell,3a} &\equiv \ReIm\{\cc{3}{lequ}{\ell\ell 3a}\},\\
\ccc{T}{lequ}{\ell,a3} &\equiv \ReIm\{\cc{3}{lequ}{\ell\ell a3}\},\\
\end{aligned}
\end{equation}
These form $9+9\text{ CPV}$ degrees of freedom.

\subsection{One-light-three-heavy operators}

We consider first operators involving three third-generation and one first- or second-generation quarks.
The $\bar LL\bar LL$ operators featuring one light field and three heavy ones
satisfy $\qq{1,3}{qq}{333a} =\qq{1,3}{qq}{3a33} =\qq{1,3}{qq}{33a3}^*
=\qq{1,3}{qq}{a333}^*$. So we impose the same equality on their coefficients and
retain only one of them as independent complex parameter. In the broken
electroweak phase, these operators are decomposed as follows:
\begin{equation}
	\begin{pmatrix}
		\qq{1}{qq}{333a}
		\\\qq{3}{qq}{333a}
	\end{pmatrix}
	=
	\begin{pmatrix}
		1	& 1	\\
		1/3	& 5/3	\\
		2	& -2	\\
		1/3	& 5/3	\\
		2	& -2	\\
		1	& 1
	\end{pmatrix}^{\!\!T}
	\left(\begin{array}{@{(}l@{\,}r@{)\:(}l@{\,}r@{)}}
	 \bar t		\gamma_\mu& P_L t	
	&\bar t		\gamma^\mu& P_L u_a	
	\\
	 \bar t		\gamma_\mu& P_L b	
	&\bar b		\gamma^\mu& P_L u_a	
	\\
	 \bar t		\gamma_\mu& T^A P_L b	
	&\bar b		\gamma^\mu& T^A P_L u_a	
	\\
	 \bar b		\gamma_\mu& P_L t	
	&\bar t		\gamma^\mu& P_L d_a	
	\\
	 \bar b		\gamma_\mu& T^A P_L t	
	&\bar t		\gamma^\mu& T^A P_L d_a	
	\\
	 \bar b		\gamma_\mu& P_L b	
	&\bar b		\gamma^\mu& P_L d_a
	\end{array}\right).
\label{eq:fcnc_lll_1l3h}
\end{equation}
Note here that only the $\bar tt\bar tu$ and $\bar bb\bar bd$ field combinations unambiguously arise from flavour-changing neutral current interactions. Both combinations arise proportional to $\cc{1}{qq}{333a}+\cc{3}{qq}{333a}$. The other field combinations could have been generated by flavour off-diagonal charged currents. The colour singlet $\bar tb\bar bu$ and $\bar bt\bar td$ operators for instance interfere with CKM-suppressed SM charged currents, while the octets do not.
The single $\bar RR\bar RR$ operator contributing to three top production also
satisfies $\qq{}{uu}{333a} =\qq{}{uu}{3a33} =\qq{}{uu}{33a3}^* =\qq{}{uu}{a333}^*$ while $\qq{}{ud}{333a} = \qq{}{ud}{33a3}^*$ and $\qq{}{ud}{3a33} = \qq{}{ud}{a333}^*$.
Similarly the $\bar LL\bar RR$ operators satisfies $\qq{1,8}{qu}{333a} =
\qq{1,8}{qu}{33a3}^*$ and $\qq{1,8}{qu}{3a33} = \qq{1,8}{qu}{a333}^*$.
The $\bar LR\bar LR$ operators have no symmetry of this kind.
So, the relevant degrees of freedom are
\begin{gather}
\begin{aligned}
	\ccc{1[I]}{qq}{333a}	&=	\ReIm\{\cc{1}{qq}{333a}\}	,\\
	\ccc{3[I]}{qq}{333a}	&=	\ReIm\{\cc{3}{qq}{333a}\}	,\\
	\ccc{ [I]}{uu}{333a}	&=	\ReIm\{\cc {}{uu}{333a}\}	,\\
\end{aligned}
\qquad
\begin{aligned}
	\ccc{1[I]}{qu}{333a}	&=	\ReIm\{\cc{1}{qu}{333a}\}	,\\
	\ccc{8[I]}{qu}{333a}	&=	\ReIm\{\cc{8}{qu}{333a}\}	,\\
	\ccc{1[I]}{qu}{3a33}	&=	\ReIm\{\cc{1}{qu}{3a33}\}	,\\
	\ccc{8[I]}{qu}{3a33}	&=	\ReIm\{\cc{8}{qu}{3a33}\}	,
\end{aligned}
\qquad
\begin{aligned}
	\ccc{1[I]}{qd}{333a}	&=	\ReIm\{\cc{1}{qd}{333a}\}	,\\
	\ccc{8[I]}{qd}{333a}	&=	\ReIm\{\cc{8}{qd}{333a}\}	,\\
	\ccc{1[I]}{qd}{3a33}	&=	\ReIm\{\cc{1}{qd}{3a33}\}	,\\
	\ccc{8[I]}{qd}{3a33}	&=	\ReIm\{\cc{8}{qd}{3a33}\}	,
\end{aligned}
\\[3mm]
\begin{aligned}
	\ccc{1[I]}{ud}{333a}	&=	\ReIm\{\cc{1}{ud}{333a}\}	,\\
	\ccc{8[I]}{ud}{333a}	&=	\ReIm\{\cc{8}{ud}{333a}\}	,\\
	\ccc{1[I]}{ud}{3a33}	&=	\ReIm\{\cc{1}{ud}{3a33}\}	,\\
	\ccc{8[I]}{ud}{3a33}	&=	\ReIm\{\cc{8}{ud}{3a33}\}	,
\end{aligned}
\qquad
\begin{aligned}
	\ccc{1[I]}{quqd}{333a}	&=	\ReIm\{\cc{1}{quqd}{333a}\}	,\\
	\ccc{1[I]}{quqd}{33a3}	&=	\ReIm\{\cc{1}{quqd}{33a3}\}	,\\
	\ccc{1[I]}{quqd}{3a33}	&=	\ReIm\{\cc{1}{quqd}{3a33}\}	,\\
	\ccc{1[I]}{quqd}{a333}	&=	\ReIm\{\cc{1}{quqd}{a333}\}	,
\end{aligned}
\qquad
\begin{aligned}
	\ccc{8[I]}{quqd}{333a}	&=	\ReIm\{\cc{8}{quqd}{333a}\}	,\\
	\ccc{8[I]}{quqd}{33a3}	&=	\ReIm\{\cc{8}{quqd}{33a3}\}	,\\
	\ccc{8[I]}{quqd}{3a33}	&=	\ReIm\{\cc{8}{quqd}{3a33}\}	,\\
	\ccc{8[I]}{quqd}{a333}	&=	\ReIm\{\cc{8}{quqd}{a333}\}	.
\end{aligned}
\end{gather}

\subsection{Three-light-one-heavy operators}
Again, we impose here the presence of a quark-antiquark pair satisfying the $U(2)_q\times U(2)_u \times U(2)_d$ symmetry. The suitable $\bar LL\bar LL$ operators satisfy the $\qq{1,3}{qq}{3aii} = \qq{1,3}{qq}{ii3a} = \qq{1,3}{qq}{a3ii}^* = \qq{1,3}{qq}{iia3}^*$, $\qq{1,3}{qq}{3iia} = \qq{1,3}{qq}{ia3i} = \qq{1,3}{qq}{i3ai}^* = \qq{1,3}{qq}{aii3}^*$ relations. Fierz transformation then lead to
\begin{equation}
	\begin{pmatrix}
		\qq{1}{qq}{3aii}	\\
		\qq{1}{qq}{3iia}	\\
		\qq{3}{qq}{3aii}	\\
		\qq{3}{qq}{3iia}	\\
	\end{pmatrix}
	=
	\begin{pmatrix}
		1	& 1/6	& 0	&  1/2	\\
		0	& 1/6	& 1	& -1/6	\\
		0	& 1	& 0	&  3	\\
		0	& 1	& 0	& -1
	\end{pmatrix}^{\!\!T}
	\left(\begin{array}{@{(}l@{\,}r@{\,}r@{)\:(}l@{\,}r@{\,}r@{)}}
	 \bar Q		\gamma_\mu& & q_a	
	&\bar q_i	\gamma^\mu& & q_i
	\\
	 \bar Q		\gamma_\mu& &\tau^I q_a	
	&\bar q_i	\gamma^\mu& &\tau^I q_i
	\\
	 \bar Q		\gamma_\mu& T^A& q_a	
	&\bar q_i	\gamma^\mu& T^A& q_i
	\\
	 \bar Q		\gamma_\mu& T^A&\tau^I q_a	
	&\bar q_i	\gamma^\mu& T^A&\tau^I q_i
	\end{array}\right)
\end{equation}
and one can thus define the following degrees of freedom:
\begin{align}
	\ccc{1,1[I]}{qq}{3a}&\equiv
	\ReIm\{
	\cc{1}{qq}{3aii}
	+\frac{1}{6}\cc{1}{qq}{3iia}
	+\frac{1}{2}\cc{3}{qq}{3iia}
	\},
	\\
	\ccc{3,1[I]}{qq}{3a}&\equiv
	\ReIm\{
	\cc{3}{qq}{3aii}
	+\frac{1}{6}(\cc{1}{qq}{3iia}-\cc{3}{qq}{3iia})
	\},
	\\
	\ccc{1,8[I]}{qq}{3a} &\equiv
	\ReIm\{
	\cc{1}{qq}{3iia}+3\cc{3}{qq}{3iia}
	\},
	\\
	\ccc{3,8[I]}{qq}{3a} &\equiv
	\ReIm\{
	\cc{1}{qq}{3iia}-\cc{3}{qq}{3iia}
	\},
\end{align}
in a way similar to the flavour conserving case. For $\bar RR\bar RR$ operators, one has first:
\begin{equation}
	\begin{pmatrix}
	\qq{}{uu}{3aii}\\
	\qq{}{uu}{3iia}
	\end{pmatrix}
	=
	\begin{pmatrix}
	1	& 1/3\\
	0	& 2
	\end{pmatrix}^{\!\!T}
	\left(\begin{array}{@{(}l@{\,}r@{\,}r@{)\:(}l@{\,}r@{\,}r@{)}}
	 \bar t		\gamma_\mu& & u_a	
	&\bar u_i	\gamma^\mu& & u_i
	\\
	 \bar t		\gamma_\mu&T^A & u_a	
	&\bar u_i	\gamma^\mu&T^A & u_i
	\end{array}\right)
\end{equation}
in the case of the $\qq{}{uu}{}$ operator whose coefficient with different flavour assignations satisfies the same relations as that of $\qq{}{qq}{}$. One thus defines
\begin{equation}
	\begin{aligned}
	\ccc{1[I]}{uu}{3a} &= \ReIm\{
		\cc{}{uu}{3aii} + \frac{1}{3} \cc{}{uu}{3iia}
		\},\\
	\ccc{8[I]}{uu}{3a} &= \ReIm\{
		2\cc{}{uu}{3iia}
		\},
	\end{aligned}
\end{equation}
while no Fierzing is required for the definition of the remaining $\bar RR\bar RR$ and $\bar LL\bar RR$ degrees of freedom:
\begin{equation}
	\begin{aligned}
	\ccc{1[I]}{ud}{3a}	&= \ReIm\{\cc{1}{ud}{3aii}\},\\
	\ccc{8[I]}{ud}{3a}	&= \ReIm\{\cc{8}{ud}{3aii}\},
	\end{aligned}
	\qquad
	\begin{aligned}
	\ccc{1[I]}{qu}{3a}	&= \ReIm\{\cc{1}{qu}{3aii}\},\\
	\ccc{8[I]}{qu}{3a}	&= \ReIm\{\cc{8}{qu}{3aii}\},\\
	\ccc{1[I]}{qu}{a3}	&= \ReIm\{\cc{1}{qu}{ii3a}\},\\
	\ccc{8[I]}{qu}{a3}	&= \ReIm\{\cc{8}{qu}{ii3a}\},
	\end{aligned}
	\qquad
	\begin{aligned}
	\ccc{1[I]}{qd}{3a}	&= \ReIm\{\cc{1}{qd}{3aii}\},\\
	\ccc{8[I]}{qd}{3a}	&= \ReIm\{\cc{8}{qd}{3aii}\}.
	\end{aligned}
\end{equation}

\subsection{Counting}
In total, the counting of FCNC degrees of freedom is as follows:
\begin{center}
\begin{tabular}{r@{\quad}l}
one-light-one-heavy	& $(15+15\,\text{CPV})\times 2$\\
one-light-one-heavy-two-leptons & $(9+9\,\text{CPV})\times 2 \times 3$\\
one-light-three-heavy	& $(23+23\,\text{CPV})\times 2$\\
three-light-one-heavy	& $(14+14\,\text{CPV})\times 2$\\
\end{tabular}
\end{center}
where the factor of two stands for the light quark flavour $a\in\{1,2\}$, and the factor of three for lepton flavours. One counts $61$ CP-even FCNC degrees of freedom without those flavour factors (and $316$ including all flavour combinations and CP-odd parameters). Global constraints deriving from direct measurements on a subset of these degrees of freedom were set in Ref.\,\cite{Durieux:2014xla}.

\bibliographystyle{apsrev4-1_title}
\raggedright\bibliography{eft_note}


\begin{table}[p]
\vspace{-3cm}\centering
\scalebox{.65}{\input{interferences.tex}}
\caption{Linear dependences on the various degrees of freedom of total top pair production rates at the $13\,$TeV LHC. For convenience, numerical values for the $S^k_j/\sum_l B^k_l$ ratios are provided in permil. Absolute SM rates (i.e.\ $\sum_l B^k_l$) are also quoted in picobarns. LO simulation at the parton level, the five-flavour scheme, NNPDF2.3LO1, running renormalization and factorization scales are used. Simple $p_T>20\,$GeV, $|\eta|<2.5$, $\Delta R>0.4$ cuts are imposed on ($b$-)jets, charged leptons and photons. As input parameters, $m_t=172\,$GeV (all other fermion masses and Yukawa couplings are set to zero) and $m_h=125\,$GeV are notably employed. Monte Carlo uncertainties are of the order of $15\%$, but outliers are expected given the number of quantity evaluated.
}
\label{tab:lin_pair}
\end{table}

\begin{table}[p]
\vspace{-3cm}\centering
\scalebox{.7}{\input{interferences_single_top.tex}}
\caption{Same as \autoref{tab:lin_pair}, for single top production processes.
}
\label{tab:lin_single}
\end{table}



\begin{sidewaystable}[p]
\centering\hspace*{-3cm}
\adjustbox{width=.99\paperheight}{\input{squared_tt.tex}}
\caption{Quadratic dependence on the various degrees of freedom of the total $pp\to t\bar t$ rate. For conveniences, numerical values for $S_{ij}^k/\sum_l B^k_l$ are quoted in permil.  Numerical values smaller than $10^{-5}$ are omitted. For more details see the legend of \autoref{tab:lin_pair}.
}
\label{tab:quad_tt}
\end{sidewaystable}




\begin{sidewaystable}[p]
\centering\hspace*{-3cm}
\adjustbox{width=.99\paperheight}{\input{squared_ttee.tex}}
\caption{Quadratic dependence on the various degrees of freedom of the total $pp\to t\bar te^+e^-$ rate. For conveniences, numerical values for $S_{ij}^k/\sum_l B^k_l$ are quoted in permil. For more details see the legend of \autoref{tab:lin_pair}.
}
\label{tab:quad_ttee}
\end{sidewaystable}

\begin{sidewaystable}[p]
\centering\hspace*{-3cm}
\adjustbox{width=.99\paperheight}{\input{squared_ttev.tex}}
\caption{Quadratic dependence on the various degrees of freedom of the total $pp\to t\bar te^+\nu$ rate. For conveniences, numerical values for $S_{ij}^k/\sum_l B^k_l$ are quoted in permil. For more details see the legend of \autoref{tab:lin_pair}.
}
\label{tab:quad_ttev}
\end{sidewaystable}

\begin{sidewaystable}[p]
\centering\hspace*{-3cm}
\adjustbox{width=.99\paperheight}{\input{squared_tta.tex}}
\caption{Quadratic dependence on the various degrees of freedom of the total $pp\to t\bar t\gamma$ rate. For conveniences, numerical values for $S_{ij}^k/\sum_l B^k_l$ are quoted in permil. For more details see the legend of \autoref{tab:lin_pair}.
}
\label{tab:quad_tta}
\end{sidewaystable}

\begin{sidewaystable}[p]
\centering\hspace*{-3cm}
\adjustbox{width=.99\paperheight}{\input{squared_tth.tex}}
\caption{Quadratic dependence on the various degrees of freedom of the total $pp\to t\bar th$ rate. For conveniences, numerical values for $S_{ij}^k/\sum_l B^k_l$ are quoted in permil. For more details see the legend of \autoref{tab:lin_pair}.
}
\label{tab:quad_tth}
\end{sidewaystable}


\begin{sidewaystable}[p]
\centering\hspace*{-3cm}
\adjustbox{width=.99\paperheight}{\input{squared_tj.tex}}
\caption{Quadratic dependence on the various degrees of freedom of the total $pp\to tj$ rate. For conveniences, numerical values for $S_{ij}^k/\sum_l B^k_l$ are quoted in permil. For more details see the legend of \autoref{tab:lin_pair}.
}
\label{tab:quad_tj}
\end{sidewaystable}

\begin{sidewaystable}[p]
\centering\hspace*{-3cm}
\adjustbox{width=.99\paperheight}{\input{squared_tev.tex}}
\caption{Quadratic dependence on the various degrees of freedom of the total $pp\to te^-\nu$ rate. For conveniences, numerical values for $S_{ij}^k/\sum_l B^k_l$ are quoted in permil. For more details see the legend of \autoref{tab:lin_pair}.
}
\label{tab:quad_tev}
\end{sidewaystable}

\begin{sidewaystable}[p]
\centering\hspace*{-3cm}
\adjustbox{width=.99\paperheight}{\input{squared_tjee.tex}}
\caption{Quadratic dependence on the various degrees of freedom of the total $pp\to tje^+e^-$ rate. For conveniences, numerical values for $S_{ij}^k/\sum_l B^k_l$ are quoted in permil. For more details see the legend of \autoref{tab:lin_pair}.
}
\label{tab:quad_tjee}
\end{sidewaystable}

\begin{sidewaystable}[p]
\centering\hspace*{-3cm}
\adjustbox{width=.99\paperheight}{\input{squared_tja.tex}}
\caption{Quadratic dependence on the various degrees of freedom of the total $pp\to tj\gamma$ rate. For conveniences, numerical values for $S_{ij}^k/\sum_l B^k_l$ are quoted in permil. For more details see the legend of \autoref{tab:lin_pair}.
}
\label{tab:quad_tja}
\end{sidewaystable}

\begin{sidewaystable}[p]
\centering\hspace*{-3cm}
\adjustbox{width=.99\paperheight}{\input{squared_tjh.tex}}
\caption{Quadratic dependence on the various degrees of freedom of the total $pp\to tjh$ rate. For conveniences, numerical values for $S_{ij}^k/\sum_l B^k_l$ are quoted in permil. For more details see the legend of \autoref{tab:lin_pair}.
}
\label{tab:quad_tjh}
\end{sidewaystable}

\end{document}

%% file: bdecays.tex

{
\let\section\subsection







\newcommand{\be}{\begin{equation}}
\newcommand{\ee}{\end{equation}}
\newcommand{\bea}{\begin{eqnarray}}
\newcommand{\eea}{\end{eqnarray}}
\newcommand{\GeV}{\textrm{ GeV}}
\newcommand{\TeV}{\textrm{ TeV}}
\newcommand{\SU}{\textrm{SU}}

\def\BK{\B(K^+\to \pi^+\nu\bar\nu)}
\def\BB{\B(B\to K^{(*)}\nu\bar\nu)}
\def\RD{R_{D^{(*)}}}

 
\section{Constraints from low-energy flavour physics}

The CKM matrix can be approximated as the identity matrix when focusing on measurements involving resonant top quarks. However, when confronting new physics models with flavour observables, suppressed in the SM by small CKM matrix elements, it is of course crucial to consistently keep all the CKM factors.
By working in the down-quark mass basis, the misalignment between the up- and down- quark masses can be described by considering the quark doublet as $q_i = (V^*_{ji} u_{L, j},  d_{L, i})^T$.
Another basis often used is the one where up quarks have diagonal mass matrix. This can be obtained by simply rotating $q_i$ with the CKM matrix: $\tilde q_i = (V q)_i = (u_{L, i},  V_{ij} d_{L, j})^T$.
This choice changes the flavour structure of the operators by the same CKM rotations, for example the term
$C^{(ij)} \bar q_i \gamma_\mu  q_j \to \tilde C^{(ij)} \bar {\tilde q}_i \gamma_\mu  {\tilde q}_j$, where $\tilde C = V C V^\dagger$.
In the spirit of this note, we assume that the $(33)$ element of such flavour matrices is the largest one. The $U(2)_q$ flavour symmetry relates the off-diagonal components to the CKM matrix \cite{Barbieri:2011ci,Barbieri:2012uh}. In this well motivated framework, in general the new physics contributions are not necessarily aligned with the up or down quarks, but will have some misalignment of the order of the relevant CKM elements.
For definiteness, in the following we mostly work in the down-quark mass basis.

Let us consider the charged-current transition $b \to c\, e_i \bar\nu_j$ and study the experimental limits we can extract on the top-quark operators. From the general effective Hamiltonian at the $B$-meson mass scale, we can consider only the operators involving the left-handed $c_L$ quark, since $c_R$ necessarily arises from a second-generation family index. We are left with
\be
	H_{\rm eff}^{b\to c e \bar\nu} = -\frac{2}{v^2} V_{cb} \left( (\delta^{i j} + c_{V_L}^{ij}) (\bar c_L \gamma^\mu b_L) (\bar e^i_L \gamma_\mu \nu_L^j ) + c_{S_R}^{ij} (\bar c_L b_R)(\bar e_R^i \nu_L^j) + h.c. \right)~,
\ee
where $v \approx 246 \GeV$. The tree-level matching to the SMEFT is given by \cite{Cirigliano:2012ab, Aebischer:2015fzz}
\be\begin{split}
	c_{V_L}^{ij} &= - \frac{v^2}{\Lambda^2} \left( \frac{\sum_{k} V_{ck} C_{l q}^{3 (ij k 3)}}{V_{cb}} \right) + \frac{v^2}{\Lambda^2} \left( \frac{\sum_{k} V_{ck} C_{\varphi q}^{3(k 3)} }{V_{cb}} \right)   \delta^{ij}~, \\
	c_{S_R}^{ij} &= - \frac{v^2}{\Lambda^2} \frac{\sum_{k} V_{ck} C_{ledq}^{(ji3k)}}{V_{cb}}~,
\end{split}\ee
where $k=1,2,3$ and for simplicity we assumed real coefficients. In general, under the $U(2)_q$ symmetry, the contribution from the $(33)$ element and from the $(23)$ one are of the same order since, for example $C_{l q}^{3 (ij 2 3)} / C_{l q}^{3 (ij 3 3)} \sim O(V_{cb})$. In realistic fits to these observables it is thus important to keep track of all contributions, which can lead to important phenomenological consequences (see e.g.\ Ref.\,\cite{Buttazzo:2017ixm}). Nevertheless, in order to extract the indicative numerical constraints, in the following we assume \emph{down-alignment} and keep as non vanishing only the $(33)$ element of these operators.

Consistently with the rest of the note, we neglect lepton-flavour-violating terms, while we allow possible deviations from lepton-flavour universality. 
In the case of operators with electron and muon one can derive the constraints from $B \to D^{(*)} \ell \nu$ decays \cite{Jung:2018lfu}, which experimentally agree with the SM prediction.
In the case of the $\tau$ lepton, instead, the corresponding decays $B \to D^{(*)} \tau \nu$ show an interesting deviation from the SM prediction with a statistical significance of more than $4\sigma$ \cite{Lees:2013uzd,Aaij:2015yra,Huschle:2015rga,Sato:2016svk,Hirose:2016wfn}. In particular, the relevant observables are ratios between the decay modes to tau and light leptons in which hadronic uncertainties largely cancel. Combining the different measurements one can express the result as (see e.g.\ Ref.\,\cite{Buttazzo:2017ixm})
\be
	R_{D^{(*)}} \equiv \frac{\mathcal{B}(B \to D^{(*)} \tau \nu)}{\mathcal{B}(B \to D^{(*)} \ell \nu)}  \left( \frac{\mathcal{B}(B \to D^{(*)} \tau \nu)_{\rm SM}}{\mathcal{B}(B \to D^{(*)} \ell \nu)_{\rm SM}} \right)^{-1} = |1 + c_{V_L}^{\tau\tau}|^2 = 1.237 \pm 0.053~,
\ee
where we showed explicitly only the contribution from the vector operator. This corresponds to $c_{V_L}^{\tau\tau} = 0.112 \pm 0.024$.
While also the scalar operator $c_{S_R}^{\tau\tau}$ contributes to the observables above, with a different weight in $R_D$ compared to $R_{D*}$, a stronger constraint can be derived from the $B_c$ lifetime \cite{Alonso:2016oyd}, due to the chiral enhancement of the scalar contribution. This gives an approximate bound $|c_{S_R}^{\tau\tau} | \lesssim 0.39$, which should be compared to the value $c_{S_R} \approx 1.5$ that would be required to fit $R_{D*}$.
In the case of down-alignment, a non-vanishing contribution to $D-\bar{D}$ mixing is generated by the CKM rotation from the $O_{qq}^{1(3333)}$ and $O_{qq}^{3(3333)}$ operators:
\be
	\mathcal{L}_{\rm eff} \supset \frac{C_{qq}^{1(3333)} + C_{qq}^{3(3333)}}{\Lambda^2} (V_{ub} V^*_{cb})^2 (\bar u_L \gamma_\mu c_L)^2 + h.c.~.
\ee
This can be used to cast a limit on the combination above \cite{Isidori:2013ez}.
The limits expressed in terms of the degrees of freedom defined in this note are reported in \autoref{tab:Blimits}.

\begin{table}
\begin{tabular}{lllll}
\multicolumn{2}{l}{Four-heavy} & 
	\\\hline\noalign{\vskip1mm}
	$\ccc{+}{QQ}{}$
	& \warsaw{\cc{1}{qq}{3333}+\cc{3}{qq}{3333}}
	& $[-21, 21]$ \cite{Isidori:2013ez}
	& \multicolumn{2}{l}{(from $D-\bar{D}$, with down-alignment)}\\
	$ \ccc[\tilde]{+}{QQ}{}$
	& \warsaw{\cc[\tilde]{1}{qq}{3333}+\cc[\tilde]{3}{qq}{3333}}
	& $[-0.03, 0.03]$ \cite{Isidori:2013ez}
	& \multicolumn{2}{l}{(from $B_s-\bar{B}_s$, with up-alignment)}
\\[3mm]
\multicolumn{2}{l}{Two-heavy-two-lepton}
	& $\ell = e$
	& $\ell = \mu$
	& $\ell = \tau$ 
	\\\hline\noalign{\vskip1mm}
$\ccc{3}{Ql}{\ell}$
	& \warsaw{\cc{3}{lq}{\ell\ell33}}
	& $[-0.57, 0.22]$ \cite{Jung:2018lfu}
	& $[-0.22, 0.57]$ \cite{Jung:2018lfu}
	& $ -1.85 \pm 0.40$ \cite{Buttazzo:2017ixm} \\
$\ccc{S}{b}{\ell}$
	& \warsaw{\Re\{\cc{}{ledq}{\ell\ell33}\}}
	& $[-10, 10]$ \cite{Jung:2018lfu}
	& $[-17, 13]$ \cite{Jung:2018lfu}
	& $[-13,13]$ \cite{Alonso:2016oyd}\\[.5mm]\hline
\end{tabular}
\caption{Indicative limits on top-quark operators arising from semileptonic $B$ decays and heavy meson oscillations and for $\Lambda=1\,$TeV.}
\label{tab:Blimits}
\end{table}

With the choice of down-alignment, contributions to $b \to s$ transitions, such as in $B \to K^{(*)} \ell^+ \ell^-$ decays or in $B_s - \bar{B}_s$ mixing, arise proportionally to the off-diagonal $(32)$ element of the quark flavour matrix of the various operators. While, as already mentioned, the $SU(2)_q$ flavour symmetry predicts that this element should be of $\mathcal{O}(V_{cb})$, in the simplified discussion above this was put to zero by hand, thus forbidding tree-level contributions to these processes. It is worthwhile to briefly discuss what would happen in the opposite scenario of \emph{up-alignment}, where the operators are written in terms of $\tilde q_i = (u_{L, i},  V_{ij} d_{L, j})^T$ and only the (33) component of the flavour matrix is left non-vanishing. In this case the $b\to d_i$ transition would arise proportionally to $V_{ib}$, while there would be no $b \to c$ charged-current transition (which, in this case, would be strictly proportional to the (32) element of the flavour matrices). $\Delta F = 2$ processes in the down sector, therefore, put strong limits at tree level on four-quark operators:
\be
	\mathcal{L}_{\rm eff} \supset \frac{\tilde C_{qq}^{1(3333)} + \tilde C_{qq}^{3(3333)}}{\Lambda^2} \left[ (V_{ts} V^*_{tb})^2 (\bar b_L \gamma_\mu s_L)^2 + (V_{td} V^*_{tb})^2 (\bar b_L \gamma_\mu d_L)^2 + (V_{td} V^*_{ts})^2 (\bar s_L \gamma_\mu d_L)^2 \right]+ h.c.~.
\ee
In particular, the strongest constraint on the overall coefficient is from $B_s - \bar B_s$ mixing \cite{Isidori:2013ez}.
The measurement of various $b \to s \ell^+ \ell^-$ transitions would instead constrain the two-quark-two-lepton operators $\tilde O_{lq}^{1(ij33)}$, $\tilde O_{lq}^{3(ij33)}$, and $\tilde O_{eq}^{(ij33)}$ (see e.g.\ Refs.\,\cite{Altmannshofer:2015sma, Descotes-Genon:2015uva}).
Loop-level contributions to $B_s$ mixing and electroweak precision measurements can instead be used to put constraints on the $\tilde O_{\varphi q}^{1(33)}$, $\tilde O_{\varphi q}^{3(33)}$, and $\tilde O_{\varphi u}^{(33)}$ operators \cite{Brod:2014hsa}.
It is clear that both up- and down-alignment are very particular cases. Experimentally, since limits from $\Delta F = 2$ processes are stronger in the down sector, the down-alignment case might be preferred. Generic models of flavour are expected to interpolate between the two scenarios, therefore a more general analysis, which is well beyond the purpose of the present note, would be advisable.

\section{Constraints from high-$p_T$ di-lepton searches}
 
\begin{table}
\begin{tabular}{l@{\;}l lll}
\multicolumn{2}{l}{Two-heavy-two-lepton} & $\ell = e$ & $\ell = \mu$ & $\ell = \tau$ 
	\\\hline\noalign{\vskip1mm}
$\ccc{-}{Ql}{\ell}+2\,\ccc{3}{Ql}{\ell}$
	& \warsaw{\cc{1}{lq}{\ell\ell33}+\cc{3}{lq}{\ell\ell33}}
	& $[-0.32, 0.20]$ \cite{Greljo:2017vvb}
	& $[-0.43, 0.34]$ \cite{Greljo:2017vvb}
	& $[-2.6, 2.6]$ \cite{Faroughy:2016osc} \\
$\ccc{}{Qe}{\ell}$
	& \warsaw{\cc{}{eq}{\ell\ell33}}
	& $[-0.24, 0.28]$ \cite{Greljo:2017vvb}
	& $[-0.38, 0.40]$ \cite{Greljo:2017vvb}  & -- \\
$\ccc{S}{b}{\ell}$
	& \warsaw{\Re\{\cc{}{ledq}{\ell\ell33}\}}
	& --
	& --
	& $[-1.9, 1.9]$ \cite{Faroughy:2016osc} \\[.5mm]\hline
\end{tabular}
\caption{Indicative limits on top-quark operators arising from dilepton pair production at the LHC and with $\Lambda=1\,$TeV.}
\label{tab:dileptonlimits}
\end{table}

It is well known that the high-energy tail of $2 \to 2$ scattering processes is very sensitive to effective operators, since it allows to use the growth with energy of the new physics amplitudes to increase the signal over background ratio.
In particular, let us focus here on fermion-fermion scattering processes at the LHC such as $q \bar q \to q \bar q$ and $q \bar q \to \ell^+ \ell^-$. These have been shown to provide very strong constraints on four-fermion operators \cite{Cirigliano:2012ab, Faroughy:2016osc, Farina:2016rws, Greljo:2017vvb, Alioli:2017jdo, Alioli:2017nzr}. Since the sensitivity comes from the high energy tail, the issue of the validity of the EFT expansion is a crucial one to be addressed in this case. In particular, the underlying assumption for these bounds to be valid is that the maximal centre of mass energy from which the sensitivity is gained, $E_{\rm max}$, has to be much smaller than the mass scale of the heavy states which have been integrated out, $E_{\rm max} \ll M_{\rm NP}$. This issue has been studied in detail in the references above and it has been shown that there exist models in which the new states are heavy enough for the EFT approach to be valid, and for which the constraints obtained are relevant. For example, in Ref.\,\cite{Greljo:2017vvb}, it has been shown that $q \bar q \to e^+ e^-, \mu^+ \mu^-$ processes can be used to put significant constraints on some models addressing the neutral-current $B$-physics anomalies. Instead, the EFT interpretation of the limits in the $\tau\tau$ final state \cite{Faroughy:2016osc} should be taken as indicative only, since the mass scale of new physics cannot be too high in that case. Nevertheless, comparing with the limits on explicit models \cite{Faroughy:2016osc} shows that the EFT ones still provide a good first-order indication.
In \autoref{tab:dileptonlimits} we report the limits from \cite{Faroughy:2016osc,Greljo:2017vvb} on the two-quark-two-lepton operators involving third generation quarks, in particular the $b$ quark since it is the only one accessible in the initial state.

}

%% file: low-energy.tex

%
%
\subsection{Indirect constraints from low-energy probes of CP violation}
In addition to limits from direct observables, complementary constraints can be derived from low-energy measurements.
Such indirect observables do not involve resonant top quarks, 
but they are affected, in some cases very significantly, by virtual top quarks and as such can be used to probe SMEFT top-quark operators.  
Here we focus on a subset of SMEFT operators, namely the operators defined in \autoref{app:dof} that violate CP, which are mainly constrained by electric-dipole-moment (EDM) experiments and asymmetry measurements in $B\to X_s\gamma$.  
Although we do not change the flavour structure of the operators themselves, we will deviate slightly from the baseline flavour assumptions of \autoref{sec:flavour}. In particular, we restore the off-diagonal CKM matrix elements and the light Yukawa couplings. Although these small SM parameters can be neglected in direct probes, they give rise to loop diagrams that, in some cases, induce the dominant contributions to low-energy probes. Indirect constraints on the CP-even parts, coming from electroweak precision data, are discussed in the next subsection.

As not all readers might be familiar with EDM phenomenology we give a brief introduction here. EDMs of leptons, nucleons, atoms, and molecules are probes of flavour-diagonal CPV that suffer from essentially no SM background. While the CKM mechanism predicts nonzero EDMs, they are orders of magnitude below the current experimental limits.  At present, the strongest EDM constraints arise from measurements on three different systems: the neutron, the ${}^{199}$Hg atom, and the polar molecule ThO. The limit on the latter can be interpreted (with care) as a limit on the electron EDM. In order to interpret measurements on these complicated systems in terms of the SMEFT Wilson coefficients, several steps need to be taken. First of all, the SMEFT operators must be evolved to lower energies, typically up to a scale of a few GeV where QCD is still perturbative. At that point, the SMEFT operators are matched to an effective Lagrangian describing the dynamics of the relevant low-energy degrees of freedom such as nucleons, pions, photons, and electrons. This effective Lagrangian is then used to calculate the EDMs of nucleons, nuclei, atoms, and molecules. A detailed discussion is beyond the scope of this note, and below we briefly describe how to connect the limit on the neutron and electron EDM to the SMEFT Wilson coefficients.
\newline

We start by discussing the limits from the neutron EDM. This observable obtains contributions from several CPV operators, namely four of the top-Higgs couplings, $c_{t\varphi}^{}$, $c_{tG}^{}$, $c_{bW}^{}$, and $c_{\varphi tb}^{}$, as well as the four-quark operator coefficients, $c_{QtQb}^{1,8}$.
To estimate the limits, we
evolve these operators, by using renormalization-group equations (RGEs), from the scale of new physics, $\Lambda$, to the scale of the top-quark mass. At this scale we integrate out the top quark. Here both $c_{t\varphi}^{}$ and $c_{tG}^{}$ induce a threshold contribution to a purely gluonic CPV operator without top quarks, the so-called Weinberg operator 
\begin{equation}
\mathcal L_W =  \frac{g_S}{6}\frac{d_W}{\Lambda^2}f^{abc} \epsilon^{\mu\nu\alpha\beta}G^a_{\alpha\beta}G^b_{\mu\rho}G^{c\,\rho}_{\nu}\,,
\end{equation}
with \cite{Dicus:1989va,Weinberg:1989dx,BraatenPRL,Boyd:1990bx,Kamenik:2011dk, Brod:2013cka}
\begin{equation}
d_W(m_t) = \frac{g_S^2}{64\pi^4}\frac{v}{\sqrt{2} m_t}h(m_t, m_h)\,c_{t\varphi}^{I}
+ \frac{g_S^2}{(4 \pi)^2}\frac{v}{\sqrt{2} m_t }\, c_{tG}^{I}\,,
\label{eq:matching}\end{equation}
where $h(m_t,m_h)\simeq 0.05$ is a finite two-loop integral. Note that $c^I_{t\varphi}$ also contributes to the EDMs of light quarks, proportional to their Yukawa couplings, through two-loop Barr-Zee diagrams \cite{Barr:1990vd,Gunion:1990iv,Abe:2013qla,Jung:2013hka}. We include this effect in the limits discussed below.

In addition, $c_{\varphi tb}^{}$, $c_{bW}^{}$, and $c_{QtQb}^{1,8}$ contribute to the Weinberg operator by first inducing  the bottom chromo-EDM, $O_{dG}^{(33)}$. For the four-quark operators and $O_{dW}^{(33)}$ the bottom chromo-EDM is induced through renormalization-group evolution \cite{Dekens:2013zca, Alonso:2013hga, Cirigliano:2016nyn} between $\mu=\Lambda$ and $\mu=m_t$, while  $O_{\varphi tb}^{}$ only provides a matching contribution
\begin{eqnarray}
\frac{d {\rm Im}\,C_{dG}^{(33)}}{d\ln\mu}&=&\frac{\sqrt{2}}{(4\pi)^2}\frac{m_t}{v}\left(c_{QtQb}^{1I}
-\frac{1}{2N_C}c_{QtQb}^{8I}\right)
+\frac{2}{(4\pi)^2}\left[3g_W^2+\frac{1}{3}g_Y^2\right] c_{bW}^{I}\dots\,,\nonumber\\
{\rm Im}\,C_{dG}^{(33)}(m_t^-)&=&{\rm Im}\,C_{dG}^{(33)}(m_t^+)+\frac{1}{\sqrt{2}(4\pi)^2}\frac{m_t}{v}f_W\,  c_{\varphi tb}^{I}
\,,
\end{eqnarray}
where $f_W\simeq 0.7$ is  a loop function \cite{Alioli:2017ces} and the ellipsis stands for the self-renormalization.  At the bottom mass scale $O_{dG}^{(33)}$ then induces a contribution to the Weinberg operator 
\begin{equation}
d_W(m_b^-) =d_W(m_b^+) + \frac{g_S^2}{(4\pi)^2}\frac{v}{\sqrt{2} m_b }\, \mathrm{Im}(C^{(33)}_{dG}(m_b^+))\,.
\end{equation}
The Weinberg operator can now be evolved to lower energies and, around the QCD scale, be matched to hadronic operators. In particular, it induces a contribution to the neutron EDM
\begin{equation}
|d_n| \simeq (50\,\mathrm{MeV})\,e g_s(1\,\mathrm{GeV})\,d_W(1\,\mathrm{GeV})/\Lambda^2\,.\label{neutronEDM}
\end{equation}
The required hadronic matrix element  suffers from large uncertainties and here we have taken the average of various estimates \cite{Pospelov_Weinberg, Weinberg:1989dx, deVries:2010ah}. The impact of hadronic and nuclear uncertainties on low-energy precision constraints on the SMEFT operators can be significant and has been discussed in detail in Refs.~\cite{Chien:2015xha, Cirigliano:2016nyn}. 
The constraints that result from employing \autoref{neutronEDM} and the experimental limit,  $d_n < 3.0 \cdot 10^{-13}$ e fm \cite{Afach:2015sja,Baker:2006ts}, are collected in \autoref{Tab:EDMlimits}.
\newline

Moving on to the electron EDM, there are again several operators that contribute, namely, three  top-Higgs couplings $c^{I}_{tA, tW,t\varphi}$, as well as the semi-leptonic operators $c_{t,b}^{S(e)}$ and  $c_{t}^{T(e)}$.
Of the semi-leptonic operators the tensor operator induces the electron EDM through a single top loop, while the scalar ones require additional loops. The relevant RGEs are given by \cite{Cirigliano:2016nyn, Alonso:2013hga},
\begin{equation}
\frac{d }{d\ln\mu}\begin{pmatrix} d_e\\
c_{t}^{SI(e)}/\Lambda^2\\c_{t}^{TI(e)}/\Lambda^2\\c_{b}^{SI(e)}/\Lambda^2\end{pmatrix}
=\frac{1}{(4\pi)^2}\begin{pmatrix}
0 &0& -16N_C Q_t m_t&0\\
0 & -6C_Fg_S^2&0&2(1-N_c)y_ty_b\\
0 & \frac{3}{8}g_W^2+\frac{5}{8} g_Y^2 &2C_F g_S^2&0\\
0&0&0&-6C_F g_S^2\\
\end{pmatrix}\cdot \begin{pmatrix} d_e\\
c_{t}^{SI(e)}/\Lambda^2\\c_{t}^{TI(e)}/\Lambda^2\\c_{b}^{SI(e)}/\Lambda^2\end{pmatrix}\,,
\end{equation}
where $C_F=(N_C^2-1)/2N_C$, $y_{b,t}=m_{b,t}\sqrt{2}/v$, $Q_f$ stands for the electric charge, and we only kept the electroweak terms that are relevant for the mixing of $c_{t,b}^{SI(e)}$ into $c_{t}^{TI(e)}$. The solution of the above equations provides the leading logarithmic contributions to the electron EDM. These, combined with $d_e\leq 8.7\cdot 10^{-16}\,e$ fm \cite{Baron:2013eja}, give stringent constraints, which are again collected in \autoref{Tab:EDMlimits}.
The same loops that induce the electron EDM also give a contribution to the electron anomalous magnetic moment, proportional to the real part of the semileptonic couplings. 
Although the resulting limits are weaker than the EDM limits they are still significant for two of the couplings, we obtain $|c_{t}^{S(e)}|\lesssim 2\cdot 10^{-2}$ and  $|c_{t}^{T(e)}|\lesssim 3\cdot 10^{-5}$.

Of the top-Higgs couplings, $c_{t\varphi}^{I}$ generates the electron EDM through two-loop Barr-Zee diagrams \cite{Barr:1990vd,Gunion:1990iv,Abe:2013qla,Jung:2013hka}, giving a stronger limit than the neutron EDM. 
In addition, when we evolve the $c^{I}_{tA, tW}$ couplings from the scale of new physics, $\Lambda$, to lower energies they first mix into CPV Higgs-gauge couplings of the form $(\varphi^\dagger \varphi)\tilde X_{\mu\nu} X^{\mu\nu}$ \cite{Cirigliano:2016nyn,Cirigliano:2016njn,Fuyuto:2017xup}, where $X$ denotes an $SU(2)$ or $U(1)$ gauge-field strength. In a second step these gauge-Higgs couplings mix into the electron and light-quark EDMs. This last step is proportional to the Yukawa couplings of the light fields leading to a strong suppression. Nevertheless, the experimental limit on the electron EDM is sufficiently strong to overwhelm the other probes of the CPV components of these two top-quark dipoles \cite{Cirigliano:2016nyn,Cirigliano:2016njn}.  
\newline

\begin{table}
\renewcommand{\arraystretch}{1.2}%
\begin{tabular}{@{}l@{\:}llllll@{}}
\multicolumn{2}{@{}l}{Four-heavy} & \multicolumn{4}{l}{}\\
\hline\noalign{\vskip 1mm}
$c_{QtQb}^{1I}$
	& \warsaw{\Im\{\cc{1}{quqd}{3333}\}} 
	& $[-3.4,\, 3.4]\cdot 10^{-3}$ &($d_n$)&&\\
$c_{QtQb}^{8I}$
	& \warsaw{\Im\{\cc{8}{quqd}{3333}\}} 
	&$[-2.2,\, 2.2]\cdot 10^{-2}$&($d_n$) &&\\[2mm]
\multicolumn{2}{@{}l}{Two-heavy} & \\\hline\noalign{\vskip 1mm}
$c_{t\varphi}^{I}$
	& \warsaw{\Im\{\cc{}{u\varphi}{33}\}}
	&$[-3.7,\,3.7]$&($d_n$) &\quad$[-0.18,\, 0.18]$&($d_e$)\\
$c_{\varphi tb}^{I}$
	& \warsaw{\Im\{\cc{}{\varphi ud}{33}\}}
	&$[-0.019,\, 0.019]$&($d_n$)&\quad $[-0.052,\, 0.052]$&($B\to X_s\gamma$)\\
$c_{ tW}^{I}$
	& \warsaw{\Im\{\cc{}{uW}{33}\}}
	&$[-8.1,\, 8.1]\cdot 10^{-3}$&($d_e$)&\quad$[-2.4,\, 4.5]$&($B\to X_s\gamma$)\\
$c_{tA}^{I}$
	& \warsaw{\Im\{\cw\cc{}{uB}{33}+\sw\cc{}{uW}{33}\}}
	&$[-6.3,\, 6.3]\cdot 10^{-3}$&($d_e$)&\quad$[-9.0,\, 5.0]$&($B\to X_s\gamma$)\\
$c_{bW}^{I}$
	& \warsaw{\Im\{\cc{}{dW}{33}\}}
	& $[-5.5,\, 5.5]\cdot 10^{-4}$&($d_n$)&\quad $[-4.3,\, 2.3]\cdot 10^{-2}$&($B\to X_s\gamma$)\\
$c_{tG}^{I}$
	& \warsaw{\Im\{\cc{}{uG}{33}\}}
	& $[-6.9,\,6.9]\cdot 10^{-3}$&($d_n$)\\[2mm]
\multicolumn{2}{@{}l}{Two-heavy-two-lepton} & \\\hline\noalign{\vskip 1mm}
$c_{t}^{SI(e)}$
	& \warsaw{\Im\{\cc{1}{lequ}{1133}\}}
	& $[-5.5,\, 5.5]\cdot 10^{-8}$&($d_e$)\\
$c_{t}^{TI(e)}$
	& \warsaw{\Im\{\cc{3}{lequ}{1133}\}}
	& $[-8.0,\, 8.0]\cdot 10^{-11}$&($d_e$)\\
$c_{b}^{SI(e)}$
	& \warsaw{\Im\{\cc{}{ledq}{1133}\}}
	& $[-2.5,\, 2.5]\cdot 10^{-4}$&($d_e$)\\\hline
\end{tabular}
\caption{Constraints from the electron and neutron EDMs as well as $A_{CP}(B\to X_s\gamma)$. Here we turn on one coupling at a time and assume  $\Lambda =1$ TeV. The source of the constraints are indicated in brackets.}\label{Tab:EDMlimits}
\end{table}

Finally, we briefly discuss limits from rare $B$ decays. 
At the one-loop level, the couplings $c^{I}_{\varphi tb, bW, tA, tW}$ give contributions to flavour-changing dipole operators that mediate $b\rightarrow s $ transitions proportional to the CKM element $V_{ts}\simeq 0.04$. 
The contributions of these flavour-changing operators to the CP asymmetry $A_{CP}(B\to X_s\gamma)$ \cite{Benzke:2010tq}, together with the experimental measurement \cite{Amhis:2016xyh}, can  be used set the limits in \autoref{Tab:EDMlimits}.
It should be noted that measurements of the branching ratio can constrain the real parts of these couplings as well. This leads to limits which are typically a factor of a few stronger than those on the imaginary parts, see, for example,  Refs.~\cite{Grzadkowski:2008mf,Drobnak:2011aa,Drobnak:2011wj,Kamenik:2011dk}.


\subsubsection*{Summary}
All the above discussed constraints are collected in \autoref{Tab:EDMlimits}. With the exception of $c^{I}_{t\varphi}$, the CPV coefficients are constrained at the percent level or stronger by EDM experiments. The semi-leptonic operators are more stringently constrained, which is mainly due to the fact that their contribution to $d_e$ is proportional to $m_t$ (where one would naively expect $m_e$). Instead, the constraints from $B\to X_s\gamma$ are particularly strong for the $c_{bW}^{I}$ and $c_{\varphi tb}^{I}$ couplings because of an $m_t/m_b$ enhancement.
In most cases, the constraints on the imaginary parts are stronger than the corresponding limits on the real parts of the top-quark couplings. As a result, it will be difficult to reach a similar sensitivity by studying CPV observables at the LHC. 

The interpretation of these low-energy constraints requires some care however. In deriving these limits we have assumed one dimension-six operator to be present at the scale $\Lambda$ at a time. This assumption is no longer valid if multiple top operators are important at the scale $\Lambda$, or if one makes less restrictive assumptions about the flavor structure, such as a non-linear flavor symmetry \cite{Kagan:2009bn}. A global analysis involving all top-quark operators, for example, would leave some combinations of operator coefficients unconstrained~\cite{Cirigliano:2016nyn}. The difference between \emph{individual} and \emph{global} constraints is typically large for the CPV components as only a handful of sensitive low-energy measurements exist, in contrast to a much larger range of high-energy measurements of the real components. In a global setting, collider constraints on the CPV Wilson coefficients are therefore necessary to bound unconstrained directions in the parameter space. An example is the recent ATLAS measurement~\cite{Aaboud:2017yqf} of a CPV phase in $t\rightarrow bW$ decays which significantly impacts the global fit of CPV top-Higgs interactions~\cite{Cirigliano:2016nyn,Alioli:2017ces}.

In addition, the low-energy observables get contributions from SMEFT operators that do not involve top quarks. For example, if an electron EDM is generated at the scale $\Lambda$ with exactly the right size, it could weaken the limits on the semi-leptonic operators significantly. 
In fact, the leading logarithmic contributions we considered here result from divergent loops and require non-top operators, such as $d_e$, to absorb the divergences. In these cases one might expect $d_e(\Lambda)\neq 0$, which could in principle lead to cancellations and weakened limits. These cancellations have to be very severe in order to avoid the strong low-energy constraints. In any case, in order to evade the low-energy limits strong correlations between SMEFT operators are required, and this would  pose highly non-trivial constraints on models of beyond-the-SM physics.

%% file: smeftsim.tex


\def\nn{\nonumber}
\def\hyp{\mathsf{y}}
\renewcommand{\O}{\mathcal{O}}
\newcommand{\Q}{\mathcal{Q}}
\renewcommand{\a}{\alpha}
\renewcommand{\b}{\beta}
\renewcommand{\d}{\delta}
\newcommand{\g}{\gamma}
\newcommand{\s}{\sigma}
\renewcommand{\dag}{\dagger}
\newcommand{\de}{\partial}
\newcommand{\des}{\slashed{\de}}
\newcommand{\bsw}{s_{\bar \theta}}
\newcommand{\bcw}{c_{\bar \theta}}
\newcommand{\hsw}{s_{\hat \theta}}
\newcommand{\hcw}{c_{\hat \theta}}
\newcommand{\dsw}{\d s_\theta}
\newcommand{\hv}{\hat{v}}
\newcommand{\aem}{\alpha_{\rm em}}
\renewcommand{\to}{\rightarrow}

\newcommand{\cHlt}{c_{Hl}^{(3)}}
\newcommand{\cHls}{c_{Hl}^{(1)}}
\newcommand{\cHqt}{c_{Hq}^{(3)}}
\newcommand{\cHqs}{c_{Hq}^{(1)}}
\newcommand{\red}[1]{{\color{red} #1}}

\let\section\subsection
\let\subsection\subsubsection
\let\subsubsection\paragraph




\label{sec:smeftsim}

The \texttt{SMEFTsim} package~\cite{Brivio:2017btx} (available at \url{http://feynrules.irmp.ucl.ac.be/wiki/SMEFT}) provides a complete \texttt{FeynRules}~\cite{Alloul:2013bka} implementation of the $B$-conserving dimension-six Lagrangian in the Warsaw basis~\cite{Grzadkowski:2010es}, that automatically performs the field and parameter redefinitions required to have canonically normalized kinetic terms and following from the choice of an input parameters set. The package contains \texttt{FeynRules} models as well as
pre-exported UFO models for six different Lagrangians, corresponding to two possible input parameters sets ($\{\hat\a_{\rm ew}, \hat m_Z, \hat G_F\}$ or $\{\hat m_W, \hat m_Z, \hat G_F\}$) and to three possible assumptions on the flavour structure of the theory:
\begin{itemize}
 \item a {\bf flavour general} case, in which all the flavour indices are explicitly kept and the Wilson coefficients of the fermionic operators are defined as tensorial parameters.
 \item a {\bf $U(3)^5$ flavour symmetric} case in which flavour contractions are fixed by the symmetry. The Yukawa couplings are treated as flavour-breaking spurions and consistently inserted in operators with chirality-flipping fermion currents.
 \item a {\bf linear MFV} case~\cite{DAmbrosio:2002vsn} that, unlike the $U(3)^5$ symmetric setup, does not contain CP violating parameters beyond the CKM phase while it includes insertions of the flavour-violating spurions $Y_f Y^\dag_f$, $Y_f^\dag Y_f$ up to linear order.
\end{itemize}
Two versions of \texttt{SMEFTsim} are available, denominated ``set~A'' and ``set~B'': these implementations are independent but completely equivalent, and their simultaneous availability is meant to help a cross-check of the results.

\subsubsection*{General comments}
\begin{list}{--}{\topsep 3pt \itemsep 0pt \parsep 3pt}
 \item Analogously to \texttt{dim6top}, \texttt{SMEFTsim} is a leading order, unitary gauge implementation.
 \item  The Wilson coefficients are dimensionless quantities and the EFT cutoff $\Lambda$ is an external parameter of the UFO, set by default to 1~TeV ({{\tt LambdaSMEFT = 1.e+03}}). 
 \item The definitions of the relevant parameters (in the Warsaw basis) are given in \autoref{table:param_defined_SMEFTsim}. The correspondence with the parameters used in \texttt{dim6top} can be deduced from the definitions summarized in \autoref{tab:limits}.\\
 A {\tt QED=-1} interaction order has been assigned to {\tt LambdaSMEFT}, in order to avoid couplings with negative {\tt QED} order.\footnote{One coupling with negative {\tt QED} order is still present nonetheless, namely the $C_\varphi$ correction to the $h^3$ vertex, which plays no role here.}

 \begin{table}[p]\scriptsize
\renewcommand{\arraystretch}{1.2}
 \begin{tabular}{@{}>{$}l<{$}>{\tt}p{5.4cm}>{$}l<{$}>{\tt}p{5.3cm}@{}}
 \hline
C_{u\varphi}^{(33)}& cuHAbs33, cuHPh33& 				C_{\varphi u}^{(33)}& cHuAbs33, cHuPh33\\
C_{\varphi q}^{1(33)}& cHq1Abs33 &					C_{\varphi q}^{3(33)}& cHq3Abs33	\\
C_{\varphi ud}^{(33)}& cHudAbs33, cHudPh33&		 &	\\    
\hline
C_{uW}^{(33)}& cuWAbs33, cuWPh33&				C_{dW}^{(33)}&	cdWAbs33, cdWPh33\\
C_{uB}^{(33)}& cuBAbs33, cuBPh33&				C_{uG}^{(33)}&	cuGAbs33, cuGPh33\\
C_{qq}^{1(3333)}& cqq1Abs3333&					C_{qq}^{3(3333)}& cqq3Abs3333	\\
C_{qq}^{1(aa33)}& cqq1Abs(aa)33&				C_{qq}^{3(aa33)}& cqq3Abs(aa)33	\\
C_{qq}^{1(a33a)}& cqq1Abs(a)33(a)&				C_{qq}^{3(aa33)}& cqq3Abs(a)33(a)\\
C_{qu}^{1(3333)}& cqu1Abs3333&					C_{qu}^{8(3333)}& cqu8Abs3333	\\
C_{qu}^{1(aa33)}& cqu1Abs(aa)33&				C_{qu}^{8(aa33)}& cqu8Abs(aa)33	\\
C_{qu}^{1(33aa)}& cqu1Abs33(aa)&				C_{qu}^{8(33aa)}& cqu8Abs33(aa)	\\
C_{qd}^{1(3333)}& cqd1Abs3333&					C_{qd}^{8(3333)}& cqdAbs3333	\\
C_{qd}^{1(33aa)}& cqd1Abs33(aa)&				C_{qd}^{8(33(aa))}& cqdAbs33(aa)\\
C_{uu}^{(3333)}& cuuAbs3333&					C_{uu}^{(aa33)}& cuuAbs(aa)33	\\  
C_{uu}^{(a33a)}& cuuAbs(a)33(a)&			 & \\
\hline
C_{ud}^{1(3333)}& cud1Abs3333&					C_{ud}^{8(3333)}& cud8Abs3333	\\
C_{ud}^{1(33aa)}& cud1Abs33(aa)&				C_{ud}^{8(33aa)}& cud8Abs33(aa)	\\
C_{quqd}^{1(3333)}& cquqd1Abs3x3x3x3, cquqd1Ph3x3x3x3&		C_{quqd}^{8(3333)}&	cquqd8Abs3x3x3x3, cquqd8Ph3x3x3x3\\
\hline
C_{lq}^{1(ll33)}& clq1Abs(ll)33&				C_{lq}^{3(ll33)}& clq3Abs(ll)33	\\
C_{lu}^{(ll33)}& cluAbs(ll)33 &					C_{eq}^{(ll33)}& cqeAbs33(ll)	\\
C_{eu}^{(ll33)}& ceuAbs(ll)33&					C_{lequ}^{1(ll33)}& clequ1Abs(l)x(l)x3x3, clequ1Ph(l)x(l)x3x3	\\
C_{lequ}^{3(ll33)}& clequ3Abs(l)x(l)x3x3, clequ3Ph(l)x(l)x3x3&	C_{ledq}^{(ll33)}&	cledqAbs(l)x(l)x3x3, cledqPh(l)x(l)x3x3\\
\hline
 \end{tabular}
\caption{Flavour conserving parameters in the \texttt{SMEFTsim-A} UFO implementation, that are in direct correspondence with the degrees of freedom defined in \autoref{app:dof}.
The index $a=\{1,2\}$ runs over the light quark generations, while $l=\{1,2,3\}$ indicates lepton flavours.
}\label{table:param_defined_SMEFTsim}
\vspace{1.5cm}
\centering\footnotesize
\renewcommand{\arraystretch}{1.3}
 \begin{tabular}{ll@{\hspace*{8mm}}l}
 \hline
  flavour case& 	relevant parameters&  total number\\\hline
  general&
  $[C_{uH},\, C_{uB},\, C_{uW},\, C_{uG}]_{i3,3j}\,,\quad[C_{dW},\, C_{Hud}]_{3j}\,\quad [C_{Hu},\,C_{Hq}^{(1),(3)}]_{i3}\,,$			 & 618 abs. values\\

  & $[C_{qq}^{(1),(3)},\,C_{uu},\,C_{qu}^{(1),(8)}]_{3jkl,i3kl,ij3l,ijk3}\,,\quad [C_{ud}^{(1),(8)},\,C_{qe},\,C_{qd}^{(1),(8)}]_{3jkl,i3kl}\,,$  & 557 phases\\
  & $[C_{lq}^{(1),(3)},\,C_{eu},\,C_{lu},\,C_{lequ}^{(1),(3)} ]_{ij3l,ijk3}\,,\quad[C_{ledq}]_{ijk3}\,,\quad $ $[C_{quqd}^{(1),(8)}]_{3jkl,i3kl,ij3l}$ & \\
  & $C_{H\square}$, $C_{HD}$, $C_{HWB}$, $[C_{ll}]_{1221}$,
  $[C_{Hl}^{(3)}]_{11,22}$						 &
  \\[3mm]
  
  $U(3)^5$&

  $C_{uH}\,,$ $C_{Hu}\,,$ $C_{Hq}^{(1),(3)}\,,$ $C_{Hud}\,,$
  $C_{uW}\,,$ $C_{uB}\,,$ $C_{uG}\,,$ $C_{dW}\,,$
  $C_{qq}^{(1),(3)}\,,$
  $C_{qq}^{(1),(3)\prime}\,,$& 34 abs. values\\
  & $C_{lq}^{(1),(3)}\,,$
    $C_{uu}\,,$ $C_{uu}'\,,$ 
    $C_{ud}^{(1),(8)}\,,$ $C_{eu}\,,$
    $C_{lu}\,,$ $C_{qe}\,,$
    $C_{qu}^{(1),(8)}\,,$ $C_{qd}^{(1),(8)}\,,$
    $C_{quqd}^{(1),(8)}\,,$& 8 phases\\
  & $C_{H\square}\,,$ $C_{HD}\,,$ $C_{HWB}\,,$ $C_{ll}\,,$
   $C'_{ll}\,,$ $C_{Hl}^{(3)}$
  \\[3mm]
  
  linear MFV&

  $C_{uH}\,,$ $C_{Hu}^*\,,$ $C_{Hq}^{(1),(3)**}\,,$ $C_{Hud}\,,$
  $C_{uW}\,,$ $C_{uB}\,,$ $C_{uG}\,,$ $C_{dW}\,,$
  $C_{qq}^{(1),(3)****}\,,$ 
  & 83 abs. values\\
  & $C_{qq}^{(1),(3)\prime\, ****}\,,$    $C_{lq}^{(1),(3)**}\,,$
    $C_{uu}^{**}\,,$ $C_{uu}^{\prime\, **}\,,$ 
    $C_{ud}^{(1),(8)**}\,,$ $C_{eu}^*\,,$
    $C_{lu}^*\,,$ $C_{qe}^{**}\,,$
    & \\
  & $C_{qu}^{(1),(8)***}\,,$ $C_{qd}^{(1),(8)***}\,,$ $C_{quqd}^{(1),(8)}\,,$
  & \\
  & $C_{H\square}\,,$ $C_{HD}\,,$ $C_{HWB}\,,$ $C_{ll}\,,$
   $C'_{ll}\,,$ $C_{Hl}^{(3)}$
  \\[2mm]\hline
 \end{tabular}
\caption{Detailed list of the parameters that are relevant for top-quark physics for each flavour assumption in \texttt{SMEFTsim}. These are all set to 1 when importing the {\tt restrict\_TopEFT} restriction in any given model. The indices $i,j,k,l$ take values in \{1,2,3\}. The terms in the last rows are included as they contribute to top couplings via parameter redefinitions. In the MFV case all the parameters are real and coefficients with $n$ asterisks admit $n$ independent insertions of flavour-violating spurions.
}\label{tab:parameters_topEFT_rst}
\end{table}
 
 \item Complex Wilson coefficients are parametrized by their absolute value and complex phase {\tt ciAbs Exp[I ciPh]} rather than by real and imaginary parts. When testing the imaginary part of a given operator it is therefore necessary to set the corresponding phase to $\pi/2 = 1.570796$.
 
 \item The SM couplings of the Higgs boson that first arise at loop level (i.e. $hgg$, $h\gamma\gamma$, $hZ\gamma$) are included and parametrized as effective vertices with a coefficient that reproduces the SM loop function. These vertices are assigned an interaction order {\tt SMHLOOP = 1}.
 
 \item All the flavour contractions are automatically implemented in the model, with a parametrization that allows to fix only the independent ones. This means that, for instance, setting $C_{qq}^{1(aa33)}=1$ introduces both the $O_{qq}^{1(aa33)}$ and $O_{qq}^{1(33aa)}$ terms in the Lagrangian.

 \item An interaction order {\tt NP=1} has been defined for the Wilson coefficients, equivalent to {\tt DIM6} in {\tt dim6top}. There is no analogue of {\tt FCNC} nor of the individual {\tt DIM6\_ci}. 
 
\end{list}

\subsubsection*{Syntax}
\begin{itemize}
 \item In \mg\ import the model, restricted by the file {\tt restrict\_XXX.dat} by\\
 {\tt   > import model SMEFTsim\_A\_general\_alphaScheme-XXX}
 
 All the restrictions available leave {\tt MB} and {\tt ymb} non-vanishing. It is possible to switch to the five flavour scheme either modifying manually the restriction card or setting\\
 {\tt > define p = p b b\~{} }\\
 {\tt > define j = p}\\
 after importing the model and then modifying the {\tt param\_card} after generating the process output. 
 
 \item Processes can be generated with\\
 {\tt > generate p p > t t\~\ NP==1 SMHLOOP=0}\\
to allow the insertion of one Wilson coefficient in the amplitude and excluding SM $hgg$, $h\gamma\gamma$, and $hZ\gamma$ vertices or, for instance, with\\
{\tt > generate p p > t t\~\ NP\^{}2==1 SMHLOOP=0}\\
to select for the interference term. {\tt QED} and {\tt QCD} interaction orders can also be specified to select for specific diagrams. To ensure the inclusion of all the tree-level SM contributions, it is advisable to specify {\tt QED<=8 QCD<=8}.

\end{itemize}

%% file: interferences.tex
\begin{tabular}{ll*{7}{c}}
&&$pp\to t\bar{t}$	& $pp\to t\bar{t}\,b\bar{b}$	& $pp\to t\bar{t}\,t\bar{t}$	& $pp\to t\bar{t}\,e^+\nu$	& $pp\to t\bar{t}\,e^+e^-$	& $pp\to t\bar{t}\,\gamma$	& $pp\to t\bar{t}\,h$\\\hline
$\text{SM}$	& \texttt{sm}	& $5.2\times 10^{2}$ pb	& $1.9$ pb	& $0.0098$ pb	& $0.02$ pb	& $0.016$ pb	& $1.4$ pb	& $0.4$ pb	\\\hline
$\ccc{1}{QQ}{}$	& \texttt{cQQ1}	& $-0.25$	& $-1.9$	& $-1\times 10^{2}$	& 	& $-1.6$	& $-0.67$	& $-0.71$	\\
$\ccc{8}{QQ}{}$	& \texttt{cQQ8}	& $-0.16$	& $-3.2$	& $-34$	& 	& $-0.91$	& $-0.5$	& $-0.27$	\\
$\ccc{1}{Qt}{}$	& \texttt{cQt1}	& $-0.15$	& $-5.6$	& $1\times 10^{2}$	& 	& $-0.76$	& $-0.19$	& $-0.55$	\\
$\ccc{8}{Qt}{}$	& \texttt{cQt8}	& $-0.053$	& $-1.8$	& $-41$	& 	& $-0.18$	& $-0.095$	& $-0.15$	\\
$\ccc{1}{Qb}{}$	& \texttt{cQb1}	& $-0.0055$	& $0.72$	& $-0.052$	& 	& $-0.015$	& $-0.007$	& $-0.026$	\\
$\ccc{8}{Qb}{}$	& \texttt{cQb8}	& $0.14$	& $3.9$	& $0.12$	& 	& $0.35$	& $0.16$	& $0.56$	\\
$\ccc{1}{tt}{}$	& \texttt{ctt1}	& 	& 	& $-1.8\times 10^{2}$	& 	& 	& 	& 	\\
$\ccc{1}{tb}{}$	& \texttt{ctb1}	& $-0.0095$	& $0.46$	& $-0.059$	& 	& $-0.02$	& $-0.026$	& $-0.039$	\\
$\ccc{8}{tb}{}$	& \texttt{ctb8}	& $0.13$	& $3.5$	& $0.11$	& 	& $0.26$	& $0.31$	& $0.56$	\\
$\ccc{1}{QtQb}{}$	& \texttt{cQtQb1}	& 	& 	& 	& 	& 	& 	& 	\\
$\ccc{8}{QtQb}{}$	& \texttt{cQtQb8}	& 	& 	& 	& 	& 	& 	& 	\\
$\ccc{1I}{QtQb}{}$	& \texttt{cQtQb1I}	& 	& 	& 	& 	& 	& 	& 	\\
$\ccc{8I}{QtQb}{}$	& \texttt{cQtQb8I}	& 	& 	& 	& 	& 	& 	& 	\\\hline
$\ccc{3,8}{Qq}{}$	& \texttt{cQq83}	& $2.7$	& $-0.11$	& $4.7$	& $-85$	& $-20$	& $8.5$	& $15$	\\
$\ccc{1,8}{Qq}{}$	& \texttt{cQq81}	& $12$	& $7.1$	& $25$	& $2.6\times 10^{2}$	& $71$	& $40$	& $75$	\\
$\ccc{8}{tq}{}$	& \texttt{ctq8}	& $13$	& $8.2$	& $27$	& $2.6\times 10^{2}$	& $62$	& $51$	& $74$	\\
$\ccc{8}{Qu}{}$	& \texttt{cQu8}	& $7.4$	& $4.4$	& $18$	& 	& $21$	& $41$	& $44$	\\
$\ccc{8}{tu}{}$	& \texttt{ctu8}	& $7.4$	& $3$	& $16$	& 	& $14$	& $22$	& $45$	\\
$\ccc{8}{Qd}{}$	& \texttt{cQd8}	& $5$	& $3$	& $11$	& 	& $17$	& $7.3$	& $29$	\\
$\ccc{8}{td}{}$	& \texttt{ctd8}	& $5$	& $2.1$	& $10$	& 	& $12$	& $10$	& $28$	\\
$\ccc{3,1}{Qq}{}$	& \texttt{cQq13}	& $3.3$	& $3$	& $5.8$	& $1.1\times 10^{2}$	& $22$	& $11$	& $18$	\\
$\ccc{1,1}{Qq}{}$	& \texttt{cQq11}	& $0.94$	& $-1.4$	& $-7.7$	& $-5.9$	& $-5$	& $3$	& $5.4$	\\
$\ccc{1}{tq}{}$	& \texttt{ctq1}	& $0.65$	& $2.4$	& $-7.9$	& $8.7$	& $0.84$	& $3.7$	& $4.8$	\\
$\ccc{1}{Qu}{}$	& \texttt{cQu1}	& $0.57$	& $1.5$	& $-5.2$	& 	& $1.5$	& $2.9$	& $4.3$	\\
$\ccc{1}{tu}{}$	& \texttt{ctu1}	& $1.1$	& $-0.29$	& $-3.8$	& 	& $2.3$	& $3.3$	& $6.6$	\\
$\ccc{1}{Qd}{}$	& \texttt{cQd1}	& $-0.19$	& $0.55$	& $-4$	& 	& $-0.66$	& $-0.3$	& $-1.4$	\\
$\ccc{1}{td}{}$	& \texttt{ctd1}	& $-0.37$	& $-1.3$	& $-5$	& 	& $-0.91$	& $-1.3$	& $-2.1$	\\\hline
$\ccc{}{t\varphi}{}$	& \texttt{ctp}	& 	& $-0.00035$	& $-9.1$	& $-0.034$	& $-0.0093$	& 	& $-1.2\times 10^{2}$	\\
$\ccc{-}{\varphi Q}{}$	& \texttt{cpQM}	& $-0.063$	& $1$	& $-41$	& $-0.76$	& $-1\times 10^{2}$	& $-0.13$	& $-0.29$	\\
$\ccc{3}{\varphi Q}{}$	& \texttt{cpQ3}	& $0.68$	& $22$	& $0.065$	& $0.46$	& $3.7$	& $1.5$	& $1.8$	\\
$\ccc{}{\varphi t}{}$	& \texttt{cpt}	& $-0.024$	& $2.8$	& $42$	& $-0.36$	& $68$	& $-0.058$	& $-0.16$	\\
$\ccc{}{\varphi tb}{}$	& \texttt{cptb}	& 	& 	& 	& 	& 	& 	& 	\\
$\ccc{}{tW}{}$	& \texttt{ctW}	& $0.98$	& $1$	& $-34$	& $13$	& $1.1$	& $69$	& $9.4$	\\
$\ccc{}{tZ}{}$	& \texttt{ctZ}	& $-0.54$	& $0.028$	& $27$	& $-0.048$	& $-3.6$	& $-55$	& $-4.3$	\\
$\ccc{}{bW}{}$	& \texttt{cbW}	& 	& 	& 	& 	& 	& 	& 	\\
$\ccc{}{tG}{}$	& \texttt{ctG}	& $2.7\times 10^{2}$	& $2.5\times 10^{2}$	& $3.8\times 10^{2}$	& $2.4\times 10^{2}$	& $3.1\times 10^{2}$	& $2.4\times 10^{2}$	& $8.4\times 10^{2}$	\\
$\ccc{I}{t\varphi}{}$	& \texttt{ctpI}	& 	& $-7.3\times 10^{-7}$	& $0.045$	& $-0.00064$	& $-0.00029$	& 	& $0.045$	\\
$\ccc{I}{\varphi tb}{}$	& \texttt{cptbI}	& 	& 	& 	& 	& 	& 	& 	\\
$\ccc{I}{tW}{}$	& \texttt{ctWI}	& $4.8\times 10^{-6}$	& $0.032$	& $-1.6$	& $-0.19$	& $0.29$	& $0.91$	& $0.031$	\\
$\ccc{I}{tZ}{}$	& \texttt{ctZI}	& $-1.4\times 10^{-6}$	& $0.1$	& $-1.2$	& $0.0098$	& $3.2$	& $-0.56$	& $-0.057$	\\
$\ccc{I}{bW}{}$	& \texttt{cbWI}	& 	& 	& 	& 	& 	& 	& 	\\
$\ccc{I}{tG}{}$	& \texttt{ctGI}	& $-0.00098$	& $0.48$	& $0.66$	& $0.031$	& $-0.7$	& $0.019$	& $-2.4$	\\\hline
$\ccc{3}{Ql}{1}$	& \texttt{cQl31}	& 	& 	& 	& $0.011$	& $0.06$	& 	& 	\\
$\ccc{-}{Ql}{1}$	& \texttt{cQlM1}	& 	& 	& 	& $-0.0062$	& $-9.8$	& 	& 	\\
$\ccc{}{Qe}{1}$	& \texttt{cQe1}	& 	& 	& 	& 	& $-1.5$	& 	& 	\\
$\ccc{}{tl}{1}$	& \texttt{ctl1}	& 	& 	& 	& $-0.0023$	& $-3.6$	& 	& 	\\
$\ccc{}{te}{1}$	& \texttt{cte1}	& 	& 	& 	& 	& $-6.7$	& 	& 	\\
$\ccc{S}{tl}{1}$	& \texttt{ctlS1}	& 	& 	& 	& 	& 	& 	& 	\\
$\ccc{T}{tl}{1}$	& \texttt{ctlT1}	& 	& 	& 	& 	& 	& 	& 	\\
$\ccc{S}{bl}{1}$	& \texttt{cblS1}	& 	& 	& 	& 	& 	& 	& 	\\
$\ccc{SI}{tl}{1}$	& \texttt{ctlSI1}	& 	& 	& 	& 	& 	& 	& 	\\
$\ccc{TI}{tl}{1}$	& \texttt{ctlTI1}	& 	& 	& 	& 	& 	& 	& 	\\
$\ccc{SI}{bl}{1}$	& \texttt{cblSI1}	& 	& 	& 	& 	& 	& 	& 	\\\hline
$\ccc{1}{tQqu}{}$	& \texttt{ctQqu1}	& 	& 	& 	& 	& 	& 	& 	\\
$\ccc{8}{tQqu}{}$	& \texttt{ctQqu8}	& 	& 	& 	& 	& 	& 	& 	\\
$\ccc{1}{bQqd}{}$	& \texttt{cbQqd1}	& 	& 	& 	& 	& 	& 	& 	\\
$\ccc{8}{bQqd}{}$	& \texttt{cbQqd8}	& 	& 	& 	& 	& 	& 	& 	\\
$\ccc{1}{Qtqd}{}$	& \texttt{cQtqd1}	& 	& 	& 	& 	& 	& 	& 	\\
$\ccc{8}{Qtqd}{}$	& \texttt{cQtqd8}	& 	& 	& 	& 	& 	& 	& 	\\
$\ccc{1}{Qbqu}{}$	& \texttt{cQbqu1}	& 	& 	& 	& 	& 	& 	& 	\\
$\ccc{8}{Qbqu}{}$	& \texttt{cQbqu8}	& 	& 	& 	& 	& 	& 	& 	\\
$\ccc{1}{btud}{}$	& \texttt{cbtud1}	& 	& 	& 	& 	& 	& 	& 	\\
$\ccc{8}{btud}{}$	& \texttt{cbtud8}	& 	& 	& 	& 	& 	& 	& 	\\
$\ccc{1I}{tQqu}{}$	& \texttt{ctQqu1I}	& 	& 	& 	& 	& 	& 	& 	\\
$\ccc{8I}{tQqu}{}$	& \texttt{ctQqu8I}	& 	& 	& 	& 	& 	& 	& 	\\
$\ccc{1I}{bQqd}{}$	& \texttt{cbQqd1I}	& 	& 	& 	& 	& 	& 	& 	\\
$\ccc{8I}{bQqd}{}$	& \texttt{cbQqd8I}	& 	& 	& 	& 	& 	& 	& 	\\
$\ccc{1I}{Qtqd}{}$	& \texttt{cQtqd1I}	& 	& 	& 	& 	& 	& 	& 	\\
$\ccc{8I}{Qtqd}{}$	& \texttt{cQtqd8I}	& 	& 	& 	& 	& 	& 	& 	\\
$\ccc{1I}{Qbqu}{}$	& \texttt{cQbqu1I}	& 	& 	& 	& 	& 	& 	& 	\\
$\ccc{8I}{Qbqu}{}$	& \texttt{cQbqu8I}	& 	& 	& 	& 	& 	& 	& 	\\
$\ccc{1I}{btud}{}$	& \texttt{cbtud1I}	& 	& 	& 	& 	& 	& 	& 	\\
$\ccc{8I}{btud}{}$	& \texttt{cbtud8I}	& 	& 	& 	& 	& 	& 	& 	\\\hline
\end{tabular}

%% file: interferences_single_top.tex
\begin{tabular}{ll*{5}{c}}
&&$pp\to tj$	& $pp\to t\,e^-\bar{\nu}$	& $pp\to tj\,e^+e^-$	& $pp\to tj\,\gamma$	& $pp\to tj\,h$\\\hline
$\text{SM}$	& \texttt{sm}	& $55$ pb	& $2.5$ pb	& $0.0054$ pb	& $0.39$ pb	& $0.016$ pb	\\\hline
$\ccc{1}{QQ}{}$	& \texttt{cQQ1}	& 	& 	& 	& 	& 	\\
$\ccc{8}{QQ}{}$	& \texttt{cQQ8}	& 	& 	& 	& 	& 	\\
$\ccc{1}{Qt}{}$	& \texttt{cQt1}	& 	& 	& 	& 	& 	\\
$\ccc{8}{Qt}{}$	& \texttt{cQt8}	& 	& 	& 	& 	& 	\\
$\ccc{1}{Qb}{}$	& \texttt{cQb1}	& 	& 	& 	& 	& 	\\
$\ccc{8}{Qb}{}$	& \texttt{cQb8}	& 	& 	& 	& 	& 	\\
$\ccc{1}{tt}{}$	& \texttt{ctt1}	& 	& 	& 	& 	& 	\\
$\ccc{1}{tb}{}$	& \texttt{ctb1}	& 	& 	& 	& 	& 	\\
$\ccc{8}{tb}{}$	& \texttt{ctb8}	& 	& 	& 	& 	& 	\\
$\ccc{1}{QtQb}{}$	& \texttt{cQtQb1}	& 	& 	& 	& 	& 	\\
$\ccc{8}{QtQb}{}$	& \texttt{cQtQb8}	& 	& 	& 	& 	& 	\\
$\ccc{1I}{QtQb}{}$	& \texttt{cQtQb1I}	& 	& 	& 	& 	& 	\\
$\ccc{8I}{QtQb}{}$	& \texttt{cQtQb8I}	& 	& 	& 	& 	& 	\\\hline
$\ccc{3,8}{Qq}{}$	& \texttt{cQq83}	& $-3.4\times 10^{-15}$	& 	& $-6.4\times 10^{-15}$	& $-5.2\times 10^{-15}$	& $-4.1\times 10^{-15}$	\\
$\ccc{1,8}{Qq}{}$	& \texttt{cQq81}	& 	& 	& 	& 	& 	\\
$\ccc{8}{tq}{}$	& \texttt{ctq8}	& 	& 	& 	& 	& 	\\
$\ccc{8}{Qu}{}$	& \texttt{cQu8}	& 	& 	& 	& 	& 	\\
$\ccc{8}{tu}{}$	& \texttt{ctu8}	& 	& 	& 	& 	& 	\\
$\ccc{8}{Qd}{}$	& \texttt{cQd8}	& 	& 	& 	& 	& 	\\
$\ccc{8}{td}{}$	& \texttt{ctd8}	& 	& 	& 	& 	& 	\\
$\ccc{3,1}{Qq}{}$	& \texttt{cQq13}	& $-3.8\times 10^{2}$	& 	& $-7.9\times 10^{2}$	& $-6.1\times 10^{2}$	& $-4.6\times 10^{2}$	\\
$\ccc{1,1}{Qq}{}$	& \texttt{cQq11}	& 	& 	& 	& 	& 	\\
$\ccc{1}{tq}{}$	& \texttt{ctq1}	& 	& 	& 	& 	& 	\\
$\ccc{1}{Qu}{}$	& \texttt{cQu1}	& 	& 	& 	& 	& 	\\
$\ccc{1}{tu}{}$	& \texttt{ctu1}	& 	& 	& 	& 	& 	\\
$\ccc{1}{Qd}{}$	& \texttt{cQd1}	& 	& 	& 	& 	& 	\\
$\ccc{1}{td}{}$	& \texttt{ctd1}	& 	& 	& 	& 	& 	\\\hline
$\ccc{}{t\varphi}{}$	& \texttt{ctp}	& 	& 	& 	& 	& $-68$	\\
$\ccc{-}{\varphi Q}{}$	& \texttt{cpQM}	& 	& 	& $21$	& 	& 	\\
$\ccc{3}{\varphi Q}{}$	& \texttt{cpQ3}	& $1.2\times 10^{2}$	& $1.2\times 10^{2}$	& $2.2\times 10^{2}$	& $1.2\times 10^{2}$	& $1.3\times 10^{2}$	\\
$\ccc{}{\varphi t}{}$	& \texttt{cpt}	& 	& 	& $5.2$	& 	& 	\\
$\ccc{}{\varphi tb}{}$	& \texttt{cptb}	& 	& 	& 	& 	& 	\\
$\ccc{}{tW}{}$	& \texttt{ctW}	& $84$	& $-76$	& $45$	& $50$	& $9.1\times 10^{2}$	\\
$\ccc{}{tZ}{}$	& \texttt{ctZ}	& 	& 	& $-10$	& $-6$	& 	\\
$\ccc{}{bW}{}$	& \texttt{cbW}	& 	& 	& 	& 	& 	\\
$\ccc{}{tG}{}$	& \texttt{ctG}	& 	& $59$	& 	& 	& 	\\
$\ccc{I}{t\varphi}{}$	& \texttt{ctpI}	& 	& 	& 	& 	& $-0.21$	\\
$\ccc{I}{\varphi tb}{}$	& \texttt{cptbI}	& 	& 	& 	& 	& 	\\
$\ccc{I}{tW}{}$	& \texttt{ctWI}	& $1.6\times 10^{-16}$	& $-1.4$	& $0.47$	& $0.022$	& $-0.13$	\\
$\ccc{I}{tZ}{}$	& \texttt{ctZI}	& 	& 	& $-0.87$	& $0.67$	& 	\\
$\ccc{I}{bW}{}$	& \texttt{cbWI}	& 	& 	& 	& 	& 	\\
$\ccc{I}{tG}{}$	& \texttt{ctGI}	& 	& $0.4$	& 	& 	& 	\\\hline
$\ccc{3}{Ql}{1}$	& \texttt{cQl31}	& 	& $4.1$	& $6$	& 	& 	\\
$\ccc{-}{Ql}{1}$	& \texttt{cQlM1}	& 	& 	& $2.2$	& 	& 	\\
$\ccc{}{Qe}{1}$	& \texttt{cQe1}	& 	& 	& $-0.39$	& 	& 	\\
$\ccc{}{tl}{1}$	& \texttt{ctl1}	& 	& 	& $-0.036$	& 	& 	\\
$\ccc{}{te}{1}$	& \texttt{cte1}	& 	& 	& $0.064$	& 	& 	\\
$\ccc{S}{tl}{1}$	& \texttt{ctlS1}	& 	& 	& 	& 	& 	\\
$\ccc{T}{tl}{1}$	& \texttt{ctlT1}	& 	& 	& 	& 	& 	\\
$\ccc{S}{bl}{1}$	& \texttt{cblS1}	& 	& 	& 	& 	& 	\\
$\ccc{SI}{tl}{1}$	& \texttt{ctlSI1}	& 	& 	& 	& 	& 	\\
$\ccc{TI}{tl}{1}$	& \texttt{ctlTI1}	& 	& 	& 	& 	& 	\\
$\ccc{SI}{bl}{1}$	& \texttt{cblSI1}	& 	& 	& 	& 	& 	\\\hline
$\ccc{1}{tQqu}{}$	& \texttt{ctQqu1}	& 	& 	& 	& 	& 	\\
$\ccc{8}{tQqu}{}$	& \texttt{ctQqu8}	& 	& 	& 	& 	& 	\\
$\ccc{1}{bQqd}{}$	& \texttt{cbQqd1}	& 	& 	& 	& 	& 	\\
$\ccc{8}{bQqd}{}$	& \texttt{cbQqd8}	& 	& 	& 	& 	& 	\\
$\ccc{1}{Qtqd}{}$	& \texttt{cQtqd1}	& 	& 	& 	& 	& 	\\
$\ccc{8}{Qtqd}{}$	& \texttt{cQtqd8}	& 	& 	& 	& 	& 	\\
$\ccc{1}{Qbqu}{}$	& \texttt{cQbqu1}	& 	& 	& 	& 	& 	\\
$\ccc{8}{Qbqu}{}$	& \texttt{cQbqu8}	& 	& 	& 	& 	& 	\\
$\ccc{1}{btud}{}$	& \texttt{cbtud1}	& 	& 	& 	& 	& 	\\
$\ccc{8}{btud}{}$	& \texttt{cbtud8}	& 	& 	& 	& 	& 	\\
$\ccc{1I}{tQqu}{}$	& \texttt{ctQqu1I}	& 	& 	& 	& 	& 	\\
$\ccc{8I}{tQqu}{}$	& \texttt{ctQqu8I}	& 	& 	& 	& 	& 	\\
$\ccc{1I}{bQqd}{}$	& \texttt{cbQqd1I}	& 	& 	& 	& 	& 	\\
$\ccc{8I}{bQqd}{}$	& \texttt{cbQqd8I}	& 	& 	& 	& 	& 	\\
$\ccc{1I}{Qtqd}{}$	& \texttt{cQtqd1I}	& 	& 	& 	& 	& 	\\
$\ccc{8I}{Qtqd}{}$	& \texttt{cQtqd8I}	& 	& 	& 	& 	& 	\\
$\ccc{1I}{Qbqu}{}$	& \texttt{cQbqu1I}	& 	& 	& 	& 	& 	\\
$\ccc{8I}{Qbqu}{}$	& \texttt{cQbqu8I}	& 	& 	& 	& 	& 	\\
$\ccc{1I}{btud}{}$	& \texttt{cbtud1I}	& 	& 	& 	& 	& 	\\
$\ccc{8I}{btud}{}$	& \texttt{cbtud8I}	& 	& 	& 	& 	& 	\\\hline
\end{tabular}

%% file: squared_tt.tex
\begin{tabular}{l|ccccccccccccc|cccccccccccccc|ccccccccccccccc|ccccccccccc|cccccccccccccccccccc|}
&$\ccc{1}{QQ}{}$	& $\ccc{8}{QQ}{}$	& $\ccc{1}{Qt}{}$	& $\ccc{8}{Qt}{}$	& $\ccc{1}{Qb}{}$	& $\ccc{8}{Qb}{}$	& $\ccc{1}{tt}{}$	& $\ccc{1}{tb}{}$	& $\ccc{8}{tb}{}$	& $\ccc{1}{QtQb}{}$	& $\ccc{8}{QtQb}{}$	& $\ccc{1I}{QtQb}{}$	& $\ccc{8I}{QtQb}{}$	& $\ccc{3,8}{Qq}{}$	& $\ccc{1,8}{Qq}{}$	& $\ccc{8}{tq}{}$	& $\ccc{8}{Qu}{}$	& $\ccc{8}{tu}{}$	& $\ccc{8}{Qd}{}$	& $\ccc{8}{td}{}$	& $\ccc{3,1}{Qq}{}$	& $\ccc{1,1}{Qq}{}$	& $\ccc{1}{tq}{}$	& $\ccc{1}{Qu}{}$	& $\ccc{1}{tu}{}$	& $\ccc{1}{Qd}{}$	& $\ccc{1}{td}{}$	& $\ccc{}{t\varphi}{}$	& $\ccc{-}{\varphi Q}{}$	& $\ccc{3}{\varphi Q}{}$	& $\ccc{}{\varphi t}{}$	& $\ccc{}{\varphi tb}{}$	& $\ccc{}{tW}{}$	& $\ccc{}{tZ}{}$	& $\ccc{}{bW}{}$	& $\ccc{}{tG}{}$	& $\ccc{I}{t\varphi}{}$	& $\ccc{I}{\varphi tb}{}$	& $\ccc{I}{tW}{}$	& $\ccc{I}{tZ}{}$	& $\ccc{I}{bW}{}$	& $\ccc{I}{tG}{}$	& $\ccc{3}{Ql}{1}$	& $\ccc{-}{Ql}{1}$	& $\ccc{}{Qe}{1}$	& $\ccc{}{tl}{1}$	& $\ccc{}{te}{1}$	& $\ccc{S}{tl}{1}$	& $\ccc{T}{tl}{1}$	& $\ccc{S}{bl}{1}$	& $\ccc{SI}{tl}{1}$	& $\ccc{TI}{tl}{1}$	& $\ccc{SI}{bl}{1}$	& $\ccc{1}{tQqu}{}$	& $\ccc{8}{tQqu}{}$	& $\ccc{1}{bQqd}{}$	& $\ccc{8}{bQqd}{}$	& $\ccc{1}{Qtqd}{}$	& $\ccc{8}{Qtqd}{}$	& $\ccc{1}{Qbqu}{}$	& $\ccc{8}{Qbqu}{}$	& $\ccc{1}{btud}{}$	& $\ccc{8}{btud}{}$	& $\ccc{1I}{tQqu}{}$	& $\ccc{8I}{tQqu}{}$	& $\ccc{1I}{bQqd}{}$	& $\ccc{8I}{bQqd}{}$	& $\ccc{1I}{Qtqd}{}$	& $\ccc{8I}{Qtqd}{}$	& $\ccc{1I}{Qbqu}{}$	& $\ccc{8I}{Qbqu}{}$	& $\ccc{1I}{btud}{}$	& $\ccc{8I}{btud}{}$
\\
\hline
\end{tabular}

%% file: squared_ttee.tex
\begin{tabular}{l|ccccccccccccc|cccccccccccccc|ccccccccccccccc|ccccccccccc|cccccccccccccccccccc|}
&$\ccc{1}{QQ}{}$	& $\ccc{8}{QQ}{}$	& $\ccc{1}{Qt}{}$	& $\ccc{8}{Qt}{}$	& $\ccc{1}{Qb}{}$	& $\ccc{8}{Qb}{}$	& $\ccc{1}{tt}{}$	& $\ccc{1}{tb}{}$	& $\ccc{8}{tb}{}$	& $\ccc{1}{QtQb}{}$	& $\ccc{8}{QtQb}{}$	& $\ccc{1I}{QtQb}{}$	& $\ccc{8I}{QtQb}{}$	& $\ccc{3,8}{Qq}{}$	& $\ccc{1,8}{Qq}{}$	& $\ccc{8}{tq}{}$	& $\ccc{8}{Qu}{}$	& $\ccc{8}{tu}{}$	& $\ccc{8}{Qd}{}$	& $\ccc{8}{td}{}$	& $\ccc{3,1}{Qq}{}$	& $\ccc{1,1}{Qq}{}$	& $\ccc{1}{tq}{}$	& $\ccc{1}{Qu}{}$	& $\ccc{1}{tu}{}$	& $\ccc{1}{Qd}{}$	& $\ccc{1}{td}{}$	& $\ccc{}{t\varphi}{}$	& $\ccc{-}{\varphi Q}{}$	& $\ccc{3}{\varphi Q}{}$	& $\ccc{}{\varphi t}{}$	& $\ccc{}{\varphi tb}{}$	& $\ccc{}{tW}{}$	& $\ccc{}{tZ}{}$	& $\ccc{}{bW}{}$	& $\ccc{}{tG}{}$	& $\ccc{I}{t\varphi}{}$	& $\ccc{I}{\varphi tb}{}$	& $\ccc{I}{tW}{}$	& $\ccc{I}{tZ}{}$	& $\ccc{I}{bW}{}$	& $\ccc{I}{tG}{}$	& $\ccc{3}{Ql}{1}$	& $\ccc{-}{Ql}{1}$	& $\ccc{}{Qe}{1}$	& $\ccc{}{tl}{1}$	& $\ccc{}{te}{1}$	& $\ccc{S}{tl}{1}$	& $\ccc{T}{tl}{1}$	& $\ccc{S}{bl}{1}$	& $\ccc{SI}{tl}{1}$	& $\ccc{TI}{tl}{1}$	& $\ccc{SI}{bl}{1}$	& $\ccc{1}{tQqu}{}$	& $\ccc{8}{tQqu}{}$	& $\ccc{1}{bQqd}{}$	& $\ccc{8}{bQqd}{}$	& $\ccc{1}{Qtqd}{}$	& $\ccc{8}{Qtqd}{}$	& $\ccc{1}{Qbqu}{}$	& $\ccc{8}{Qbqu}{}$	& $\ccc{1}{btud}{}$	& $\ccc{8}{btud}{}$	& $\ccc{1I}{tQqu}{}$	& $\ccc{8I}{tQqu}{}$	& $\ccc{1I}{bQqd}{}$	& $\ccc{8I}{bQqd}{}$	& $\ccc{1I}{Qtqd}{}$	& $\ccc{8I}{Qtqd}{}$	& $\ccc{1I}{Qbqu}{}$	& $\ccc{8I}{Qbqu}{}$	& $\ccc{1I}{btud}{}$	& $\ccc{8I}{btud}{}$
\\
\hline
\end{tabular}

%% file: squared_ttev.tex
\begin{tabular}{l|ccccccccccccc|cccccccccccccc|ccccccccccccccc|ccccccccccc|cccccccccccccccccccc|}
&$\ccc{1}{QQ}{}$	& $\ccc{8}{QQ}{}$	& $\ccc{1}{Qt}{}$	& $\ccc{8}{Qt}{}$	& $\ccc{1}{Qb}{}$	& $\ccc{8}{Qb}{}$	& $\ccc{1}{tt}{}$	& $\ccc{1}{tb}{}$	& $\ccc{8}{tb}{}$	& $\ccc{1}{QtQb}{}$	& $\ccc{8}{QtQb}{}$	& $\ccc{1I}{QtQb}{}$	& $\ccc{8I}{QtQb}{}$	& $\ccc{3,8}{Qq}{}$	& $\ccc{1,8}{Qq}{}$	& $\ccc{8}{tq}{}$	& $\ccc{8}{Qu}{}$	& $\ccc{8}{tu}{}$	& $\ccc{8}{Qd}{}$	& $\ccc{8}{td}{}$	& $\ccc{3,1}{Qq}{}$	& $\ccc{1,1}{Qq}{}$	& $\ccc{1}{tq}{}$	& $\ccc{1}{Qu}{}$	& $\ccc{1}{tu}{}$	& $\ccc{1}{Qd}{}$	& $\ccc{1}{td}{}$	& $\ccc{}{t\varphi}{}$	& $\ccc{-}{\varphi Q}{}$	& $\ccc{3}{\varphi Q}{}$	& $\ccc{}{\varphi t}{}$	& $\ccc{}{\varphi tb}{}$	& $\ccc{}{tW}{}$	& $\ccc{}{tZ}{}$	& $\ccc{}{bW}{}$	& $\ccc{}{tG}{}$	& $\ccc{I}{t\varphi}{}$	& $\ccc{I}{\varphi tb}{}$	& $\ccc{I}{tW}{}$	& $\ccc{I}{tZ}{}$	& $\ccc{I}{bW}{}$	& $\ccc{I}{tG}{}$	& $\ccc{3}{Ql}{1}$	& $\ccc{-}{Ql}{1}$	& $\ccc{}{Qe}{1}$	& $\ccc{}{tl}{1}$	& $\ccc{}{te}{1}$	& $\ccc{S}{tl}{1}$	& $\ccc{T}{tl}{1}$	& $\ccc{S}{bl}{1}$	& $\ccc{SI}{tl}{1}$	& $\ccc{TI}{tl}{1}$	& $\ccc{SI}{bl}{1}$	& $\ccc{1}{tQqu}{}$	& $\ccc{8}{tQqu}{}$	& $\ccc{1}{bQqd}{}$	& $\ccc{8}{bQqd}{}$	& $\ccc{1}{Qtqd}{}$	& $\ccc{8}{Qtqd}{}$	& $\ccc{1}{Qbqu}{}$	& $\ccc{8}{Qbqu}{}$	& $\ccc{1}{btud}{}$	& $\ccc{8}{btud}{}$	& $\ccc{1I}{tQqu}{}$	& $\ccc{8I}{tQqu}{}$	& $\ccc{1I}{bQqd}{}$	& $\ccc{8I}{bQqd}{}$	& $\ccc{1I}{Qtqd}{}$	& $\ccc{8I}{Qtqd}{}$	& $\ccc{1I}{Qbqu}{}$	& $\ccc{8I}{Qbqu}{}$	& $\ccc{1I}{btud}{}$	& $\ccc{8I}{btud}{}$
\\
\hline
\end{tabular}

%% file: squared_tta.tex
\begin{tabular}{l|ccccccccccccc|cccccccccccccc|ccccccccccccccc|ccccccccccc|cccccccccccccccccccc|}
&$\ccc{1}{QQ}{}$	& $\ccc{8}{QQ}{}$	& $\ccc{1}{Qt}{}$	& $\ccc{8}{Qt}{}$	& $\ccc{1}{Qb}{}$	& $\ccc{8}{Qb}{}$	& $\ccc{1}{tt}{}$	& $\ccc{1}{tb}{}$	& $\ccc{8}{tb}{}$	& $\ccc{1}{QtQb}{}$	& $\ccc{8}{QtQb}{}$	& $\ccc{1I}{QtQb}{}$	& $\ccc{8I}{QtQb}{}$	& $\ccc{3,8}{Qq}{}$	& $\ccc{1,8}{Qq}{}$	& $\ccc{8}{tq}{}$	& $\ccc{8}{Qu}{}$	& $\ccc{8}{tu}{}$	& $\ccc{8}{Qd}{}$	& $\ccc{8}{td}{}$	& $\ccc{3,1}{Qq}{}$	& $\ccc{1,1}{Qq}{}$	& $\ccc{1}{tq}{}$	& $\ccc{1}{Qu}{}$	& $\ccc{1}{tu}{}$	& $\ccc{1}{Qd}{}$	& $\ccc{1}{td}{}$	& $\ccc{}{t\varphi}{}$	& $\ccc{-}{\varphi Q}{}$	& $\ccc{3}{\varphi Q}{}$	& $\ccc{}{\varphi t}{}$	& $\ccc{}{\varphi tb}{}$	& $\ccc{}{tW}{}$	& $\ccc{}{tZ}{}$	& $\ccc{}{bW}{}$	& $\ccc{}{tG}{}$	& $\ccc{I}{t\varphi}{}$	& $\ccc{I}{\varphi tb}{}$	& $\ccc{I}{tW}{}$	& $\ccc{I}{tZ}{}$	& $\ccc{I}{bW}{}$	& $\ccc{I}{tG}{}$	& $\ccc{3}{Ql}{1}$	& $\ccc{-}{Ql}{1}$	& $\ccc{}{Qe}{1}$	& $\ccc{}{tl}{1}$	& $\ccc{}{te}{1}$	& $\ccc{S}{tl}{1}$	& $\ccc{T}{tl}{1}$	& $\ccc{S}{bl}{1}$	& $\ccc{SI}{tl}{1}$	& $\ccc{TI}{tl}{1}$	& $\ccc{SI}{bl}{1}$	& $\ccc{1}{tQqu}{}$	& $\ccc{8}{tQqu}{}$	& $\ccc{1}{bQqd}{}$	& $\ccc{8}{bQqd}{}$	& $\ccc{1}{Qtqd}{}$	& $\ccc{8}{Qtqd}{}$	& $\ccc{1}{Qbqu}{}$	& $\ccc{8}{Qbqu}{}$	& $\ccc{1}{btud}{}$	& $\ccc{8}{btud}{}$	& $\ccc{1I}{tQqu}{}$	& $\ccc{8I}{tQqu}{}$	& $\ccc{1I}{bQqd}{}$	& $\ccc{8I}{bQqd}{}$	& $\ccc{1I}{Qtqd}{}$	& $\ccc{8I}{Qtqd}{}$	& $\ccc{1I}{Qbqu}{}$	& $\ccc{8I}{Qbqu}{}$	& $\ccc{1I}{btud}{}$	& $\ccc{8I}{btud}{}$
\\
\hline
\end{tabular}

%% file: squared_tth.tex
\begin{tabular}{l|ccccccccccccc|cccccccccccccc|ccccccccccccccc|ccccccccccc|cccccccccccccccccccc|}
&$\ccc{1}{QQ}{}$	& $\ccc{8}{QQ}{}$	& $\ccc{1}{Qt}{}$	& $\ccc{8}{Qt}{}$	& $\ccc{1}{Qb}{}$	& $\ccc{8}{Qb}{}$	& $\ccc{1}{tt}{}$	& $\ccc{1}{tb}{}$	& $\ccc{8}{tb}{}$	& $\ccc{1}{QtQb}{}$	& $\ccc{8}{QtQb}{}$	& $\ccc{1I}{QtQb}{}$	& $\ccc{8I}{QtQb}{}$	& $\ccc{3,8}{Qq}{}$	& $\ccc{1,8}{Qq}{}$	& $\ccc{8}{tq}{}$	& $\ccc{8}{Qu}{}$	& $\ccc{8}{tu}{}$	& $\ccc{8}{Qd}{}$	& $\ccc{8}{td}{}$	& $\ccc{3,1}{Qq}{}$	& $\ccc{1,1}{Qq}{}$	& $\ccc{1}{tq}{}$	& $\ccc{1}{Qu}{}$	& $\ccc{1}{tu}{}$	& $\ccc{1}{Qd}{}$	& $\ccc{1}{td}{}$	& $\ccc{}{t\varphi}{}$	& $\ccc{-}{\varphi Q}{}$	& $\ccc{3}{\varphi Q}{}$	& $\ccc{}{\varphi t}{}$	& $\ccc{}{\varphi tb}{}$	& $\ccc{}{tW}{}$	& $\ccc{}{tZ}{}$	& $\ccc{}{bW}{}$	& $\ccc{}{tG}{}$	& $\ccc{I}{t\varphi}{}$	& $\ccc{I}{\varphi tb}{}$	& $\ccc{I}{tW}{}$	& $\ccc{I}{tZ}{}$	& $\ccc{I}{bW}{}$	& $\ccc{I}{tG}{}$	& $\ccc{3}{Ql}{1}$	& $\ccc{-}{Ql}{1}$	& $\ccc{}{Qe}{1}$	& $\ccc{}{tl}{1}$	& $\ccc{}{te}{1}$	& $\ccc{S}{tl}{1}$	& $\ccc{T}{tl}{1}$	& $\ccc{S}{bl}{1}$	& $\ccc{SI}{tl}{1}$	& $\ccc{TI}{tl}{1}$	& $\ccc{SI}{bl}{1}$	& $\ccc{1}{tQqu}{}$	& $\ccc{8}{tQqu}{}$	& $\ccc{1}{bQqd}{}$	& $\ccc{8}{bQqd}{}$	& $\ccc{1}{Qtqd}{}$	& $\ccc{8}{Qtqd}{}$	& $\ccc{1}{Qbqu}{}$	& $\ccc{8}{Qbqu}{}$	& $\ccc{1}{btud}{}$	& $\ccc{8}{btud}{}$	& $\ccc{1I}{tQqu}{}$	& $\ccc{8I}{tQqu}{}$	& $\ccc{1I}{bQqd}{}$	& $\ccc{8I}{bQqd}{}$	& $\ccc{1I}{Qtqd}{}$	& $\ccc{8I}{Qtqd}{}$	& $\ccc{1I}{Qbqu}{}$	& $\ccc{8I}{Qbqu}{}$	& $\ccc{1I}{btud}{}$	& $\ccc{8I}{btud}{}$
\\
\hline
\end{tabular}

%% file: squared_tj.tex
\begin{tabular}{l|ccccccccccccc|cccccccccccccc|ccccccccccccccc|ccccccccccc|cccccccccccccccccccc|}
&$\ccc{1}{QQ}{}$	& $\ccc{8}{QQ}{}$	& $\ccc{1}{Qt}{}$	& $\ccc{8}{Qt}{}$	& $\ccc{1}{Qb}{}$	& $\ccc{8}{Qb}{}$	& $\ccc{1}{tt}{}$	& $\ccc{1}{tb}{}$	& $\ccc{8}{tb}{}$	& $\ccc{1}{QtQb}{}$	& $\ccc{8}{QtQb}{}$	& $\ccc{1I}{QtQb}{}$	& $\ccc{8I}{QtQb}{}$	& $\ccc{3,8}{Qq}{}$	& $\ccc{1,8}{Qq}{}$	& $\ccc{8}{tq}{}$	& $\ccc{8}{Qu}{}$	& $\ccc{8}{tu}{}$	& $\ccc{8}{Qd}{}$	& $\ccc{8}{td}{}$	& $\ccc{3,1}{Qq}{}$	& $\ccc{1,1}{Qq}{}$	& $\ccc{1}{tq}{}$	& $\ccc{1}{Qu}{}$	& $\ccc{1}{tu}{}$	& $\ccc{1}{Qd}{}$	& $\ccc{1}{td}{}$	& $\ccc{}{t\varphi}{}$	& $\ccc{-}{\varphi Q}{}$	& $\ccc{3}{\varphi Q}{}$	& $\ccc{}{\varphi t}{}$	& $\ccc{}{\varphi tb}{}$	& $\ccc{}{tW}{}$	& $\ccc{}{tZ}{}$	& $\ccc{}{bW}{}$	& $\ccc{}{tG}{}$	& $\ccc{I}{t\varphi}{}$	& $\ccc{I}{\varphi tb}{}$	& $\ccc{I}{tW}{}$	& $\ccc{I}{tZ}{}$	& $\ccc{I}{bW}{}$	& $\ccc{I}{tG}{}$	& $\ccc{3}{Ql}{1}$	& $\ccc{-}{Ql}{1}$	& $\ccc{}{Qe}{1}$	& $\ccc{}{tl}{1}$	& $\ccc{}{te}{1}$	& $\ccc{S}{tl}{1}$	& $\ccc{T}{tl}{1}$	& $\ccc{S}{bl}{1}$	& $\ccc{SI}{tl}{1}$	& $\ccc{TI}{tl}{1}$	& $\ccc{SI}{bl}{1}$	& $\ccc{1}{tQqu}{}$	& $\ccc{8}{tQqu}{}$	& $\ccc{1}{bQqd}{}$	& $\ccc{8}{bQqd}{}$	& $\ccc{1}{Qtqd}{}$	& $\ccc{8}{Qtqd}{}$	& $\ccc{1}{Qbqu}{}$	& $\ccc{8}{Qbqu}{}$	& $\ccc{1}{btud}{}$	& $\ccc{8}{btud}{}$	& $\ccc{1I}{tQqu}{}$	& $\ccc{8I}{tQqu}{}$	& $\ccc{1I}{bQqd}{}$	& $\ccc{8I}{bQqd}{}$	& $\ccc{1I}{Qtqd}{}$	& $\ccc{8I}{Qtqd}{}$	& $\ccc{1I}{Qbqu}{}$	& $\ccc{8I}{Qbqu}{}$	& $\ccc{1I}{btud}{}$	& $\ccc{8I}{btud}{}$
\\
\hline
\end{tabular}

%% file: squared_tev.tex
\begin{tabular}{l|ccccccccccccc|cccccccccccccc|ccccccccccccccc|ccccccccccc|cccccccccccccccccccc|}
&$\ccc{1}{QQ}{}$	& $\ccc{8}{QQ}{}$	& $\ccc{1}{Qt}{}$	& $\ccc{8}{Qt}{}$	& $\ccc{1}{Qb}{}$	& $\ccc{8}{Qb}{}$	& $\ccc{1}{tt}{}$	& $\ccc{1}{tb}{}$	& $\ccc{8}{tb}{}$	& $\ccc{1}{QtQb}{}$	& $\ccc{8}{QtQb}{}$	& $\ccc{1I}{QtQb}{}$	& $\ccc{8I}{QtQb}{}$	& $\ccc{3,8}{Qq}{}$	& $\ccc{1,8}{Qq}{}$	& $\ccc{8}{tq}{}$	& $\ccc{8}{Qu}{}$	& $\ccc{8}{tu}{}$	& $\ccc{8}{Qd}{}$	& $\ccc{8}{td}{}$	& $\ccc{3,1}{Qq}{}$	& $\ccc{1,1}{Qq}{}$	& $\ccc{1}{tq}{}$	& $\ccc{1}{Qu}{}$	& $\ccc{1}{tu}{}$	& $\ccc{1}{Qd}{}$	& $\ccc{1}{td}{}$	& $\ccc{}{t\varphi}{}$	& $\ccc{-}{\varphi Q}{}$	& $\ccc{3}{\varphi Q}{}$	& $\ccc{}{\varphi t}{}$	& $\ccc{}{\varphi tb}{}$	& $\ccc{}{tW}{}$	& $\ccc{}{tZ}{}$	& $\ccc{}{bW}{}$	& $\ccc{}{tG}{}$	& $\ccc{I}{t\varphi}{}$	& $\ccc{I}{\varphi tb}{}$	& $\ccc{I}{tW}{}$	& $\ccc{I}{tZ}{}$	& $\ccc{I}{bW}{}$	& $\ccc{I}{tG}{}$	& $\ccc{3}{Ql}{1}$	& $\ccc{-}{Ql}{1}$	& $\ccc{}{Qe}{1}$	& $\ccc{}{tl}{1}$	& $\ccc{}{te}{1}$	& $\ccc{S}{tl}{1}$	& $\ccc{T}{tl}{1}$	& $\ccc{S}{bl}{1}$	& $\ccc{SI}{tl}{1}$	& $\ccc{TI}{tl}{1}$	& $\ccc{SI}{bl}{1}$	& $\ccc{1}{tQqu}{}$	& $\ccc{8}{tQqu}{}$	& $\ccc{1}{bQqd}{}$	& $\ccc{8}{bQqd}{}$	& $\ccc{1}{Qtqd}{}$	& $\ccc{8}{Qtqd}{}$	& $\ccc{1}{Qbqu}{}$	& $\ccc{8}{Qbqu}{}$	& $\ccc{1}{btud}{}$	& $\ccc{8}{btud}{}$	& $\ccc{1I}{tQqu}{}$	& $\ccc{8I}{tQqu}{}$	& $\ccc{1I}{bQqd}{}$	& $\ccc{8I}{bQqd}{}$	& $\ccc{1I}{Qtqd}{}$	& $\ccc{8I}{Qtqd}{}$	& $\ccc{1I}{Qbqu}{}$	& $\ccc{8I}{Qbqu}{}$	& $\ccc{1I}{btud}{}$	& $\ccc{8I}{btud}{}$
\\
\hline
\end{tabular}

%% file: squared_tjee.tex
\begin{tabular}{l|ccccccccccccc|cccccccccccccc|ccccccccccccccc|ccccccccccc|cccccccccccccccccccc|}
&$\ccc{1}{QQ}{}$	& $\ccc{8}{QQ}{}$	& $\ccc{1}{Qt}{}$	& $\ccc{8}{Qt}{}$	& $\ccc{1}{Qb}{}$	& $\ccc{8}{Qb}{}$	& $\ccc{1}{tt}{}$	& $\ccc{1}{tb}{}$	& $\ccc{8}{tb}{}$	& $\ccc{1}{QtQb}{}$	& $\ccc{8}{QtQb}{}$	& $\ccc{1I}{QtQb}{}$	& $\ccc{8I}{QtQb}{}$	& $\ccc{3,8}{Qq}{}$	& $\ccc{1,8}{Qq}{}$	& $\ccc{8}{tq}{}$	& $\ccc{8}{Qu}{}$	& $\ccc{8}{tu}{}$	& $\ccc{8}{Qd}{}$	& $\ccc{8}{td}{}$	& $\ccc{3,1}{Qq}{}$	& $\ccc{1,1}{Qq}{}$	& $\ccc{1}{tq}{}$	& $\ccc{1}{Qu}{}$	& $\ccc{1}{tu}{}$	& $\ccc{1}{Qd}{}$	& $\ccc{1}{td}{}$	& $\ccc{}{t\varphi}{}$	& $\ccc{-}{\varphi Q}{}$	& $\ccc{3}{\varphi Q}{}$	& $\ccc{}{\varphi t}{}$	& $\ccc{}{\varphi tb}{}$	& $\ccc{}{tW}{}$	& $\ccc{}{tZ}{}$	& $\ccc{}{bW}{}$	& $\ccc{}{tG}{}$	& $\ccc{I}{t\varphi}{}$	& $\ccc{I}{\varphi tb}{}$	& $\ccc{I}{tW}{}$	& $\ccc{I}{tZ}{}$	& $\ccc{I}{bW}{}$	& $\ccc{I}{tG}{}$	& $\ccc{3}{Ql}{1}$	& $\ccc{-}{Ql}{1}$	& $\ccc{}{Qe}{1}$	& $\ccc{}{tl}{1}$	& $\ccc{}{te}{1}$	& $\ccc{S}{tl}{1}$	& $\ccc{T}{tl}{1}$	& $\ccc{S}{bl}{1}$	& $\ccc{SI}{tl}{1}$	& $\ccc{TI}{tl}{1}$	& $\ccc{SI}{bl}{1}$	& $\ccc{1}{tQqu}{}$	& $\ccc{8}{tQqu}{}$	& $\ccc{1}{bQqd}{}$	& $\ccc{8}{bQqd}{}$	& $\ccc{1}{Qtqd}{}$	& $\ccc{8}{Qtqd}{}$	& $\ccc{1}{Qbqu}{}$	& $\ccc{8}{Qbqu}{}$	& $\ccc{1}{btud}{}$	& $\ccc{8}{btud}{}$	& $\ccc{1I}{tQqu}{}$	& $\ccc{8I}{tQqu}{}$	& $\ccc{1I}{bQqd}{}$	& $\ccc{8I}{bQqd}{}$	& $\ccc{1I}{Qtqd}{}$	& $\ccc{8I}{Qtqd}{}$	& $\ccc{1I}{Qbqu}{}$	& $\ccc{8I}{Qbqu}{}$	& $\ccc{1I}{btud}{}$	& $\ccc{8I}{btud}{}$
\\
\hline
\end{tabular}

%% file: squared_tja.tex
\begin{tabular}{l|ccccccccccccc|cccccccccccccc|ccccccccccccccc|ccccccccccc|cccccccccccccccccccc|}
&$\ccc{1}{QQ}{}$	& $\ccc{8}{QQ}{}$	& $\ccc{1}{Qt}{}$	& $\ccc{8}{Qt}{}$	& $\ccc{1}{Qb}{}$	& $\ccc{8}{Qb}{}$	& $\ccc{1}{tt}{}$	& $\ccc{1}{tb}{}$	& $\ccc{8}{tb}{}$	& $\ccc{1}{QtQb}{}$	& $\ccc{8}{QtQb}{}$	& $\ccc{1I}{QtQb}{}$	& $\ccc{8I}{QtQb}{}$	& $\ccc{3,8}{Qq}{}$	& $\ccc{1,8}{Qq}{}$	& $\ccc{8}{tq}{}$	& $\ccc{8}{Qu}{}$	& $\ccc{8}{tu}{}$	& $\ccc{8}{Qd}{}$	& $\ccc{8}{td}{}$	& $\ccc{3,1}{Qq}{}$	& $\ccc{1,1}{Qq}{}$	& $\ccc{1}{tq}{}$	& $\ccc{1}{Qu}{}$	& $\ccc{1}{tu}{}$	& $\ccc{1}{Qd}{}$	& $\ccc{1}{td}{}$	& $\ccc{}{t\varphi}{}$	& $\ccc{-}{\varphi Q}{}$	& $\ccc{3}{\varphi Q}{}$	& $\ccc{}{\varphi t}{}$	& $\ccc{}{\varphi tb}{}$	& $\ccc{}{tW}{}$	& $\ccc{}{tZ}{}$	& $\ccc{}{bW}{}$	& $\ccc{}{tG}{}$	& $\ccc{I}{t\varphi}{}$	& $\ccc{I}{\varphi tb}{}$	& $\ccc{I}{tW}{}$	& $\ccc{I}{tZ}{}$	& $\ccc{I}{bW}{}$	& $\ccc{I}{tG}{}$	& $\ccc{3}{Ql}{1}$	& $\ccc{-}{Ql}{1}$	& $\ccc{}{Qe}{1}$	& $\ccc{}{tl}{1}$	& $\ccc{}{te}{1}$	& $\ccc{S}{tl}{1}$	& $\ccc{T}{tl}{1}$	& $\ccc{S}{bl}{1}$	& $\ccc{SI}{tl}{1}$	& $\ccc{TI}{tl}{1}$	& $\ccc{SI}{bl}{1}$	& $\ccc{1}{tQqu}{}$	& $\ccc{8}{tQqu}{}$	& $\ccc{1}{bQqd}{}$	& $\ccc{8}{bQqd}{}$	& $\ccc{1}{Qtqd}{}$	& $\ccc{8}{Qtqd}{}$	& $\ccc{1}{Qbqu}{}$	& $\ccc{8}{Qbqu}{}$	& $\ccc{1}{btud}{}$	& $\ccc{8}{btud}{}$	& $\ccc{1I}{tQqu}{}$	& $\ccc{8I}{tQqu}{}$	& $\ccc{1I}{bQqd}{}$	& $\ccc{8I}{bQqd}{}$	& $\ccc{1I}{Qtqd}{}$	& $\ccc{8I}{Qtqd}{}$	& $\ccc{1I}{Qbqu}{}$	& $\ccc{8I}{Qbqu}{}$	& $\ccc{1I}{btud}{}$	& $\ccc{8I}{btud}{}$
\\
\hline
\end{tabular}

%% file: squared_tjh.tex
\begin{tabular}{l|ccccccccccccc|cccccccccccccc|ccccccccccccccc|ccccccccccc|cccccccccccccccccccc|}
&$\ccc{1}{QQ}{}$	& $\ccc{8}{QQ}{}$	& $\ccc{1}{Qt}{}$	& $\ccc{8}{Qt}{}$	& $\ccc{1}{Qb}{}$	& $\ccc{8}{Qb}{}$	& $\ccc{1}{tt}{}$	& $\ccc{1}{tb}{}$	& $\ccc{8}{tb}{}$	& $\ccc{1}{QtQb}{}$	& $\ccc{8}{QtQb}{}$	& $\ccc{1I}{QtQb}{}$	& $\ccc{8I}{QtQb}{}$	& $\ccc{3,8}{Qq}{}$	& $\ccc{1,8}{Qq}{}$	& $\ccc{8}{tq}{}$	& $\ccc{8}{Qu}{}$	& $\ccc{8}{tu}{}$	& $\ccc{8}{Qd}{}$	& $\ccc{8}{td}{}$	& $\ccc{3,1}{Qq}{}$	& $\ccc{1,1}{Qq}{}$	& $\ccc{1}{tq}{}$	& $\ccc{1}{Qu}{}$	& $\ccc{1}{tu}{}$	& $\ccc{1}{Qd}{}$	& $\ccc{1}{td}{}$	& $\ccc{}{t\varphi}{}$	& $\ccc{-}{\varphi Q}{}$	& $\ccc{3}{\varphi Q}{}$	& $\ccc{}{\varphi t}{}$	& $\ccc{}{\varphi tb}{}$	& $\ccc{}{tW}{}$	& $\ccc{}{tZ}{}$	& $\ccc{}{bW}{}$	& $\ccc{}{tG}{}$	& $\ccc{I}{t\varphi}{}$	& $\ccc{I}{\varphi tb}{}$	& $\ccc{I}{tW}{}$	& $\ccc{I}{tZ}{}$	& $\ccc{I}{bW}{}$	& $\ccc{I}{tG}{}$	& $\ccc{3}{Ql}{1}$	& $\ccc{-}{Ql}{1}$	& $\ccc{}{Qe}{1}$	& $\ccc{}{tl}{1}$	& $\ccc{}{te}{1}$	& $\ccc{S}{tl}{1}$	& $\ccc{T}{tl}{1}$	& $\ccc{S}{bl}{1}$	& $\ccc{SI}{tl}{1}$	& $\ccc{TI}{tl}{1}$	& $\ccc{SI}{bl}{1}$	& $\ccc{1}{tQqu}{}$	& $\ccc{8}{tQqu}{}$	& $\ccc{1}{bQqd}{}$	& $\ccc{8}{bQqd}{}$	& $\ccc{1}{Qtqd}{}$	& $\ccc{8}{Qtqd}{}$	& $\ccc{1}{Qbqu}{}$	& $\ccc{8}{Qbqu}{}$	& $\ccc{1}{btud}{}$	& $\ccc{8}{btud}{}$	& $\ccc{1I}{tQqu}{}$	& $\ccc{8I}{tQqu}{}$	& $\ccc{1I}{bQqd}{}$	& $\ccc{8I}{bQqd}{}$	& $\ccc{1I}{Qtqd}{}$	& $\ccc{8I}{Qtqd}{}$	& $\ccc{1I}{Qbqu}{}$	& $\ccc{8I}{Qbqu}{}$	& $\ccc{1I}{btud}{}$	& $\ccc{8I}{btud}{}$
\\
\hline
\end{tabular}

%% file: eft_note.bbl
\begin{thebibliography}{107}%
\makeatletter
\providecommand \@ifxundefined [1]{%
 \@ifx{#1\undefined}
}%
\providecommand \@ifnum [1]{%
 \ifnum #1\expandafter \@firstoftwo
 \else \expandafter \@secondoftwo
 \fi
}%
\providecommand \@ifx [1]{%
 \ifx #1\expandafter \@firstoftwo
 \else \expandafter \@secondoftwo
 \fi
}%
\providecommand \natexlab [1]{#1}%
\providecommand \enquote  [1]{``#1''}%
\providecommand \bibnamefont  [1]{#1}%
\providecommand \bibfnamefont [1]{#1}%
\providecommand \citenamefont [1]{#1}%
\providecommand \href@noop [0]{\@secondoftwo}%
\providecommand \href [0]{\begingroup \@sanitize@url \@href}%
\providecommand \@href[1]{\@@startlink{#1}\@@href}%
\providecommand \@@href[1]{\endgroup#1\@@endlink}%
\providecommand \@sanitize@url [0]{\catcode `\\12\catcode `\$12\catcode
  `\&12\catcode `\#12\catcode `\^12\catcode `\_12\catcode `\%12\relax}%
\providecommand \@@startlink[1]{}%
\providecommand \@@endlink[0]{}%
\providecommand \url  [0]{\begingroup\@sanitize@url \@url }%
\providecommand \@url [1]{\endgroup\@href {#1}{\urlprefix }}%
\providecommand \urlprefix  [0]{URL }%
\providecommand \Eprint [0]{\href }%
\providecommand \doibase [0]{http://dx.doi.org/}%
\providecommand \selectlanguage [0]{\@gobble}%
\providecommand \bibinfo  [0]{\@secondoftwo}%
\providecommand \bibfield  [0]{\@secondoftwo}%
\providecommand \translation [1]{[#1]}%
\providecommand \BibitemOpen [0]{}%
\providecommand \bibitemStop [0]{}%
\providecommand \bibitemNoStop [0]{.\EOS\space}%
\providecommand \EOS [0]{\spacefactor3000\relax}%
\providecommand \BibitemShut  [1]{\csname bibitem#1\endcsname}%
\let\auto@bib@innerbib\@empty
\bibitem [{\citenamefont {Grzadkowski} \emph {et~al.}(2010)\citenamefont
  {Grzadkowski}, \citenamefont {Iskrzynski}, \citenamefont {Misiak}, and
  \citenamefont {Rosiek}}]{Grzadkowski:2010es}%
  \BibitemOpen
  \bibfield  {author} {\bibinfo {author} {\bibfnamefont {B.}~\bibnamefont
  {Grzadkowski}}, \bibinfo {author} {\bibfnamefont {M.}~\bibnamefont
  {Iskrzynski}}, \bibinfo {author} {\bibfnamefont {M.}~\bibnamefont {Misiak}},
  and \bibinfo {author} {\bibfnamefont {J.}~\bibnamefont {Rosiek}}, }\bibfield
  {title} {\emph {\bibinfo {title} {{Dimension-Six Terms in the Standard Model
  Lagrangian}}}, }\href {\doibase 10.1007/JHEP10(2010)085} {\bibfield
  {journal} {\bibinfo  {journal} {JHEP} }\textbf {\bibinfo {volume} {10}}
  (\bibinfo {year} {2010}) \bibinfo {pages} {085}}, \Eprint
  {http://arxiv.org/abs/1008.4884}{arXiv:1008.4884 [hep-ph]}\BibitemShut
  {NoStop}%
\bibitem [{\citenamefont {Aguilar-Saavedra}(2009)}]{AguilarSaavedra:2008zc}%
  \BibitemOpen
  \bibfield  {author} {\bibinfo {author} {\bibfnamefont {J.~A.} \bibnamefont
  {Aguilar-Saavedra}}, }\bibfield  {title} {\emph {\bibinfo {title} {{A Minimal
  set of top anomalous couplings}}}, }\href {\doibase
  10.1016/j.nuclphysb.2008.12.012} {\bibfield  {journal} {\bibinfo  {journal}
  {Nucl. Phys.} }\textbf {\bibinfo {volume} {B812}} (\bibinfo {year} {2009})
  \bibinfo {pages} {181}}, \Eprint
  {http://arxiv.org/abs/0811.3842}{arXiv:0811.3842 [hep-ph]}\BibitemShut
  {NoStop}%
\bibitem [{\citenamefont {Zhang} and \citenamefont
  {Willenbrock}(2011)}]{Zhang:2010dr}%
  \BibitemOpen
  \bibfield  {author} {\bibinfo {author} {\bibfnamefont {C.}~\bibnamefont
  {Zhang}} and \bibinfo {author} {\bibfnamefont {S.}~\bibnamefont
  {Willenbrock}}, }\bibfield  {title} {\emph {\bibinfo {title}
  {{Effective-Field-Theory Approach to Top-Quark Production and Decay}}},
  }\href {\doibase 10.1103/PhysRevD.83.034006} {\bibfield  {journal} {\bibinfo
  {journal} {Phys. Rev.} }\textbf {\bibinfo {volume} {D83}} (\bibinfo {year}
  {2011}) \bibinfo {pages} {034006}}, \Eprint
  {http://arxiv.org/abs/1008.3869}{arXiv:1008.3869 [hep-ph]}\BibitemShut
  {NoStop}%
\bibitem [{\citenamefont {Krauss} \emph {et~al.}(2017)\citenamefont {Krauss},
  \citenamefont {Kuttimalai}, and \citenamefont {Plehn}}]{Krauss:2016ely}%
  \BibitemOpen
  \bibfield  {author} {\bibinfo {author} {\bibfnamefont {F.}~\bibnamefont
  {Krauss}}, \bibinfo {author} {\bibfnamefont {S.}~\bibnamefont {Kuttimalai}},
  and \bibinfo {author} {\bibfnamefont {T.}~\bibnamefont {Plehn}}, }\bibfield
  {title} {\emph {\bibinfo {title} {{LHC multijet events as a probe for
  anomalous dimension-six gluon interactions}}}, }\href {\doibase
  10.1103/PhysRevD.95.035024} {\bibfield  {journal} {\bibinfo  {journal} {Phys.
  Rev.} }\textbf {\bibinfo {volume} {D95}} (\bibinfo {year} {2017}) \bibinfo
  {pages} {035024}}, \Eprint {http://arxiv.org/abs/1611.00767}{arXiv:1611.00767
  [hep-ph]}\BibitemShut {NoStop}%
\bibitem [{\citenamefont {Aguilar-Saavedra}(2011)}]{AguilarSaavedra:2010zi}%
  \BibitemOpen
  \bibfield  {author} {\bibinfo {author} {\bibfnamefont {J.~A.} \bibnamefont
  {Aguilar-Saavedra}}, }\bibfield  {title} {\emph {\bibinfo {title} {{Effective
  four-fermion operators in top physics: A Roadmap}}}, }\href {\doibase
  10.1016/j.nuclphysb.2010.10.015} {\bibfield  {journal} {\bibinfo  {journal}
  {Nucl. Phys.} }\textbf {\bibinfo {volume} {B843}} (\bibinfo {year} {2011})
  \bibinfo {pages} {638}}, \bibinfo {note} {[Erratum: Nucl.
  Phys.B851,443(2011)]}, \Eprint
  {http://arxiv.org/abs/1008.3562}{arXiv:1008.3562 [hep-ph]}\BibitemShut
  {NoStop}%
\bibitem [{\citenamefont {Zhang}(2014)}]{Zhang:2014rja}%
  \BibitemOpen
  \bibfield  {author} {\bibinfo {author} {\bibfnamefont {C.}~\bibnamefont
  {Zhang}}, }\bibfield  {title} {\emph {\bibinfo {title} {{Effective field
  theory approach to top-quark decay at next-to-leading order in QCD}}}, }\href
  {\doibase 10.1103/PhysRevD.90.014008} {\bibfield  {journal} {\bibinfo
  {journal} {Phys. Rev.} }\textbf {\bibinfo {volume} {D90}} (\bibinfo {year}
  {2014}) \bibinfo {pages} {014008}}, \Eprint
  {http://arxiv.org/abs/1404.1264}{arXiv:1404.1264 [hep-ph]}\BibitemShut
  {NoStop}%
\bibitem [{\citenamefont {Degrande} \emph {et~al.}(2015)\citenamefont
  {Degrande}, \citenamefont {Maltoni}, \citenamefont {Wang}, and \citenamefont
  {Zhang}}]{Degrande:2014tta}%
  \BibitemOpen
  \bibfield  {author} {\bibinfo {author} {\bibfnamefont {C.}~\bibnamefont
  {Degrande}}, \bibinfo {author} {\bibfnamefont {F.}~\bibnamefont {Maltoni}},
  \bibinfo {author} {\bibfnamefont {J.}~\bibnamefont {Wang}},  and \bibinfo
  {author} {\bibfnamefont {C.}~\bibnamefont {Zhang}}, }\bibfield  {title}
  {\emph {\bibinfo {title} {{Automatic computations at next-to-leading order in
  QCD for top-quark flavor-changing neutral processes}}}, }\href {\doibase
  10.1103/PhysRevD.91.034024} {\bibfield  {journal} {\bibinfo  {journal} {Phys.
  Rev.} }\textbf {\bibinfo {volume} {D91}} (\bibinfo {year} {2015}) \bibinfo
  {pages} {034024}}, \Eprint {http://arxiv.org/abs/1412.5594}{arXiv:1412.5594
  [hep-ph]}\BibitemShut {NoStop}%
\bibitem [{\citenamefont {Röntsch} and \citenamefont
  {Schulze}(2014)}]{Rontsch:2014cca}%
  \BibitemOpen
  \bibfield  {author} {\bibinfo {author} {\bibfnamefont {R.}~\bibnamefont
  {Röntsch}} and \bibinfo {author} {\bibfnamefont {M.}~\bibnamefont
  {Schulze}}, }\bibfield  {title} {\emph {\bibinfo {title} {{Constraining
  couplings of top quarks to the Z boson in $ t\overline{t} $ + Z production at
  the LHC}}}, }\href {\doibase 10.1007/JHEP09(2015)132,
  10.1007/JHEP07(2014)091} {\bibfield  {journal} {\bibinfo  {journal} {JHEP}
  }\textbf {\bibinfo {volume} {07}} (\bibinfo {year} {2014}) \bibinfo {pages}
  {091}}, \bibinfo {note} {[Erratum: JHEP09,132(2015)]}, \Eprint
  {http://arxiv.org/abs/1404.1005}{arXiv:1404.1005 [hep-ph]}\BibitemShut
  {NoStop}%
\bibitem [{\citenamefont {Buarque~Franzosi} and \citenamefont
  {Zhang}(2015)}]{Franzosi:2015osa}%
  \BibitemOpen
  \bibfield  {author} {\bibinfo {author} {\bibfnamefont {D.}~\bibnamefont
  {Buarque~Franzosi}} and \bibinfo {author} {\bibfnamefont {C.}~\bibnamefont
  {Zhang}}, }\bibfield  {title} {\emph {\bibinfo {title} {{Probing the
  top-quark chromomagnetic dipole moment at next-to-leading order in QCD}}},
  }\href {\doibase 10.1103/PhysRevD.91.114010} {\bibfield  {journal} {\bibinfo
  {journal} {Phys. Rev.} }\textbf {\bibinfo {volume} {D91}} (\bibinfo {year}
  {2015}) \bibinfo {pages} {114010}}, \Eprint
  {http://arxiv.org/abs/1503.08841}{arXiv:1503.08841 [hep-ph]}\BibitemShut
  {NoStop}%
\bibitem [{\citenamefont {Röntsch} and \citenamefont
  {Schulze}(2015)}]{Rontsch:2015una}%
  \BibitemOpen
  \bibfield  {author} {\bibinfo {author} {\bibfnamefont {R.}~\bibnamefont
  {Röntsch}} and \bibinfo {author} {\bibfnamefont {M.}~\bibnamefont
  {Schulze}}, }\bibfield  {title} {\emph {\bibinfo {title} {{Probing top-Z
  dipole moments at the LHC and ILC}}}, }\href {\doibase
  10.1007/JHEP08(2015)044} {\bibfield  {journal} {\bibinfo  {journal} {JHEP}
  }\textbf {\bibinfo {volume} {08}} (\bibinfo {year} {2015}) \bibinfo {pages}
  {044}}, \Eprint {http://arxiv.org/abs/1501.05939}{arXiv:1501.05939
  [hep-ph]}\BibitemShut {NoStop}%
\bibitem [{\citenamefont {Zhang}(2016)}]{Zhang:2016omx}%
  \BibitemOpen
  \bibfield  {author} {\bibinfo {author} {\bibfnamefont {C.}~\bibnamefont
  {Zhang}}, }\bibfield  {title} {\emph {\bibinfo {title} {{Single Top
  Production at Next-to-Leading Order in the Standard Model Effective Field
  Theory}}}, }\href {\doibase 10.1103/PhysRevLett.116.162002} {\bibfield
  {journal} {\bibinfo  {journal} {Phys. Rev. Lett.} }\textbf {\bibinfo {volume}
  {116}} (\bibinfo {year} {2016}) \bibinfo {pages} {162002}}, \Eprint
  {http://arxiv.org/abs/1601.06163}{arXiv:1601.06163 [hep-ph]}\BibitemShut
  {NoStop}%
\bibitem [{\citenamefont {Bessidskaia~Bylund} \emph {et~al.}(2016)\citenamefont
  {Bessidskaia~Bylund}, \citenamefont {Maltoni}, \citenamefont {Tsinikos},
  \citenamefont {Vryonidou}, and \citenamefont {Zhang}}]{Bylund:2016phk}%
  \BibitemOpen
  \bibfield  {author} {\bibinfo {author} {\bibfnamefont {O.}~\bibnamefont
  {Bessidskaia~Bylund}}, \bibinfo {author} {\bibfnamefont {F.}~\bibnamefont
  {Maltoni}}, \bibinfo {author} {\bibfnamefont {I.}~\bibnamefont {Tsinikos}},
  \bibinfo {author} {\bibfnamefont {E.}~\bibnamefont {Vryonidou}},  and
  \bibinfo {author} {\bibfnamefont {C.}~\bibnamefont {Zhang}}, }\bibfield
  {title} {\emph {\bibinfo {title} {{Probing top quark neutral couplings in the
  Standard Model Effective Field Theory at NLO in QCD}}}, }\href {\doibase
  10.1007/JHEP05(2016)052} {\bibfield  {journal} {\bibinfo  {journal} {JHEP}
  }\textbf {\bibinfo {volume} {05}} (\bibinfo {year} {2016}) \bibinfo {pages}
  {052}}, \Eprint {http://arxiv.org/abs/1601.08193}{arXiv:1601.08193
  [hep-ph]}\BibitemShut {NoStop}%
\bibitem [{\citenamefont {Maltoni} \emph {et~al.}(2016)\citenamefont {Maltoni},
  \citenamefont {Vryonidou}, and \citenamefont {Zhang}}]{Maltoni:2016yxb}%
  \BibitemOpen
  \bibfield  {author} {\bibinfo {author} {\bibfnamefont {F.}~\bibnamefont
  {Maltoni}}, \bibinfo {author} {\bibfnamefont {E.}~\bibnamefont {Vryonidou}},
  and \bibinfo {author} {\bibfnamefont {C.}~\bibnamefont {Zhang}}, }\bibfield
  {title} {\emph {\bibinfo {title} {{Higgs production in association with a
  top-antitop pair in the Standard Model Effective Field Theory at NLO in
  QCD}}}, }\href {\doibase 10.1007/JHEP10(2016)123} {\bibfield  {journal}
  {\bibinfo  {journal} {JHEP} }\textbf {\bibinfo {volume} {10}} (\bibinfo
  {year} {2016}) \bibinfo {pages} {123}}, \Eprint
  {http://arxiv.org/abs/1607.05330}{arXiv:1607.05330 [hep-ph]}\BibitemShut
  {NoStop}%
\bibitem [{\citenamefont {Abbott} and \citenamefont
  {Wise}(1980)}]{Abbott:1980zj}%
  \BibitemOpen
  \bibfield  {author} {\bibinfo {author} {\bibfnamefont {L.~F.} \bibnamefont
  {Abbott}} and \bibinfo {author} {\bibfnamefont {M.~B.} \bibnamefont {Wise}},
  }\bibfield  {title} {\emph {\bibinfo {title} {{The Effective Hamiltonian for
  Nucleon Decay}}}, }\href {\doibase 10.1103/PhysRevD.22.2208} {\bibfield
  {journal} {\bibinfo  {journal} {Phys. Rev.} }\textbf {\bibinfo {volume}
  {D22}} (\bibinfo {year} {1980}) \bibinfo {pages} {2208}}\BibitemShut
  {NoStop}%
\bibitem [{\citenamefont {Chivukula} and \citenamefont
  {Georgi}(1987)}]{Chivukula:1987py}%
  \BibitemOpen
  \bibfield  {author} {\bibinfo {author} {\bibfnamefont {R.~S.} \bibnamefont
  {Chivukula}} and \bibinfo {author} {\bibfnamefont {H.}~\bibnamefont
  {Georgi}}, }\bibfield  {title} {\emph {\bibinfo {title} {{Composite
  Technicolor Standard Model}}}, }\href {\doibase 10.1016/0370-2693(87)90713-1}
  {\bibfield  {journal} {\bibinfo  {journal} {Phys. Lett.} }\textbf {\bibinfo
  {volume} {B188}} (\bibinfo {year} {1987}) \bibinfo {pages} {99}}\BibitemShut
  {NoStop}%
\bibitem [{\citenamefont {Hall} and \citenamefont
  {Randall}(1990)}]{Hall:1990ac}%
  \BibitemOpen
  \bibfield  {author} {\bibinfo {author} {\bibfnamefont {L.~J.} \bibnamefont
  {Hall}} and \bibinfo {author} {\bibfnamefont {L.}~\bibnamefont {Randall}},
  }\bibfield  {title} {\emph {\bibinfo {title} {{Weak scale effective
  supersymmetry}}}, }\href {\doibase 10.1103/PhysRevLett.65.2939} {\bibfield
  {journal} {\bibinfo  {journal} {Phys. Rev. Lett.} }\textbf {\bibinfo {volume}
  {65}} (\bibinfo {year} {1990}) \bibinfo {pages} {2939}}\BibitemShut {NoStop}%
\bibitem [{\citenamefont {D'Ambrosio} \emph {et~al.}(2002)\citenamefont
  {D'Ambrosio}, \citenamefont {Giudice}, \citenamefont {Isidori}, and
  \citenamefont {Strumia}}]{DAmbrosio:2002vsn}%
  \BibitemOpen
  \bibfield  {author} {\bibinfo {author} {\bibfnamefont {G.}~\bibnamefont
  {D'Ambrosio}}, \bibinfo {author} {\bibfnamefont {G.~F.} \bibnamefont
  {Giudice}}, \bibinfo {author} {\bibfnamefont {G.}~\bibnamefont {Isidori}},
  and \bibinfo {author} {\bibfnamefont {A.}~\bibnamefont {Strumia}}, }\bibfield
   {title} {\emph {\bibinfo {title} {{Minimal flavor violation: An Effective
  field theory approach}}}, }\href {\doibase 10.1016/S0550-3213(02)00836-2}
  {\bibfield  {journal} {\bibinfo  {journal} {Nucl. Phys.} }\textbf {\bibinfo
  {volume} {B645}} (\bibinfo {year} {2002}) \bibinfo {pages} {155}}, \Eprint
  {http://arxiv.org/abs/hep-ph/0207036}{arXiv:hep-ph/0207036}\BibitemShut
  {NoStop}%
\bibitem [{\citenamefont {Kagan} \emph {et~al.}(2009)\citenamefont {Kagan},
  \citenamefont {Perez}, \citenamefont {Volansky}, and \citenamefont
  {Zupan}}]{Kagan:2009bn}%
  \BibitemOpen
  \bibfield  {author} {\bibinfo {author} {\bibfnamefont {A.~L.} \bibnamefont
  {Kagan}}, \bibinfo {author} {\bibfnamefont {G.}~\bibnamefont {Perez}},
  \bibinfo {author} {\bibfnamefont {T.}~\bibnamefont {Volansky}},  and \bibinfo
  {author} {\bibfnamefont {J.}~\bibnamefont {Zupan}}, }\bibfield  {title}
  {\emph {\bibinfo {title} {{General Minimal Flavor Violation}}}, }\href
  {\doibase 10.1103/PhysRevD.80.076002} {\bibfield  {journal} {\bibinfo
  {journal} {Phys. Rev.} }\textbf {\bibinfo {volume} {D80}} (\bibinfo {year}
  {2009}) \bibinfo {pages} {076002}}, \Eprint
  {http://arxiv.org/abs/0903.1794}{arXiv:0903.1794 [hep-ph]}\BibitemShut
  {NoStop}%
\bibitem [{\citenamefont {Gedalia} \emph {et~al.}(2010)\citenamefont {Gedalia},
  \citenamefont {Mannelli}, and \citenamefont {Perez}}]{Gedalia:2010zs}%
  \BibitemOpen
  \bibfield  {author} {\bibinfo {author} {\bibfnamefont {O.}~\bibnamefont
  {Gedalia}}, \bibinfo {author} {\bibfnamefont {L.}~\bibnamefont {Mannelli}},
  and \bibinfo {author} {\bibfnamefont {G.}~\bibnamefont {Perez}}, }\bibfield
  {title} {\emph {\bibinfo {title} {{Covariant Description of Flavor Violation
  at the LHC}}}, }\href {\doibase 10.1016/j.physletb.2010.08.064} {\bibfield
  {journal} {\bibinfo  {journal} {Phys. Lett.} }\textbf {\bibinfo {volume}
  {B693}} (\bibinfo {year} {2010}) \bibinfo {pages} {301}}, \Eprint
  {http://arxiv.org/abs/1002.0778}{arXiv:1002.0778 [hep-ph]}\BibitemShut
  {NoStop}%
\bibitem [{\citenamefont {Butter} \emph {et~al.}(2016)\citenamefont {Butter},
  \citenamefont {Éboli}, \citenamefont {Gonzalez-Fraile}, \citenamefont
  {Gonzalez-Garcia}, \citenamefont {Plehn}, and \citenamefont
  {Rauch}}]{Butter:2016cvz}%
  \BibitemOpen
  \bibfield  {author} {\bibinfo {author} {\bibfnamefont {A.}~\bibnamefont
  {Butter}}, \bibinfo {author} {\bibfnamefont {O.~J.~P.} \bibnamefont
  {Éboli}}, \bibinfo {author} {\bibfnamefont {J.}~\bibnamefont
  {Gonzalez-Fraile}}, \bibinfo {author} {\bibfnamefont {M.~C.} \bibnamefont
  {Gonzalez-Garcia}}, \bibinfo {author} {\bibfnamefont {T.}~\bibnamefont
  {Plehn}},  and \bibinfo {author} {\bibfnamefont {M.}~\bibnamefont {Rauch}},
  }\bibfield  {title} {\emph {\bibinfo {title} {{The Gauge-Higgs Legacy of the
  LHC Run I}}}, }\href {\doibase 10.1007/JHEP07(2016)152} {\bibfield  {journal}
  {\bibinfo  {journal} {JHEP} }\textbf {\bibinfo {volume} {07}} (\bibinfo
  {year} {2016}) \bibinfo {pages} {152}}, \Eprint
  {http://arxiv.org/abs/1604.03105}{arXiv:1604.03105 [hep-ph]}\BibitemShut
  {NoStop}%
\bibitem [{\citenamefont {Murphy}(2018)}]{Murphy:2017omb}%
  \BibitemOpen
  \bibfield  {author} {\bibinfo {author} {\bibfnamefont {C.~W.} \bibnamefont
  {Murphy}}, }\bibfield  {title} {\emph {\bibinfo {title} {{Statistical
  approach to Higgs boson couplings in the standard model effective field
  theory}}}, }\href {\doibase 10.1103/PhysRevD.97.015007} {\bibfield  {journal}
  {\bibinfo  {journal} {Phys. Rev.} }\textbf {\bibinfo {volume} {D97}}
  (\bibinfo {year} {2018}) \bibinfo {pages} {015007}}, \Eprint
  {http://arxiv.org/abs/1710.02008}{arXiv:1710.02008 [hep-ph]}\BibitemShut
  {NoStop}%
\bibitem [{\citenamefont {Di~Vita} \emph {et~al.}(2017)\citenamefont {Di~Vita},
  \citenamefont {Grojean}, \citenamefont {Panico}, \citenamefont {Riembau}, and
  \citenamefont {Vantalon}}]{DiVita:2017eyz}%
  \BibitemOpen
  \bibfield  {author} {\bibinfo {author} {\bibfnamefont {S.}~\bibnamefont
  {Di~Vita}}, \bibinfo {author} {\bibfnamefont {C.}~\bibnamefont {Grojean}},
  \bibinfo {author} {\bibfnamefont {G.}~\bibnamefont {Panico}}, \bibinfo
  {author} {\bibfnamefont {M.}~\bibnamefont {Riembau}},  and \bibinfo {author}
  {\bibfnamefont {T.}~\bibnamefont {Vantalon}}, }\bibfield  {title} {\emph
  {\bibinfo {title} {{A global view on the Higgs self-coupling}}}, }\href
  {\doibase 10.1007/JHEP09(2017)069} {\bibfield  {journal} {\bibinfo  {journal}
  {JHEP} }\textbf {\bibinfo {volume} {09}} (\bibinfo {year} {2017}) \bibinfo
  {pages} {069}}, \Eprint {http://arxiv.org/abs/1704.01953}{arXiv:1704.01953
  [hep-ph]}\BibitemShut {NoStop}%
\bibitem [{\citenamefont {Falkowski} \emph {et~al.}(2017)\citenamefont
  {Falkowski}, \citenamefont {Gonzalez-Alonso}, \citenamefont {Greljo},
  \citenamefont {Marzocca}, and \citenamefont {Son}}]{Falkowski:2016cxu}%
  \BibitemOpen
  \bibfield  {author} {\bibinfo {author} {\bibfnamefont {A.}~\bibnamefont
  {Falkowski}}, \bibinfo {author} {\bibfnamefont {M.}~\bibnamefont
  {Gonzalez-Alonso}}, \bibinfo {author} {\bibfnamefont {A.}~\bibnamefont
  {Greljo}}, \bibinfo {author} {\bibfnamefont {D.}~\bibnamefont {Marzocca}},
  and \bibinfo {author} {\bibfnamefont {M.}~\bibnamefont {Son}}, }\bibfield
  {title} {\emph {\bibinfo {title} {{Anomalous Triple Gauge Couplings in the
  Effective Field Theory Approach at the LHC}}}, }\href {\doibase
  10.1007/JHEP02(2017)115} {\bibfield  {journal} {\bibinfo  {journal} {JHEP}
  }\textbf {\bibinfo {volume} {02}} (\bibinfo {year} {2017}) \bibinfo {pages}
  {115}}, \Eprint {http://arxiv.org/abs/1609.06312}{arXiv:1609.06312
  [hep-ph]}\BibitemShut {NoStop}%
\bibitem [{\citenamefont {Brivio} \emph {et~al.}(2016)\citenamefont {Brivio},
  \citenamefont {Gonzalez-Fraile}, \citenamefont {Gonzalez-Garcia}, and
  \citenamefont {Merlo}}]{Brivio:2016fzo}%
  \BibitemOpen
  \bibfield  {author} {\bibinfo {author} {\bibfnamefont {I.}~\bibnamefont
  {Brivio}}, \bibinfo {author} {\bibfnamefont {J.}~\bibnamefont
  {Gonzalez-Fraile}}, \bibinfo {author} {\bibfnamefont {M.~C.} \bibnamefont
  {Gonzalez-Garcia}},  and \bibinfo {author} {\bibfnamefont {L.}~\bibnamefont
  {Merlo}}, }\bibfield  {title} {\emph {\bibinfo {title} {{The complete HEFT
  Lagrangian after the LHC Run I}}}, }\href {\doibase
  10.1140/epjc/s10052-016-4211-9} {\bibfield  {journal} {\bibinfo  {journal}
  {Eur. Phys. J.} }\textbf {\bibinfo {volume} {C76}} (\bibinfo {year} {2016})
  \bibinfo {pages} {416}}, \Eprint
  {http://arxiv.org/abs/1604.06801}{arXiv:1604.06801 [hep-ph]}\BibitemShut
  {NoStop}%
\bibitem [{\citenamefont {Englert} \emph {et~al.}(2016)\citenamefont {Englert},
  \citenamefont {Kogler}, \citenamefont {Schulz}, and \citenamefont
  {Spannowsky}}]{Englert:2015hrx}%
  \BibitemOpen
  \bibfield  {author} {\bibinfo {author} {\bibfnamefont {C.}~\bibnamefont
  {Englert}}, \bibinfo {author} {\bibfnamefont {R.}~\bibnamefont {Kogler}},
  \bibinfo {author} {\bibfnamefont {H.}~\bibnamefont {Schulz}},  and \bibinfo
  {author} {\bibfnamefont {M.}~\bibnamefont {Spannowsky}}, }\bibfield  {title}
  {\emph {\bibinfo {title} {{Higgs coupling measurements at the LHC}}}, }\href
  {\doibase 10.1140/epjc/s10052-016-4227-1} {\bibfield  {journal} {\bibinfo
  {journal} {Eur. Phys. J.} }\textbf {\bibinfo {volume} {C76}} (\bibinfo {year}
  {2016}) \bibinfo {pages} {393}}, \Eprint
  {http://arxiv.org/abs/1511.05170}{arXiv:1511.05170 [hep-ph]}\BibitemShut
  {NoStop}%
\bibitem [{\citenamefont {Mangano} \emph {et~al.}(2016)\citenamefont {Mangano},
  \citenamefont {Plehn}, \citenamefont {Reimitz}, \citenamefont {Schell}, and
  \citenamefont {Shao}}]{Plehn:2015cta}%
  \BibitemOpen
  \bibfield  {author} {\bibinfo {author} {\bibfnamefont {M.~L.} \bibnamefont
  {Mangano}}, \bibinfo {author} {\bibfnamefont {T.}~\bibnamefont {Plehn}},
  \bibinfo {author} {\bibfnamefont {P.}~\bibnamefont {Reimitz}}, \bibinfo
  {author} {\bibfnamefont {T.}~\bibnamefont {Schell}},  and \bibinfo {author}
  {\bibfnamefont {H.-S.} \bibnamefont {Shao}}, }\bibfield  {title} {\emph
  {\bibinfo {title} {{Measuring the Top Yukawa Coupling at 100 TeV}}}, }\href
  {\doibase 10.1088/0954-3899/43/3/035001} {\bibfield  {journal} {\bibinfo
  {journal} {J. Phys.} }\textbf {\bibinfo {volume} {G43}} (\bibinfo {year}
  {2016}) \bibinfo {pages} {035001}}, \Eprint
  {http://arxiv.org/abs/1507.08169}{arXiv:1507.08169 [hep-ph]}\BibitemShut
  {NoStop}%
\bibitem [{\citenamefont {Schulze} and \citenamefont
  {Soreq}(2016)}]{Schulze:2016qas}%
  \BibitemOpen
  \bibfield  {author} {\bibinfo {author} {\bibfnamefont {M.}~\bibnamefont
  {Schulze}} and \bibinfo {author} {\bibfnamefont {Y.}~\bibnamefont {Soreq}},
  }\bibfield  {title} {\emph {\bibinfo {title} {{Pinning down electroweak
  dipole operators of the top quark}}}, }\href {\doibase
  10.1140/epjc/s10052-016-4263-x} {\bibfield  {journal} {\bibinfo  {journal}
  {Eur. Phys. J.} }\textbf {\bibinfo {volume} {C76}} (\bibinfo {year} {2016})
  \bibinfo {pages} {466}}, \Eprint
  {http://arxiv.org/abs/1603.08911}{arXiv:1603.08911 [hep-ph]}\BibitemShut
  {NoStop}%
\bibitem [{\citenamefont {Atwood} and \citenamefont
  {Soni}(1992)}]{Atwood:1991ka}%
  \BibitemOpen
  \bibfield  {author} {\bibinfo {author} {\bibfnamefont {D.}~\bibnamefont
  {Atwood}} and \bibinfo {author} {\bibfnamefont {A.}~\bibnamefont {Soni}},
  }\bibfield  {title} {\emph {\bibinfo {title} {{Analysis for magnetic moment
  and electric dipole moment form-factors of the top quark via $e^+e^-\to
  t\bar{t}$}}}, }\href {\doibase 10.1103/PhysRevD.45.2405} {\bibfield
  {journal} {\bibinfo  {journal} {Phys. Rev.} }\textbf {\bibinfo {volume}
  {D45}} (\bibinfo {year} {1992}) \bibinfo {pages} {2405}}\BibitemShut
  {NoStop}%
\bibitem [{\citenamefont {Davier} \emph {et~al.}(1993)\citenamefont {Davier},
  \citenamefont {Duflot}, \citenamefont {Le~Diberder}, and \citenamefont
  {Rouge}}]{Davier:1992nw}%
  \BibitemOpen
  \bibfield  {author} {\bibinfo {author} {\bibfnamefont {M.}~\bibnamefont
  {Davier}}, \bibinfo {author} {\bibfnamefont {L.}~\bibnamefont {Duflot}},
  \bibinfo {author} {\bibfnamefont {F.}~\bibnamefont {Le~Diberder}},  and
  \bibinfo {author} {\bibfnamefont {A.}~\bibnamefont {Rouge}}, }\bibfield
  {title} {\emph {\bibinfo {title} {{The Optimal method for the measurement of
  tau polarization}}}, }\href {\doibase 10.1016/0370-2693(93)90101-M}
  {\bibfield  {journal} {\bibinfo  {journal} {Phys. Lett.} }\textbf {\bibinfo
  {volume} {B306}} (\bibinfo {year} {1993}) \bibinfo {pages} {411}}\BibitemShut
  {NoStop}%
\bibitem [{\citenamefont {Diehl} and \citenamefont
  {Nachtmann}(1994)}]{Diehl:1993br}%
  \BibitemOpen
  \bibfield  {author} {\bibinfo {author} {\bibfnamefont {M.}~\bibnamefont
  {Diehl}} and \bibinfo {author} {\bibfnamefont {O.}~\bibnamefont {Nachtmann}},
  }\bibfield  {title} {\emph {\bibinfo {title} {{Optimal observables for the
  measurement of three gauge boson couplings in $e^+e^-\to W^+W^-$}}}, }\href
  {\doibase 10.1007/BF01555899} {\bibfield  {journal} {\bibinfo  {journal} {Z.
  Phys.} }\textbf {\bibinfo {volume} {C62}} (\bibinfo {year} {1994}) \bibinfo
  {pages} {397}}\BibitemShut {NoStop}%
\bibitem [{\citenamefont {Brehmer} \emph
  {et~al.}(2017{\natexlab{a}})\citenamefont {Brehmer}, \citenamefont {Cranmer},
  \citenamefont {Kling}, and \citenamefont {Plehn}}]{Brehmer:2016nyr}%
  \BibitemOpen
  \bibfield  {author} {\bibinfo {author} {\bibfnamefont {J.}~\bibnamefont
  {Brehmer}}, \bibinfo {author} {\bibfnamefont {K.}~\bibnamefont {Cranmer}},
  \bibinfo {author} {\bibfnamefont {F.}~\bibnamefont {Kling}},  and \bibinfo
  {author} {\bibfnamefont {T.}~\bibnamefont {Plehn}}, }\bibfield  {title}
  {\emph {\bibinfo {title} {{Better Higgs boson measurements through
  information geometry}}}, }\href {\doibase 10.1103/PhysRevD.95.073002}
  {\bibfield  {journal} {\bibinfo  {journal} {Phys. Rev.} }\textbf {\bibinfo
  {volume} {D95}} (\bibinfo {year} {2017}{\natexlab{a}}) \bibinfo {pages}
  {073002}}, \Eprint {http://arxiv.org/abs/1612.05261}{arXiv:1612.05261
  [hep-ph]}\BibitemShut {NoStop}%
\bibitem [{\citenamefont {Brehmer} \emph
  {et~al.}(2017{\natexlab{b}})\citenamefont {Brehmer}, \citenamefont {Kling},
  \citenamefont {Plehn}, and \citenamefont {Tait}}]{Brehmer:2017lrt}%
  \BibitemOpen
  \bibfield  {author} {\bibinfo {author} {\bibfnamefont {J.}~\bibnamefont
  {Brehmer}}, \bibinfo {author} {\bibfnamefont {F.}~\bibnamefont {Kling}},
  \bibinfo {author} {\bibfnamefont {T.}~\bibnamefont {Plehn}},  and \bibinfo
  {author} {\bibfnamefont {T.~M.~P.} \bibnamefont {Tait}}, }\bibfield  {title}
  {\emph {\bibinfo {title} {{Better Higgs-CP Tests Through Information
  Geometry}}}, }\href@noop {} { (\bibinfo {year} {2017}{\natexlab{b}})},
  \Eprint {http://arxiv.org/abs/1712.02350}{arXiv:1712.02350
  [hep-ph]}\BibitemShut {NoStop}%
\bibitem [{\citenamefont {Gritsan} \emph {et~al.}(2016)\citenamefont {Gritsan},
  \citenamefont {Röntsch}, \citenamefont {Schulze}, and \citenamefont
  {Xiao}}]{Gritsan:2016hjl}%
  \BibitemOpen
  \bibfield  {author} {\bibinfo {author} {\bibfnamefont {A.~V.} \bibnamefont
  {Gritsan}}, \bibinfo {author} {\bibfnamefont {R.}~\bibnamefont {Röntsch}},
  \bibinfo {author} {\bibfnamefont {M.}~\bibnamefont {Schulze}},  and \bibinfo
  {author} {\bibfnamefont {M.}~\bibnamefont {Xiao}}, }\bibfield  {title} {\emph
  {\bibinfo {title} {{Constraining anomalous Higgs boson couplings to the heavy
  flavor fermions using matrix element techniques}}}, }\href {\doibase
  10.1103/PhysRevD.94.055023} {\bibfield  {journal} {\bibinfo  {journal} {Phys.
  Rev.} }\textbf {\bibinfo {volume} {D94}} (\bibinfo {year} {2016}) \bibinfo
  {pages} {055023}}, \Eprint {http://arxiv.org/abs/1606.03107}{arXiv:1606.03107
  [hep-ph]}\BibitemShut {NoStop}%
\bibitem [{\citenamefont {Abbiendi} \emph {et~al.}(2004)}]{Abbiendi:2003mk}%
  \BibitemOpen
  \bibfield  {author} {\bibinfo {author} {\bibfnamefont {G.}~\bibnamefont
  {Abbiendi}} \emph {et~al.} (\bibinfo {collaboration} {OPAL}), }\bibfield
  {title} {\emph {\bibinfo {title} {{Measurement of charged current triple
  gauge boson couplings using $W$ pairs at LEP}}}, }\href {\doibase
  10.1140/epjc/s2003-01524-6} {\bibfield  {journal} {\bibinfo  {journal} {Eur.
  Phys. J.} }\textbf {\bibinfo {volume} {C33}} (\bibinfo {year} {2004})
  \bibinfo {pages} {463}}, \Eprint
  {http://arxiv.org/abs/hep-ex/0308067}{arXiv:hep-ex/0308067}\BibitemShut
  {NoStop}%
\bibitem [{\citenamefont {Achard} \emph {et~al.}(2004)}]{Achard:2004ji}%
  \BibitemOpen
  \bibfield  {author} {\bibinfo {author} {\bibfnamefont {P.}~\bibnamefont
  {Achard}} \emph {et~al.} (\bibinfo {collaboration} {L3}), }\bibfield  {title}
  {\emph {\bibinfo {title} {{Measurement of triple gauge boson couplings of the
  $W$ boson at LEP}}}, }\href {\doibase 10.1016/j.physletb.2004.02.045}
  {\bibfield  {journal} {\bibinfo  {journal} {Phys. Lett.} }\textbf {\bibinfo
  {volume} {B586}} (\bibinfo {year} {2004}) \bibinfo {pages} {151}}, \Eprint
  {http://arxiv.org/abs/hep-ex/0402036}{arXiv:hep-ex/0402036}\BibitemShut
  {NoStop}%
\bibitem [{\citenamefont {Schael} \emph {et~al.}(2005)}]{Schael:2004tq}%
  \BibitemOpen
  \bibfield  {author} {\bibinfo {author} {\bibfnamefont {S.}~\bibnamefont
  {Schael}} \emph {et~al.} (\bibinfo {collaboration} {ALEPH}), }\bibfield
  {title} {\emph {\bibinfo {title} {{Improved measurement of the triple
  gauge-boson couplings gamma $W W$ and $Z W W$ in $e^+ e^-$ collisions}}},
  }\href {\doibase 10.1016/j.physletb.2005.03.058} {\bibfield  {journal}
  {\bibinfo  {journal} {Phys. Lett.} }\textbf {\bibinfo {volume} {B614}}
  (\bibinfo {year} {2005}) \bibinfo {pages} {7}}\BibitemShut {NoStop}%
\bibitem [{\citenamefont {Abdallah} \emph {et~al.}(2010)}]{Abdallah:2010zj}%
  \BibitemOpen
  \bibfield  {author} {\bibinfo {author} {\bibfnamefont {J.}~\bibnamefont
  {Abdallah}} \emph {et~al.} (\bibinfo {collaboration} {DELPHI}), }\bibfield
  {title} {\emph {\bibinfo {title} {{Measurements of CP-conserving Trilinear
  Gauge Boson Couplings $WWV$ ($V = \gamma,Z$) in $e^+e^-$ Collisions at
  LEP2}}}, }\href {\doibase 10.1140/epjc/s10052-010-1254-1} {\bibfield
  {journal} {\bibinfo  {journal} {Eur. Phys. J.} }\textbf {\bibinfo {volume}
  {C66}} (\bibinfo {year} {2010}) \bibinfo {pages} {35}}, \Eprint
  {http://arxiv.org/abs/1002.0752}{arXiv:1002.0752 [hep-ex]}\BibitemShut
  {NoStop}%
\bibitem [{\citenamefont {Aad} \emph {et~al.}(2016)}]{Aad:2016nal}%
  \BibitemOpen
  \bibfield  {author} {\bibinfo {author} {\bibfnamefont {G.}~\bibnamefont
  {Aad}} \emph {et~al.} (\bibinfo {collaboration} {ATLAS}), }\bibfield  {title}
  {\emph {\bibinfo {title} {{Test of CP Invariance in vector-boson fusion
  production of the Higgs boson using the Optimal Observable method in the
  ditau decay channel with the ATLAS detector}}}, }\href {\doibase
  10.1140/epjc/s10052-016-4499-5} {\bibfield  {journal} {\bibinfo  {journal}
  {Eur. Phys. J.} }\textbf {\bibinfo {volume} {C76}} (\bibinfo {year} {2016})
  \bibinfo {pages} {658}}, \Eprint
  {http://arxiv.org/abs/1602.04516}{arXiv:1602.04516 [hep-ex]}\BibitemShut
  {NoStop}%
\bibitem [{\citenamefont {Azatov} \emph {et~al.}(2017)\citenamefont {Azatov},
  \citenamefont {Contino}, \citenamefont {Machado}, and \citenamefont
  {Riva}}]{Azatov:2016sqh}%
  \BibitemOpen
  \bibfield  {author} {\bibinfo {author} {\bibfnamefont {A.}~\bibnamefont
  {Azatov}}, \bibinfo {author} {\bibfnamefont {R.}~\bibnamefont {Contino}},
  \bibinfo {author} {\bibfnamefont {C.~S.} \bibnamefont {Machado}},  and
  \bibinfo {author} {\bibfnamefont {F.}~\bibnamefont {Riva}}, }\bibfield
  {title} {\emph {\bibinfo {title} {{Helicity selection rules and
  noninterference for BSM amplitudes}}}, }\href {\doibase
  10.1103/PhysRevD.95.065014} {\bibfield  {journal} {\bibinfo  {journal} {Phys.
  Rev.} }\textbf {\bibinfo {volume} {D95}} (\bibinfo {year} {2017}) \bibinfo
  {pages} {065014}}, \Eprint {http://arxiv.org/abs/1607.05236}{arXiv:1607.05236
  [hep-ph]}\BibitemShut {NoStop}%
\bibitem [{\citenamefont {Alwall} \emph {et~al.}(2014)\citenamefont {Alwall},
  \citenamefont {Frederix}, \citenamefont {Frixione}, \citenamefont {Hirschi},
  \citenamefont {Maltoni}, \citenamefont {Mattelaer}, \citenamefont {Shao},
  \citenamefont {Stelzer}, \citenamefont {Torrielli}, and \citenamefont
  {Zaro}}]{Alwall:2014hca}%
  \BibitemOpen
  \bibfield  {author} {\bibinfo {author} {\bibfnamefont {J.}~\bibnamefont
  {Alwall}}, \bibinfo {author} {\bibfnamefont {R.}~\bibnamefont {Frederix}},
  \bibinfo {author} {\bibfnamefont {S.}~\bibnamefont {Frixione}}, \bibinfo
  {author} {\bibfnamefont {V.}~\bibnamefont {Hirschi}}, \bibinfo {author}
  {\bibfnamefont {F.}~\bibnamefont {Maltoni}}, \bibinfo {author} {\bibfnamefont
  {O.}~\bibnamefont {Mattelaer}}, \bibinfo {author} {\bibfnamefont {H.~S.}
  \bibnamefont {Shao}}, \bibinfo {author} {\bibfnamefont {T.}~\bibnamefont
  {Stelzer}}, \bibinfo {author} {\bibfnamefont {P.}~\bibnamefont {Torrielli}},
  and \bibinfo {author} {\bibfnamefont {M.}~\bibnamefont {Zaro}}, }\bibfield
  {title} {\emph {\bibinfo {title} {{The automated computation of tree-level
  and next-to-leading order differential cross sections, and their matching to
  parton shower simulations}}}, }\href {\doibase 10.1007/JHEP07(2014)079}
  {\bibfield  {journal} {\bibinfo  {journal} {JHEP} }\textbf {\bibinfo {volume}
  {07}} (\bibinfo {year} {2014}) \bibinfo {pages} {079}}, \Eprint
  {http://arxiv.org/abs/1405.0301}{arXiv:1405.0301 [hep-ph]}\BibitemShut
  {NoStop}%
\bibitem [{\citenamefont {Brehmer} \emph {et~al.}(2016)\citenamefont {Brehmer},
  \citenamefont {Freitas}, \citenamefont {Lopez-Val}, and \citenamefont
  {Plehn}}]{Brehmer:2015rna}%
  \BibitemOpen
  \bibfield  {author} {\bibinfo {author} {\bibfnamefont {J.}~\bibnamefont
  {Brehmer}}, \bibinfo {author} {\bibfnamefont {A.}~\bibnamefont {Freitas}},
  \bibinfo {author} {\bibfnamefont {D.}~\bibnamefont {Lopez-Val}},  and
  \bibinfo {author} {\bibfnamefont {T.}~\bibnamefont {Plehn}}, }\bibfield
  {title} {\emph {\bibinfo {title} {{Pushing Higgs Effective Theory to its
  Limits}}}, }\href {\doibase 10.1103/PhysRevD.93.075014} {\bibfield  {journal}
  {\bibinfo  {journal} {Phys. Rev.} }\textbf {\bibinfo {volume} {D93}}
  (\bibinfo {year} {2016}) \bibinfo {pages} {075014}}, \Eprint
  {http://arxiv.org/abs/1510.03443}{arXiv:1510.03443 [hep-ph]}\BibitemShut
  {NoStop}%
\bibitem [{\citenamefont {Contino} \emph {et~al.}(2016)\citenamefont {Contino},
  \citenamefont {Falkowski}, \citenamefont {Goertz}, \citenamefont {Grojean},
  and \citenamefont {Riva}}]{Contino:2016jqw}%
  \BibitemOpen
  \bibfield  {author} {\bibinfo {author} {\bibfnamefont {R.}~\bibnamefont
  {Contino}}, \bibinfo {author} {\bibfnamefont {A.}~\bibnamefont {Falkowski}},
  \bibinfo {author} {\bibfnamefont {F.}~\bibnamefont {Goertz}}, \bibinfo
  {author} {\bibfnamefont {C.}~\bibnamefont {Grojean}},  and \bibinfo {author}
  {\bibfnamefont {F.}~\bibnamefont {Riva}}, }\bibfield  {title} {\emph
  {\bibinfo {title} {{On the Validity of the Effective Field Theory Approach to
  SM Precision Tests}}}, }\href {\doibase 10.1007/JHEP07(2016)144} {\bibfield
  {journal} {\bibinfo  {journal} {JHEP} }\textbf {\bibinfo {volume} {07}}
  (\bibinfo {year} {2016}) \bibinfo {pages} {144}}, \Eprint
  {http://arxiv.org/abs/1604.06444}{arXiv:1604.06444 [hep-ph]}\BibitemShut
  {NoStop}%
\bibitem [{\citenamefont {de~Florian} \emph
  {et~al.}(2016)}]{deFlorian:2016spz}%
  \BibitemOpen
  \bibfield  {author} {\bibinfo {author} {\bibfnamefont {D.}~\bibnamefont
  {de~Florian}} \emph {et~al.} (\bibinfo {collaboration} {LHC Higgs Cross
  Section Working Group}), }\bibfield  {title} {\emph {\bibinfo {title}
  {{Handbook of LHC Higgs Cross Sections: 4. Deciphering the Nature of the
  Higgs Sector}}}, }\href {\doibase 10.23731/CYRM-2017-002} { (\bibinfo {year}
  {2016}), 10.23731/CYRM-2017-002}, \Eprint
  {http://arxiv.org/abs/1610.07922}{arXiv:1610.07922 [hep-ph]}\BibitemShut
  {NoStop}%
\bibitem [{\citenamefont {Zhang}(2018)}]{Zhang:2017mls}%
  \BibitemOpen
  \bibfield  {author} {\bibinfo {author} {\bibfnamefont {C.}~\bibnamefont
  {Zhang}}, }\bibfield  {title} {\emph {\bibinfo {title} {{Constraining $qqtt$
  operators from four-top production: a case for enhanced EFT sensitivity}}},
  }\href {\doibase 10.1088/1674-1137/42/2/023104} {\bibfield  {journal}
  {\bibinfo  {journal} {Chin. Phys.} }\textbf {\bibinfo {volume} {C42}}
  (\bibinfo {year} {2018}) \bibinfo {pages} {023104}}, \Eprint
  {http://arxiv.org/abs/1708.05928}{arXiv:1708.05928 [hep-ph]}\BibitemShut
  {NoStop}%
\bibitem [{\citenamefont {Buckley} \emph {et~al.}(2016)\citenamefont {Buckley},
  \citenamefont {Englert}, \citenamefont {Ferrando}, \citenamefont {Miller},
  \citenamefont {Moore}, \citenamefont {Russell}, and \citenamefont
  {White}}]{Buckley:2015lku}%
  \BibitemOpen
  \bibfield  {author} {\bibinfo {author} {\bibfnamefont {A.}~\bibnamefont
  {Buckley}}, \bibinfo {author} {\bibfnamefont {C.}~\bibnamefont {Englert}},
  \bibinfo {author} {\bibfnamefont {J.}~\bibnamefont {Ferrando}}, \bibinfo
  {author} {\bibfnamefont {D.~J.} \bibnamefont {Miller}}, \bibinfo {author}
  {\bibfnamefont {L.}~\bibnamefont {Moore}}, \bibinfo {author} {\bibfnamefont
  {M.}~\bibnamefont {Russell}},  and \bibinfo {author} {\bibfnamefont {C.~D.}
  \bibnamefont {White}}, }\bibfield  {title} {\emph {\bibinfo {title}
  {{Constraining top quark effective theory in the LHC Run II era}}}, }\href
  {\doibase 10.1007/JHEP04(2016)015} {\bibfield  {journal} {\bibinfo  {journal}
  {JHEP} }\textbf {\bibinfo {volume} {04}} (\bibinfo {year} {2016}) \bibinfo
  {pages} {015}}, \Eprint {http://arxiv.org/abs/1512.03360}{arXiv:1512.03360
  [hep-ph]}\BibitemShut {NoStop}%
\bibitem [{\citenamefont {Englert} and \citenamefont
  {Russell}(2017)}]{Englert:2017dev}%
  \BibitemOpen
  \bibfield  {author} {\bibinfo {author} {\bibfnamefont {C.}~\bibnamefont
  {Englert}} and \bibinfo {author} {\bibfnamefont {M.}~\bibnamefont {Russell}},
  }\bibfield  {title} {\emph {\bibinfo {title} {{Top quark electroweak
  couplings at future lepton colliders}}}, }\href {\doibase
  10.1140/epjc/s10052-017-5095-z} {\bibfield  {journal} {\bibinfo  {journal}
  {Eur. Phys. J.} }\textbf {\bibinfo {volume} {C77}} (\bibinfo {year} {2017})
  \bibinfo {pages} {535}}, \Eprint
  {http://arxiv.org/abs/1704.01782}{arXiv:1704.01782 [hep-ph]}\BibitemShut
  {NoStop}%
\bibitem [{\citenamefont {Buckley} \emph {et~al.}(2015)\citenamefont {Buckley},
  \citenamefont {Englert}, \citenamefont {Ferrando}, \citenamefont {Miller},
  \citenamefont {Moore}, \citenamefont {Russell}, and \citenamefont
  {White}}]{Buckley:2015nca}%
  \BibitemOpen
  \bibfield  {author} {\bibinfo {author} {\bibfnamefont {A.}~\bibnamefont
  {Buckley}}, \bibinfo {author} {\bibfnamefont {C.}~\bibnamefont {Englert}},
  \bibinfo {author} {\bibfnamefont {J.}~\bibnamefont {Ferrando}}, \bibinfo
  {author} {\bibfnamefont {D.~J.} \bibnamefont {Miller}}, \bibinfo {author}
  {\bibfnamefont {L.}~\bibnamefont {Moore}}, \bibinfo {author} {\bibfnamefont
  {M.}~\bibnamefont {Russell}},  and \bibinfo {author} {\bibfnamefont {C.~D.}
  \bibnamefont {White}}, }\bibfield  {title} {\emph {\bibinfo {title} {{Global
  fit of top quark effective theory to data}}}, }\href {\doibase
  10.1103/PhysRevD.92.091501} {\bibfield  {journal} {\bibinfo  {journal} {Phys.
  Rev.} }\textbf {\bibinfo {volume} {D92}} (\bibinfo {year} {2015}) \bibinfo
  {pages} {091501}}, \Eprint {http://arxiv.org/abs/1506.08845}{arXiv:1506.08845
  [hep-ph]}\BibitemShut {NoStop}%
\bibitem [{\citenamefont {Barbieri} \emph {et~al.}(2011)\citenamefont
  {Barbieri}, \citenamefont {Isidori}, \citenamefont {Jones-Perez},
  \citenamefont {Lodone}, and \citenamefont {Straub}}]{Barbieri:2011ci}%
  \BibitemOpen
  \bibfield  {author} {\bibinfo {author} {\bibfnamefont {R.}~\bibnamefont
  {Barbieri}}, \bibinfo {author} {\bibfnamefont {G.}~\bibnamefont {Isidori}},
  \bibinfo {author} {\bibfnamefont {J.}~\bibnamefont {Jones-Perez}}, \bibinfo
  {author} {\bibfnamefont {P.}~\bibnamefont {Lodone}},  and \bibinfo {author}
  {\bibfnamefont {D.~M.} \bibnamefont {Straub}}, }\bibfield  {title} {\emph
  {\bibinfo {title} {{$U(2)$ and Minimal Flavour Violation in Supersymmetry}}},
  }\href {\doibase 10.1140/epjc/s10052-011-1725-z} {\bibfield  {journal}
  {\bibinfo  {journal} {Eur. Phys. J.} }\textbf {\bibinfo {volume} {C71}}
  (\bibinfo {year} {2011}) \bibinfo {pages} {1725}}, \Eprint
  {http://arxiv.org/abs/1105.2296}{arXiv:1105.2296 [hep-ph]}\BibitemShut
  {NoStop}%
\bibitem [{\citenamefont {Barbieri} \emph {et~al.}(2012)\citenamefont
  {Barbieri}, \citenamefont {Buttazzo}, \citenamefont {Sala}, and \citenamefont
  {Straub}}]{Barbieri:2012uh}%
  \BibitemOpen
  \bibfield  {author} {\bibinfo {author} {\bibfnamefont {R.}~\bibnamefont
  {Barbieri}}, \bibinfo {author} {\bibfnamefont {D.}~\bibnamefont {Buttazzo}},
  \bibinfo {author} {\bibfnamefont {F.}~\bibnamefont {Sala}},  and \bibinfo
  {author} {\bibfnamefont {D.~M.} \bibnamefont {Straub}}, }\bibfield  {title}
  {\emph {\bibinfo {title} {{Flavour physics from an approximate $U(2)^3$
  symmetry}}}, }\href {\doibase 10.1007/JHEP07(2012)181} {\bibfield  {journal}
  {\bibinfo  {journal} {JHEP} }\textbf {\bibinfo {volume} {07}} (\bibinfo
  {year} {2012}) \bibinfo {pages} {181}}, \Eprint
  {http://arxiv.org/abs/1203.4218}{arXiv:1203.4218 [hep-ph]}\BibitemShut
  {NoStop}%
\bibitem [{\citenamefont {Cirigliano} \emph {et~al.}(2013)\citenamefont
  {Cirigliano}, \citenamefont {Gonzalez-Alonso}, and \citenamefont
  {Graesser}}]{Cirigliano:2012ab}%
  \BibitemOpen
  \bibfield  {author} {\bibinfo {author} {\bibfnamefont {V.}~\bibnamefont
  {Cirigliano}}, \bibinfo {author} {\bibfnamefont {M.}~\bibnamefont
  {Gonzalez-Alonso}},  and \bibinfo {author} {\bibfnamefont {M.~L.}
  \bibnamefont {Graesser}}, }\bibfield  {title} {\emph {\bibinfo {title}
  {{Non-standard Charged Current Interactions: beta decays versus the LHC}}},
  }\href {\doibase 10.1007/JHEP02(2013)046} {\bibfield  {journal} {\bibinfo
  {journal} {JHEP} }\textbf {\bibinfo {volume} {02}} (\bibinfo {year} {2013})
  \bibinfo {pages} {046}}, \Eprint
  {http://arxiv.org/abs/1210.4553}{arXiv:1210.4553 [hep-ph]}\BibitemShut
  {NoStop}%
\bibitem [{\citenamefont {Aebischer} \emph {et~al.}(2016)\citenamefont
  {Aebischer}, \citenamefont {Crivellin}, \citenamefont {Fael}, and
  \citenamefont {Greub}}]{Aebischer:2015fzz}%
  \BibitemOpen
  \bibfield  {author} {\bibinfo {author} {\bibfnamefont {J.}~\bibnamefont
  {Aebischer}}, \bibinfo {author} {\bibfnamefont {A.}~\bibnamefont
  {Crivellin}}, \bibinfo {author} {\bibfnamefont {M.}~\bibnamefont {Fael}},
  and \bibinfo {author} {\bibfnamefont {C.}~\bibnamefont {Greub}}, }\bibfield
  {title} {\emph {\bibinfo {title} {{Matching of gauge invariant dimension-six
  operators for $b\to s$ and $b\to c$ transitions}}}, }\href {\doibase
  10.1007/JHEP05(2016)037} {\bibfield  {journal} {\bibinfo  {journal} {JHEP}
  }\textbf {\bibinfo {volume} {05}} (\bibinfo {year} {2016}) \bibinfo {pages}
  {037}}, \Eprint {http://arxiv.org/abs/1512.02830}{arXiv:1512.02830
  [hep-ph]}\BibitemShut {NoStop}%
\bibitem [{\citenamefont {Buttazzo} \emph {et~al.}(2017)\citenamefont
  {Buttazzo}, \citenamefont {Greljo}, \citenamefont {Isidori}, and
  \citenamefont {Marzocca}}]{Buttazzo:2017ixm}%
  \BibitemOpen
  \bibfield  {author} {\bibinfo {author} {\bibfnamefont {D.}~\bibnamefont
  {Buttazzo}}, \bibinfo {author} {\bibfnamefont {A.}~\bibnamefont {Greljo}},
  \bibinfo {author} {\bibfnamefont {G.}~\bibnamefont {Isidori}},  and \bibinfo
  {author} {\bibfnamefont {D.}~\bibnamefont {Marzocca}}, }\bibfield  {title}
  {\emph {\bibinfo {title} {{B-physics anomalies: a guide to combined
  explanations}}}, }\href {\doibase 10.1007/JHEP11(2017)044} {\bibfield
  {journal} {\bibinfo  {journal} {JHEP} }\textbf {\bibinfo {volume} {11}}
  (\bibinfo {year} {2017}) \bibinfo {pages} {044}}, \Eprint
  {http://arxiv.org/abs/1706.07808}{arXiv:1706.07808 [hep-ph]}\BibitemShut
  {NoStop}%
\bibitem [{\citenamefont {Jung} and \citenamefont
  {Straub}(2018)}]{Jung:2018lfu}%
  \BibitemOpen
  \bibfield  {author} {\bibinfo {author} {\bibfnamefont {M.}~\bibnamefont
  {Jung}} and \bibinfo {author} {\bibfnamefont {D.~M.} \bibnamefont {Straub}},
  }\bibfield  {title} {\emph {\bibinfo {title} {{Constraining new physics in
  $b\to c\ell\nu$ transitions}}}, }\href@noop {} { (\bibinfo {year} {2018})},
  \Eprint {http://arxiv.org/abs/1801.01112}{arXiv:1801.01112
  [hep-ph]}\BibitemShut {NoStop}%
\bibitem [{\citenamefont {Lees} \emph {et~al.}(2013)}]{Lees:2013uzd}%
  \BibitemOpen
  \bibfield  {author} {\bibinfo {author} {\bibfnamefont {J.~P.} \bibnamefont
  {Lees}} \emph {et~al.} (\bibinfo {collaboration} {BaBar}), }\bibfield
  {title} {\emph {\bibinfo {title} {{Measurement of an Excess of $\bar{B} \to
  D^{(*)}\tau^- \bar{\nu}_\tau$ Decays and Implications for Charged Higgs
  Bosons}}}, }\href {\doibase 10.1103/PhysRevD.88.072012} {\bibfield  {journal}
  {\bibinfo  {journal} {Phys. Rev.} }\textbf {\bibinfo {volume} {D88}}
  (\bibinfo {year} {2013}) \bibinfo {pages} {072012}}, \Eprint
  {http://arxiv.org/abs/1303.0571}{arXiv:1303.0571 [hep-ex]}\BibitemShut
  {NoStop}%
\bibitem [{\citenamefont {Aaij} \emph {et~al.}(2015)}]{Aaij:2015yra}%
  \BibitemOpen
  \bibfield  {author} {\bibinfo {author} {\bibfnamefont {R.}~\bibnamefont
  {Aaij}} \emph {et~al.} (\bibinfo {collaboration} {LHCb}), }\bibfield  {title}
  {\emph {\bibinfo {title} {{Measurement of the ratio of branching fractions
  $\mathcal{B}(\bar{B}^0 \to
  D^{*+}\tau^{-}\bar{\nu}_{\tau})/\mathcal{B}(\bar{B}^0 \to
  D^{*+}\mu^{-}\bar{\nu}_{\mu})$}}}, }\href {\doibase
  10.1103/PhysRevLett.115.159901, 10.1103/PhysRevLett.115.111803} {\bibfield
  {journal} {\bibinfo  {journal} {Phys. Rev. Lett.} }\textbf {\bibinfo {volume}
  {115}} (\bibinfo {year} {2015}) \bibinfo {pages} {111803}}, \bibinfo {note}
  {[Erratum: Phys. Rev. Lett.115,no.15,159901(2015)]}, \Eprint
  {http://arxiv.org/abs/1506.08614}{arXiv:1506.08614 [hep-ex]}\BibitemShut
  {NoStop}%
\bibitem [{\citenamefont {Huschle} \emph {et~al.}(2015)}]{Huschle:2015rga}%
  \BibitemOpen
  \bibfield  {author} {\bibinfo {author} {\bibfnamefont {M.}~\bibnamefont
  {Huschle}} \emph {et~al.} (\bibinfo {collaboration} {Belle}), }\bibfield
  {title} {\emph {\bibinfo {title} {{Measurement of the branching ratio of
  $\bar{B} \to D^{(\ast)} \tau^- \bar{\nu}_\tau$ relative to $\bar{B} \to
  D^{(\ast)} \ell^- \bar{\nu}_\ell$ decays with hadronic tagging at Belle}}},
  }\href {\doibase 10.1103/PhysRevD.92.072014} {\bibfield  {journal} {\bibinfo
  {journal} {Phys. Rev.} }\textbf {\bibinfo {volume} {D92}} (\bibinfo {year}
  {2015}) \bibinfo {pages} {072014}}, \Eprint
  {http://arxiv.org/abs/1507.03233}{arXiv:1507.03233 [hep-ex]}\BibitemShut
  {NoStop}%
\bibitem [{\citenamefont {Sato} \emph {et~al.}(2016)}]{Sato:2016svk}%
  \BibitemOpen
  \bibfield  {author} {\bibinfo {author} {\bibfnamefont {Y.}~\bibnamefont
  {Sato}} \emph {et~al.} (\bibinfo {collaboration} {Belle}), }\bibfield
  {title} {\emph {\bibinfo {title} {{Measurement of the branching ratio of
  $\bar{B}^0 \rightarrow D^{*+} \tau^- \bar{\nu}_{\tau}$ relative to $\bar{B}^0
  \rightarrow D^{*+} \ell^- \bar{\nu}_{\ell}$ decays with a semileptonic
  tagging method}}}, }\href {\doibase 10.1103/PhysRevD.94.072007} {\bibfield
  {journal} {\bibinfo  {journal} {Phys. Rev.} }\textbf {\bibinfo {volume}
  {D94}} (\bibinfo {year} {2016}) \bibinfo {pages} {072007}}, \Eprint
  {http://arxiv.org/abs/1607.07923}{arXiv:1607.07923 [hep-ex]}\BibitemShut
  {NoStop}%
\bibitem [{\citenamefont {Hirose} \emph {et~al.}(2017)}]{Hirose:2016wfn}%
  \BibitemOpen
  \bibfield  {author} {\bibinfo {author} {\bibfnamefont {S.}~\bibnamefont
  {Hirose}} \emph {et~al.} (\bibinfo {collaboration} {Belle}), }\bibfield
  {title} {\emph {\bibinfo {title} {{Measurement of the $\tau$ lepton
  polarization and $R(D^*)$ in the decay $\bar{B} \to D^* \tau^-
  \bar{\nu}_\tau$}}}, }\href {\doibase 10.1103/PhysRevLett.118.211801}
  {\bibfield  {journal} {\bibinfo  {journal} {Phys. Rev. Lett.} }\textbf
  {\bibinfo {volume} {118}} (\bibinfo {year} {2017}) \bibinfo {pages}
  {211801}}, \Eprint {http://arxiv.org/abs/1612.00529}{arXiv:1612.00529
  [hep-ex]}\BibitemShut {NoStop}%
\bibitem [{\citenamefont {Alonso} \emph {et~al.}(2017)\citenamefont {Alonso},
  \citenamefont {Grinstein}, and \citenamefont
  {Martin~Camalich}}]{Alonso:2016oyd}%
  \BibitemOpen
  \bibfield  {author} {\bibinfo {author} {\bibfnamefont {R.}~\bibnamefont
  {Alonso}}, \bibinfo {author} {\bibfnamefont {B.}~\bibnamefont {Grinstein}},
  and \bibinfo {author} {\bibfnamefont {J.}~\bibnamefont {Martin~Camalich}},
  }\bibfield  {title} {\emph {\bibinfo {title} {{Lifetime of $B_c^-$ Constrains
  Explanations for Anomalies in $B\to D^{(*)}\tau\nu$}}}, }\href {\doibase
  10.1103/PhysRevLett.118.081802} {\bibfield  {journal} {\bibinfo  {journal}
  {Phys. Rev. Lett.} }\textbf {\bibinfo {volume} {118}} (\bibinfo {year}
  {2017}) \bibinfo {pages} {081802}}, \Eprint
  {http://arxiv.org/abs/1611.06676}{arXiv:1611.06676 [hep-ph]}\BibitemShut
  {NoStop}%
\bibitem [{\citenamefont {Isidori}(2014)}]{Isidori:2013ez}%
  \BibitemOpen
  \bibfield  {author} {\bibinfo {author} {\bibfnamefont {G.}~\bibnamefont
  {Isidori}}, }\bibfield  {title} {\emph {\bibinfo {title} {{Flavor physics and
  CP violation}}}, }in \href {\doibase 10.5170/CERN-2014-008.69} {\emph
  {\bibinfo {booktitle} {{Proceedings, 2012 European School of High-Energy
  Physics (ESHEP 2012): La Pommeraye, Anjou, France, June 06-19, 2012}}}}
  (\bibinfo {year} {2014}) pp. \bibinfo {pages} {69--105}, \Eprint
  {http://arxiv.org/abs/1302.0661}{arXiv:1302.0661 [hep-ph]}\BibitemShut
  {NoStop}%
\bibitem [{\citenamefont {Altmannshofer} and \citenamefont
  {Straub}(2015)}]{Altmannshofer:2015sma}%
  \BibitemOpen
  \bibfield  {author} {\bibinfo {author} {\bibfnamefont {W.}~\bibnamefont
  {Altmannshofer}} and \bibinfo {author} {\bibfnamefont {D.~M.} \bibnamefont
  {Straub}}, }\bibfield  {title} {\emph {\bibinfo {title} {{Implications of
  $b\to s$ measurements}}}, }in \href
  {http://inspirehep.net/record/1353682/files/arXiv:1503.06199.pdf} {\emph
  {\bibinfo {booktitle} {{Proceedings, 50th Rencontres de Moriond Electroweak
  Interactions and Unified Theories: La Thuile, Italy, March 14-21, 2015}}}}
  (\bibinfo {year} {2015}) pp. \bibinfo {pages} {333--338}, \Eprint
  {http://arxiv.org/abs/1503.06199}{arXiv:1503.06199 [hep-ph]}\BibitemShut
  {NoStop}%
\bibitem [{\citenamefont {Descotes-Genon} \emph {et~al.}(2016)\citenamefont
  {Descotes-Genon}, \citenamefont {Hofer}, \citenamefont {Matias}, and
  \citenamefont {Virto}}]{Descotes-Genon:2015uva}%
  \BibitemOpen
  \bibfield  {author} {\bibinfo {author} {\bibfnamefont {S.}~\bibnamefont
  {Descotes-Genon}}, \bibinfo {author} {\bibfnamefont {L.}~\bibnamefont
  {Hofer}}, \bibinfo {author} {\bibfnamefont {J.}~\bibnamefont {Matias}},  and
  \bibinfo {author} {\bibfnamefont {J.}~\bibnamefont {Virto}}, }\bibfield
  {title} {\emph {\bibinfo {title} {{Global analysis of $b\to s\ell\ell$
  anomalies}}}, }\href {\doibase 10.1007/JHEP06(2016)092} {\bibfield  {journal}
  {\bibinfo  {journal} {JHEP} }\textbf {\bibinfo {volume} {06}} (\bibinfo
  {year} {2016}) \bibinfo {pages} {092}}, \Eprint
  {http://arxiv.org/abs/1510.04239}{arXiv:1510.04239 [hep-ph]}\BibitemShut
  {NoStop}%
\bibitem [{\citenamefont {Brod} \emph {et~al.}(2015)\citenamefont {Brod},
  \citenamefont {Greljo}, \citenamefont {Stamou}, and \citenamefont
  {Uttayarat}}]{Brod:2014hsa}%
  \BibitemOpen
  \bibfield  {author} {\bibinfo {author} {\bibfnamefont {J.}~\bibnamefont
  {Brod}}, \bibinfo {author} {\bibfnamefont {A.}~\bibnamefont {Greljo}},
  \bibinfo {author} {\bibfnamefont {E.}~\bibnamefont {Stamou}},  and \bibinfo
  {author} {\bibfnamefont {P.}~\bibnamefont {Uttayarat}}, }\bibfield  {title}
  {\emph {\bibinfo {title} {{Probing anomalous $ t\overline{t}Z $ interactions
  with rare meson decays}}}, }\href {\doibase 10.1007/JHEP02(2015)141}
  {\bibfield  {journal} {\bibinfo  {journal} {JHEP} }\textbf {\bibinfo {volume}
  {02}} (\bibinfo {year} {2015}) \bibinfo {pages} {141}}, \Eprint
  {http://arxiv.org/abs/1408.0792}{arXiv:1408.0792 [hep-ph]}\BibitemShut
  {NoStop}%
\bibitem [{\citenamefont {Greljo} and \citenamefont
  {Marzocca}(2017)}]{Greljo:2017vvb}%
  \BibitemOpen
  \bibfield  {author} {\bibinfo {author} {\bibfnamefont {A.}~\bibnamefont
  {Greljo}} and \bibinfo {author} {\bibfnamefont {D.}~\bibnamefont {Marzocca}},
  }\bibfield  {title} {\emph {\bibinfo {title} {{High-$p_T$ dilepton tails and
  flavor physics}}}, }\href {\doibase 10.1140/epjc/s10052-017-5119-8}
  {\bibfield  {journal} {\bibinfo  {journal} {Eur. Phys. J.} }\textbf {\bibinfo
  {volume} {C77}} (\bibinfo {year} {2017}) \bibinfo {pages} {548}}, \Eprint
  {http://arxiv.org/abs/1704.09015}{arXiv:1704.09015 [hep-ph]}\BibitemShut
  {NoStop}%
\bibitem [{\citenamefont {Faroughy} \emph {et~al.}(2017)\citenamefont
  {Faroughy}, \citenamefont {Greljo}, and \citenamefont
  {Kamenik}}]{Faroughy:2016osc}%
  \BibitemOpen
  \bibfield  {author} {\bibinfo {author} {\bibfnamefont {D.~A.} \bibnamefont
  {Faroughy}}, \bibinfo {author} {\bibfnamefont {A.}~\bibnamefont {Greljo}},
  and \bibinfo {author} {\bibfnamefont {J.~F.} \bibnamefont {Kamenik}},
  }\bibfield  {title} {\emph {\bibinfo {title} {{Confronting lepton flavor
  universality violation in B decays with high-$p_T$ tau lepton searches at
  LHC}}}, }\href {\doibase 10.1016/j.physletb.2016.11.011} {\bibfield
  {journal} {\bibinfo  {journal} {Phys. Lett.} }\textbf {\bibinfo {volume}
  {B764}} (\bibinfo {year} {2017}) \bibinfo {pages} {126}}, \Eprint
  {http://arxiv.org/abs/1609.07138}{arXiv:1609.07138 [hep-ph]}\BibitemShut
  {NoStop}%
\bibitem [{\citenamefont {Farina} \emph {et~al.}(2017)\citenamefont {Farina},
  \citenamefont {Panico}, \citenamefont {Pappadopulo}, \citenamefont
  {Ruderman}, \citenamefont {Torre}, and \citenamefont
  {Wulzer}}]{Farina:2016rws}%
  \BibitemOpen
  \bibfield  {author} {\bibinfo {author} {\bibfnamefont {M.}~\bibnamefont
  {Farina}}, \bibinfo {author} {\bibfnamefont {G.}~\bibnamefont {Panico}},
  \bibinfo {author} {\bibfnamefont {D.}~\bibnamefont {Pappadopulo}}, \bibinfo
  {author} {\bibfnamefont {J.~T.} \bibnamefont {Ruderman}}, \bibinfo {author}
  {\bibfnamefont {R.}~\bibnamefont {Torre}},  and \bibinfo {author}
  {\bibfnamefont {A.}~\bibnamefont {Wulzer}}, }\bibfield  {title} {\emph
  {\bibinfo {title} {{Energy helps accuracy: electroweak precision tests at
  hadron colliders}}}, }\href {\doibase 10.1016/j.physletb.2017.06.043}
  {\bibfield  {journal} {\bibinfo  {journal} {Phys. Lett.} }\textbf {\bibinfo
  {volume} {B772}} (\bibinfo {year} {2017}) \bibinfo {pages} {210}}, \Eprint
  {http://arxiv.org/abs/1609.08157}{arXiv:1609.08157 [hep-ph]}\BibitemShut
  {NoStop}%
\bibitem [{\citenamefont {Alioli} \emph
  {et~al.}(2017{\natexlab{a}})\citenamefont {Alioli}, \citenamefont {Farina},
  \citenamefont {Pappadopulo}, and \citenamefont {Ruderman}}]{Alioli:2017jdo}%
  \BibitemOpen
  \bibfield  {author} {\bibinfo {author} {\bibfnamefont {S.}~\bibnamefont
  {Alioli}}, \bibinfo {author} {\bibfnamefont {M.}~\bibnamefont {Farina}},
  \bibinfo {author} {\bibfnamefont {D.}~\bibnamefont {Pappadopulo}},  and
  \bibinfo {author} {\bibfnamefont {J.~T.} \bibnamefont {Ruderman}}, }\bibfield
   {title} {\emph {\bibinfo {title} {{Precision Probes of QCD at High
  Energies}}}, }\href {\doibase 10.1007/JHEP07(2017)097} {\bibfield  {journal}
  {\bibinfo  {journal} {JHEP} }\textbf {\bibinfo {volume} {07}} (\bibinfo
  {year} {2017}{\natexlab{a}}) \bibinfo {pages} {097}}, \Eprint
  {http://arxiv.org/abs/1706.03068}{arXiv:1706.03068 [hep-ph]}\BibitemShut
  {NoStop}%
\bibitem [{\citenamefont {Alioli} \emph
  {et~al.}(2017{\natexlab{b}})\citenamefont {Alioli}, \citenamefont {Farina},
  \citenamefont {Pappadopulo}, and \citenamefont {Ruderman}}]{Alioli:2017nzr}%
  \BibitemOpen
  \bibfield  {author} {\bibinfo {author} {\bibfnamefont {S.}~\bibnamefont
  {Alioli}}, \bibinfo {author} {\bibfnamefont {M.}~\bibnamefont {Farina}},
  \bibinfo {author} {\bibfnamefont {D.}~\bibnamefont {Pappadopulo}},  and
  \bibinfo {author} {\bibfnamefont {J.~T.} \bibnamefont {Ruderman}}, }\bibfield
   {title} {\emph {\bibinfo {title} {{Catching a New Force by the Tail}}},
  }\href@noop {} { (\bibinfo {year} {2017}{\natexlab{b}})}, \Eprint
  {http://arxiv.org/abs/1712.02347}{arXiv:1712.02347 [hep-ph]}\BibitemShut
  {NoStop}%
\bibitem [{\citenamefont {Greiner} \emph {et~al.}(2011)\citenamefont {Greiner},
  \citenamefont {Willenbrock}, and \citenamefont {Zhang}}]{Greiner:2011tt}%
  \BibitemOpen
  \bibfield  {author} {\bibinfo {author} {\bibfnamefont {N.}~\bibnamefont
  {Greiner}}, \bibinfo {author} {\bibfnamefont {S.}~\bibnamefont
  {Willenbrock}},  and \bibinfo {author} {\bibfnamefont {C.}~\bibnamefont
  {Zhang}}, }\bibfield  {title} {\emph {\bibinfo {title} {{Effective Field
  Theory for Nonstandard Top Quark Couplings}}}, }\href {\doibase
  10.1016/j.physletb.2011.09.026} {\bibfield  {journal} {\bibinfo  {journal}
  {Phys. Lett.} }\textbf {\bibinfo {volume} {B704}} (\bibinfo {year} {2011})
  \bibinfo {pages} {218}}, \Eprint
  {http://arxiv.org/abs/1104.3122}{arXiv:1104.3122 [hep-ph]}\BibitemShut
  {NoStop}%
\bibitem [{\citenamefont {Zhang} \emph {et~al.}(2012)\citenamefont {Zhang},
  \citenamefont {Greiner}, and \citenamefont {Willenbrock}}]{Zhang:2012cd}%
  \BibitemOpen
  \bibfield  {author} {\bibinfo {author} {\bibfnamefont {C.}~\bibnamefont
  {Zhang}}, \bibinfo {author} {\bibfnamefont {N.}~\bibnamefont {Greiner}},  and
  \bibinfo {author} {\bibfnamefont {S.}~\bibnamefont {Willenbrock}}, }\bibfield
   {title} {\emph {\bibinfo {title} {{Constraints on Non-standard Top Quark
  Couplings}}}, }\href {\doibase 10.1103/PhysRevD.86.014024} {\bibfield
  {journal} {\bibinfo  {journal} {Phys. Rev.} }\textbf {\bibinfo {volume}
  {D86}} (\bibinfo {year} {2012}) \bibinfo {pages} {014024}}, \Eprint
  {http://arxiv.org/abs/1201.6670}{arXiv:1201.6670 [hep-ph]}\BibitemShut
  {NoStop}%
\bibitem [{\citenamefont {Hagiwara} \emph {et~al.}(1993)\citenamefont
  {Hagiwara}, \citenamefont {Ishihara}, \citenamefont {Szalapski}, and
  \citenamefont {Zeppenfeld}}]{Hagiwara:1993ck}%
  \BibitemOpen
  \bibfield  {author} {\bibinfo {author} {\bibfnamefont {K.}~\bibnamefont
  {Hagiwara}}, \bibinfo {author} {\bibfnamefont {S.}~\bibnamefont {Ishihara}},
  \bibinfo {author} {\bibfnamefont {R.}~\bibnamefont {Szalapski}},  and
  \bibinfo {author} {\bibfnamefont {D.}~\bibnamefont {Zeppenfeld}}, }\bibfield
  {title} {\emph {\bibinfo {title} {{Low-energy effects of new interactions in
  the electroweak boson sector}}}, }\href {\doibase 10.1103/PhysRevD.48.2182}
  {\bibfield  {journal} {\bibinfo  {journal} {Phys. Rev.} }\textbf {\bibinfo
  {volume} {D48}} (\bibinfo {year} {1993}) \bibinfo {pages} {2182}}\BibitemShut
  {NoStop}%
\bibitem [{\citenamefont {de~Blas} \emph {et~al.}(2015)\citenamefont {de~Blas},
  \citenamefont {Chala}, and \citenamefont {Santiago}}]{deBlas:2015aea}%
  \BibitemOpen
  \bibfield  {author} {\bibinfo {author} {\bibfnamefont {J.}~\bibnamefont
  {de~Blas}}, \bibinfo {author} {\bibfnamefont {M.}~\bibnamefont {Chala}},  and
  \bibinfo {author} {\bibfnamefont {J.}~\bibnamefont {Santiago}}, }\bibfield
  {title} {\emph {\bibinfo {title} {{Renormalization Group Constraints on New
  Top Interactions from Electroweak Precision Data}}}, }\href {\doibase
  10.1007/JHEP09(2015)189} {\bibfield  {journal} {\bibinfo  {journal} {JHEP}
  }\textbf {\bibinfo {volume} {09}} (\bibinfo {year} {2015}) \bibinfo {pages}
  {189}}, \Eprint {http://arxiv.org/abs/1507.00757}{arXiv:1507.00757
  [hep-ph]}\BibitemShut {NoStop}%
\bibitem [{\citenamefont {Feruglio} \emph
  {et~al.}(2017{\natexlab{a}})\citenamefont {Feruglio}, \citenamefont
  {Paradisi}, and \citenamefont {Pattori}}]{Feruglio:2016gvd}%
  \BibitemOpen
  \bibfield  {author} {\bibinfo {author} {\bibfnamefont {F.}~\bibnamefont
  {Feruglio}}, \bibinfo {author} {\bibfnamefont {P.}~\bibnamefont {Paradisi}},
  and \bibinfo {author} {\bibfnamefont {A.}~\bibnamefont {Pattori}}, }\bibfield
   {title} {\emph {\bibinfo {title} {{Revisiting Lepton Flavor Universality in
  B Decays}}}, }\href {\doibase 10.1103/PhysRevLett.118.011801} {\bibfield
  {journal} {\bibinfo  {journal} {Phys. Rev. Lett.} }\textbf {\bibinfo {volume}
  {118}} (\bibinfo {year} {2017}{\natexlab{a}}) \bibinfo {pages} {011801}},
  \Eprint {http://arxiv.org/abs/1606.00524}{arXiv:1606.00524
  [hep-ph]}\BibitemShut {NoStop}%
\bibitem [{\citenamefont {Feruglio} \emph
  {et~al.}(2017{\natexlab{b}})\citenamefont {Feruglio}, \citenamefont
  {Paradisi}, and \citenamefont {Pattori}}]{Feruglio:2017rjo}%
  \BibitemOpen
  \bibfield  {author} {\bibinfo {author} {\bibfnamefont {F.}~\bibnamefont
  {Feruglio}}, \bibinfo {author} {\bibfnamefont {P.}~\bibnamefont {Paradisi}},
  and \bibinfo {author} {\bibfnamefont {A.}~\bibnamefont {Pattori}}, }\bibfield
   {title} {\emph {\bibinfo {title} {{On the Importance of Electroweak
  Corrections for B Anomalies}}}, }\href {\doibase 10.1007/JHEP09(2017)061}
  {\bibfield  {journal} {\bibinfo  {journal} {JHEP} }\textbf {\bibinfo {volume}
  {09}} (\bibinfo {year} {2017}{\natexlab{b}}) \bibinfo {pages} {061}}, \Eprint
  {http://arxiv.org/abs/1705.00929}{arXiv:1705.00929 [hep-ph]}\BibitemShut
  {NoStop}%
\bibitem [{\citenamefont {Efrati} \emph {et~al.}(2015)\citenamefont {Efrati},
  \citenamefont {Falkowski}, and \citenamefont {Soreq}}]{Efrati:2015eaa}%
  \BibitemOpen
  \bibfield  {author} {\bibinfo {author} {\bibfnamefont {A.}~\bibnamefont
  {Efrati}}, \bibinfo {author} {\bibfnamefont {A.}~\bibnamefont {Falkowski}},
  and \bibinfo {author} {\bibfnamefont {Y.}~\bibnamefont {Soreq}}, }\bibfield
  {title} {\emph {\bibinfo {title} {{Electroweak constraints on flavorful
  effective theories}}}, }\href {\doibase 10.1007/JHEP07(2015)018} {\bibfield
  {journal} {\bibinfo  {journal} {JHEP} }\textbf {\bibinfo {volume} {07}}
  (\bibinfo {year} {2015}) \bibinfo {pages} {018}}, \Eprint
  {http://arxiv.org/abs/1503.07872}{arXiv:1503.07872 [hep-ph]}\BibitemShut
  {NoStop}%
\bibitem [{\citenamefont {Dicus}(1990)}]{Dicus:1989va}%
  \BibitemOpen
  \bibfield  {author} {\bibinfo {author} {\bibfnamefont {D.~A.} \bibnamefont
  {Dicus}}, }\bibfield  {title} {\emph {\bibinfo {title} {{Neutron Electric
  Dipole Moment From Charged Higgs Exchange}}}, }\href {\doibase
  10.1103/PhysRevD.41.999} {\bibfield  {journal} {\bibinfo  {journal}
  {Phys.Rev.} }\textbf {\bibinfo {volume} {D41}} (\bibinfo {year} {1990})
  \bibinfo {pages} {999}}\BibitemShut {NoStop}%
\bibitem [{\citenamefont {Weinberg}(1989)}]{Weinberg:1989dx}%
  \BibitemOpen
  \bibfield  {author} {\bibinfo {author} {\bibfnamefont {S.}~\bibnamefont
  {Weinberg}}, }\bibfield  {title} {\emph {\bibinfo {title} {{Larger Higgs
  Exchange Terms in the Neutron Electric Dipole Moment}}}, }\href {\doibase
  10.1103/PhysRevLett.63.2333} {\bibfield  {journal} {\bibinfo  {journal}
  {Phys. Rev. Lett.} }\textbf {\bibinfo {volume} {63}} (\bibinfo {year} {1989})
  \bibinfo {pages} {2333}}\BibitemShut {NoStop}%
\bibitem [{\citenamefont {Braaten} \emph {et~al.}(1990)\citenamefont {Braaten},
  \citenamefont {Li}, and \citenamefont {Yuan}}]{BraatenPRL}%
  \BibitemOpen
  \bibfield  {author} {\bibinfo {author} {\bibfnamefont {E.}~\bibnamefont
  {Braaten}}, \bibinfo {author} {\bibfnamefont {C.-S.} \bibnamefont {Li}},  and
  \bibinfo {author} {\bibfnamefont {T.-C.} \bibnamefont {Yuan}}, }\bibfield
  {title} {\emph {\bibinfo {title} {{The Evolution of Weinberg's Gluonic {CP}
  Violation Operator}}}, }\href {\doibase 10.1103/PhysRevLett.64.1709}
  {\bibfield  {journal} {\bibinfo  {journal} {Phys. Rev. Lett.} }\textbf
  {\bibinfo {volume} {64}} (\bibinfo {year} {1990}) \bibinfo {pages}
  {1709}}\BibitemShut {NoStop}%
\bibitem [{\citenamefont {Boyd} \emph {et~al.}(1990)\citenamefont {Boyd},
  \citenamefont {Gupta}, \citenamefont {Trivedi}, and \citenamefont
  {Wise}}]{Boyd:1990bx}%
  \BibitemOpen
  \bibfield  {author} {\bibinfo {author} {\bibfnamefont {G.}~\bibnamefont
  {Boyd}}, \bibinfo {author} {\bibfnamefont {A.~K.} \bibnamefont {Gupta}},
  \bibinfo {author} {\bibfnamefont {S.~P.} \bibnamefont {Trivedi}},  and
  \bibinfo {author} {\bibfnamefont {M.~B.} \bibnamefont {Wise}}, }\bibfield
  {title} {\emph {\bibinfo {title} {{Effective Hamiltonian for the Electric
  Dipole Moment of the Neutron}}}, }\href {\doibase
  10.1016/0370-2693(90)91874-B} {\bibfield  {journal} {\bibinfo  {journal}
  {Phys.Lett.} }\textbf {\bibinfo {volume} {B241}} (\bibinfo {year} {1990})
  \bibinfo {pages} {584}}\BibitemShut {NoStop}%
\bibitem [{\citenamefont {Kamenik} \emph {et~al.}(2012)\citenamefont {Kamenik},
  \citenamefont {Papucci}, and \citenamefont {Weiler}}]{Kamenik:2011dk}%
  \BibitemOpen
  \bibfield  {author} {\bibinfo {author} {\bibfnamefont {J.~F.} \bibnamefont
  {Kamenik}}, \bibinfo {author} {\bibfnamefont {M.}~\bibnamefont {Papucci}},
  and \bibinfo {author} {\bibfnamefont {A.}~\bibnamefont {Weiler}}, }\bibfield
  {title} {\emph {\bibinfo {title} {{Constraining the dipole moments of the top
  quark}}}, }\href {\doibase 10.1103/PhysRevD.88.039903,
  10.1103/PhysRevD.85.071501} {\bibfield  {journal} {\bibinfo  {journal} {Phys.
  Rev.} }\textbf {\bibinfo {volume} {D85}} (\bibinfo {year} {2012}) \bibinfo
  {pages} {071501}}, \bibinfo {note} {[Erratum: Phys.
  Rev.D88,no.3,039903(2013)]}, \Eprint
  {http://arxiv.org/abs/1107.3143}{arXiv:1107.3143 [hep-ph]}\BibitemShut
  {NoStop}%
\bibitem [{\citenamefont {Brod} \emph {et~al.}(2013)\citenamefont {Brod},
  \citenamefont {Haisch}, and \citenamefont {Zupan}}]{Brod:2013cka}%
  \BibitemOpen
  \bibfield  {author} {\bibinfo {author} {\bibfnamefont {J.}~\bibnamefont
  {Brod}}, \bibinfo {author} {\bibfnamefont {U.}~\bibnamefont {Haisch}},  and
  \bibinfo {author} {\bibfnamefont {J.}~\bibnamefont {Zupan}}, }\bibfield
  {title} {\emph {\bibinfo {title} {{Constraints on CP-violating Higgs
  couplings to the third generation}}}, }\href {\doibase
  10.1007/JHEP11(2013)180} {\bibfield  {journal} {\bibinfo  {journal} {JHEP}
  }\textbf {\bibinfo {volume} {11}} (\bibinfo {year} {2013}) \bibinfo {pages}
  {180}}, \Eprint {http://arxiv.org/abs/1310.1385}{arXiv:1310.1385
  [hep-ph]}\BibitemShut {NoStop}%
\bibitem [{\citenamefont {Barr} and \citenamefont {Zee}(1990)}]{Barr:1990vd}%
  \BibitemOpen
  \bibfield  {author} {\bibinfo {author} {\bibfnamefont {S.~M.} \bibnamefont
  {Barr}} and \bibinfo {author} {\bibfnamefont {A.}~\bibnamefont {Zee}},
  }\bibfield  {title} {\emph {\bibinfo {title} {{Electric Dipole Moment of the
  Electron and of the Neutron}}}, }\href {\doibase 10.1103/PhysRevLett.65.21}
  {\bibfield  {journal} {\bibinfo  {journal} {Phys. Rev. Lett.} }\textbf
  {\bibinfo {volume} {65}} (\bibinfo {year} {1990}) \bibinfo {pages}
  {21}}\BibitemShut {NoStop}%
\bibitem [{\citenamefont {Gunion} and \citenamefont
  {Wyler}(1990)}]{Gunion:1990iv}%
  \BibitemOpen
  \bibfield  {author} {\bibinfo {author} {\bibfnamefont {J.}~\bibnamefont
  {Gunion}} and \bibinfo {author} {\bibfnamefont {D.}~\bibnamefont {Wyler}},
  }\bibfield  {title} {\emph {\bibinfo {title} {{Inducing a large neutron
  electric dipole moment via a quark chromoelectric dipole moment}}}, }\href
  {\doibase 10.1016/0370-2693(90)90034-4} {\bibfield  {journal} {\bibinfo
  {journal} {Phys.Lett.} }\textbf {\bibinfo {volume} {B248}} (\bibinfo {year}
  {1990}) \bibinfo {pages} {170}}\BibitemShut {NoStop}%
\bibitem [{\citenamefont {Abe} \emph {et~al.}(2014)\citenamefont {Abe},
  \citenamefont {Hisano}, \citenamefont {Kitahara}, and \citenamefont
  {Tobioka}}]{Abe:2013qla}%
  \BibitemOpen
  \bibfield  {author} {\bibinfo {author} {\bibfnamefont {T.}~\bibnamefont
  {Abe}}, \bibinfo {author} {\bibfnamefont {J.}~\bibnamefont {Hisano}},
  \bibinfo {author} {\bibfnamefont {T.}~\bibnamefont {Kitahara}},  and \bibinfo
  {author} {\bibfnamefont {K.}~\bibnamefont {Tobioka}}, }\bibfield  {title}
  {\emph {\bibinfo {title} {{Gauge invariant Barr-Zee type contributions to
  fermionic EDMs in the two-Higgs doublet models}}}, }\href {\doibase
  10.1007/JHEP01(2014)106} {\bibfield  {journal} {\bibinfo  {journal} {JHEP}
  }\textbf {\bibinfo {volume} {1401}} (\bibinfo {year} {2014}) \bibinfo {pages}
  {106}}, \Eprint {http://arxiv.org/abs/1311.4704}{arXiv:1311.4704
  [hep-ph]}\BibitemShut {NoStop}%
\bibitem [{\citenamefont {Jung} and \citenamefont {Pich}(2014)}]{Jung:2013hka}%
  \BibitemOpen
  \bibfield  {author} {\bibinfo {author} {\bibfnamefont {M.}~\bibnamefont
  {Jung}} and \bibinfo {author} {\bibfnamefont {A.}~\bibnamefont {Pich}},
  }\bibfield  {title} {\emph {\bibinfo {title} {{Electric Dipole Moments in
  Two-Higgs-Doublet Models}}}, }\href {\doibase 10.1007/JHEP04(2014)076}
  {\bibfield  {journal} {\bibinfo  {journal} {JHEP} }\textbf {\bibinfo {volume}
  {04}} (\bibinfo {year} {2014}) \bibinfo {pages} {076}}, \Eprint
  {http://arxiv.org/abs/1308.6283}{arXiv:1308.6283 [hep-ph]}\BibitemShut
  {NoStop}%
\bibitem [{\citenamefont {Dekens} and \citenamefont
  {de~Vries}(2013)}]{Dekens:2013zca}%
  \BibitemOpen
  \bibfield  {author} {\bibinfo {author} {\bibfnamefont {W.}~\bibnamefont
  {Dekens}} and \bibinfo {author} {\bibfnamefont {J.}~\bibnamefont {de~Vries}},
  }\bibfield  {title} {\emph {\bibinfo {title} {{Renormalization Group Running
  of Dimension-Six Sources of Parity and Time-Reversal Violation}}}, }\href
  {\doibase 10.1007/JHEP05(2013)149} {\bibfield  {journal} {\bibinfo  {journal}
  {JHEP} }\textbf {\bibinfo {volume} {1305}} (\bibinfo {year} {2013}) \bibinfo
  {pages} {149}}, \Eprint {http://arxiv.org/abs/1303.3156}{arXiv:1303.3156
  [hep-ph]}\BibitemShut {NoStop}%
\bibitem [{\citenamefont {Alonso} \emph {et~al.}(2014)\citenamefont {Alonso},
  \citenamefont {Jenkins}, \citenamefont {Manohar}, and \citenamefont
  {Trott}}]{Alonso:2013hga}%
  \BibitemOpen
  \bibfield  {author} {\bibinfo {author} {\bibfnamefont {R.}~\bibnamefont
  {Alonso}}, \bibinfo {author} {\bibfnamefont {E.~E.} \bibnamefont {Jenkins}},
  \bibinfo {author} {\bibfnamefont {A.~V.} \bibnamefont {Manohar}},  and
  \bibinfo {author} {\bibfnamefont {M.}~\bibnamefont {Trott}}, }\bibfield
  {title} {\emph {\bibinfo {title} {{Renormalization Group Evolution of the
  Standard Model Dimension Six Operators III: Gauge Coupling Dependence and
  Phenomenology}}}, }\href {\doibase 10.1007/JHEP04(2014)159} {\bibfield
  {journal} {\bibinfo  {journal} {JHEP} }\textbf {\bibinfo {volume} {04}}
  (\bibinfo {year} {2014}) \bibinfo {pages} {159}}, \Eprint
  {http://arxiv.org/abs/1312.2014}{arXiv:1312.2014 [hep-ph]}\BibitemShut
  {NoStop}%
\bibitem [{\citenamefont {Cirigliano} \emph
  {et~al.}(2016{\natexlab{a}})\citenamefont {Cirigliano}, \citenamefont
  {Dekens}, \citenamefont {de~Vries}, and \citenamefont
  {Mereghetti}}]{Cirigliano:2016nyn}%
  \BibitemOpen
  \bibfield  {author} {\bibinfo {author} {\bibfnamefont {V.}~\bibnamefont
  {Cirigliano}}, \bibinfo {author} {\bibfnamefont {W.}~\bibnamefont {Dekens}},
  \bibinfo {author} {\bibfnamefont {J.}~\bibnamefont {de~Vries}},  and \bibinfo
  {author} {\bibfnamefont {E.}~\bibnamefont {Mereghetti}}, }\bibfield  {title}
  {\emph {\bibinfo {title} {{Constraining the top-Higgs sector of the Standard
  Model Effective Field Theory}}}, }\href {\doibase 10.1103/PhysRevD.94.034031}
  {\bibfield  {journal} {\bibinfo  {journal} {Phys. Rev.} }\textbf {\bibinfo
  {volume} {D94}} (\bibinfo {year} {2016}{\natexlab{a}}) \bibinfo {pages}
  {034031}}, \Eprint {http://arxiv.org/abs/1605.04311}{arXiv:1605.04311
  [hep-ph]}\BibitemShut {NoStop}%
\bibitem [{\citenamefont {Alioli} \emph
  {et~al.}(2017{\natexlab{c}})\citenamefont {Alioli}, \citenamefont
  {Cirigliano}, \citenamefont {Dekens}, \citenamefont {de~Vries}, and
  \citenamefont {Mereghetti}}]{Alioli:2017ces}%
  \BibitemOpen
  \bibfield  {author} {\bibinfo {author} {\bibfnamefont {S.}~\bibnamefont
  {Alioli}}, \bibinfo {author} {\bibfnamefont {V.}~\bibnamefont {Cirigliano}},
  \bibinfo {author} {\bibfnamefont {W.}~\bibnamefont {Dekens}}, \bibinfo
  {author} {\bibfnamefont {J.}~\bibnamefont {de~Vries}},  and \bibinfo {author}
  {\bibfnamefont {E.}~\bibnamefont {Mereghetti}}, }\bibfield  {title} {\emph
  {\bibinfo {title} {{Right-handed charged currents in the era of the Large
  Hadron Collider}}}, }\href {\doibase 10.1007/JHEP05(2017)086} {\bibfield
  {journal} {\bibinfo  {journal} {JHEP} }\textbf {\bibinfo {volume} {05}}
  (\bibinfo {year} {2017}{\natexlab{c}}) \bibinfo {pages} {086}}, \Eprint
  {http://arxiv.org/abs/1703.04751}{arXiv:1703.04751 [hep-ph]}\BibitemShut
  {NoStop}%
\bibitem [{\citenamefont {Demir} \emph {et~al.}(2003)\citenamefont {Demir},
  \citenamefont {Pospelov}, and \citenamefont {Ritz}}]{Pospelov_Weinberg}%
  \BibitemOpen
  \bibfield  {author} {\bibinfo {author} {\bibfnamefont {D.~A.} \bibnamefont
  {Demir}}, \bibinfo {author} {\bibfnamefont {M.}~\bibnamefont {Pospelov}},
  and \bibinfo {author} {\bibfnamefont {A.}~\bibnamefont {Ritz}}, }\bibfield
  {title} {\emph {\bibinfo {title} {{Hadronic EDMs, the Weinberg operator, and
  light gluinos}}}, }\href {\doibase 10.1103/PhysRevD.67.015007} {\bibfield
  {journal} {\bibinfo  {journal} {Phys. Rev. D} }\textbf {\bibinfo {volume}
  {67}} (\bibinfo {year} {2003}) \bibinfo {pages} {015007}}, \Eprint
  {http://arxiv.org/abs/hep-ph/0208257}{arXiv:hep-ph/0208257}\BibitemShut
  {NoStop}%
\bibitem [{\citenamefont {de~Vries} \emph {et~al.}(2011)\citenamefont
  {de~Vries}, \citenamefont {Timmermans}, \citenamefont {Mereghetti}, and
  \citenamefont {van Kolck}}]{deVries:2010ah}%
  \BibitemOpen
  \bibfield  {author} {\bibinfo {author} {\bibfnamefont {J.}~\bibnamefont
  {de~Vries}}, \bibinfo {author} {\bibfnamefont {R.~G.~E.} \bibnamefont
  {Timmermans}}, \bibinfo {author} {\bibfnamefont {E.}~\bibnamefont
  {Mereghetti}},  and \bibinfo {author} {\bibfnamefont {U.}~\bibnamefont {van
  Kolck}}, }\bibfield  {title} {\emph {\bibinfo {title} {{The Nucleon Electric
  Dipole Form Factor From Dimension-Six Time-Reversal Violation}}}, }\href
  {\doibase 10.1016/j.physletb.2010.11.042} {\bibfield  {journal} {\bibinfo
  {journal} {Phys. Lett.} }\textbf {\bibinfo {volume} {B695}} (\bibinfo {year}
  {2011}) \bibinfo {pages} {268}}, \Eprint
  {http://arxiv.org/abs/1006.2304}{arXiv:1006.2304 [hep-ph]}\BibitemShut
  {NoStop}%
\bibitem [{\citenamefont {Chien} \emph {et~al.}(2016)\citenamefont {Chien},
  \citenamefont {Cirigliano}, \citenamefont {Dekens}, \citenamefont {de~Vries},
  and \citenamefont {Mereghetti}}]{Chien:2015xha}%
  \BibitemOpen
  \bibfield  {author} {\bibinfo {author} {\bibfnamefont {Y.~T.} \bibnamefont
  {Chien}}, \bibinfo {author} {\bibfnamefont {V.}~\bibnamefont {Cirigliano}},
  \bibinfo {author} {\bibfnamefont {W.}~\bibnamefont {Dekens}}, \bibinfo
  {author} {\bibfnamefont {J.}~\bibnamefont {de~Vries}},  and \bibinfo {author}
  {\bibfnamefont {E.}~\bibnamefont {Mereghetti}}, }\bibfield  {title} {\emph
  {\bibinfo {title} {{Direct and indirect constraints on CP-violating
  Higgs-quark and Higgs-gluon interactions}}}, }\href {\doibase
  10.1007/JHEP02(2016)011} {\bibfield  {journal} {\bibinfo  {journal} {JHEP}
  }\textbf {\bibinfo {volume} {02}} (\bibinfo {year} {2016}) \bibinfo {pages}
  {011}}, \bibinfo {note} {[JHEP02,011(2016)]}, \Eprint
  {http://arxiv.org/abs/1510.00725}{arXiv:1510.00725 [hep-ph]}\BibitemShut
  {NoStop}%
\bibitem [{\citenamefont {Pendlebury} \emph {et~al.}(2015)}]{Afach:2015sja}%
  \BibitemOpen
  \bibfield  {author} {\bibinfo {author} {\bibfnamefont {J.}~\bibnamefont
  {Pendlebury}} \emph {et~al.}, }\bibfield  {title} {\emph {\bibinfo {title}
  {{Revised experimental upper limit on the electric dipole moment of the
  neutron}}}, }\href {\doibase 10.1103/PhysRevD.92.092003} {\bibfield
  {journal} {\bibinfo  {journal} {Phys. Rev.} }\textbf {\bibinfo {volume}
  {D92}} (\bibinfo {year} {2015}) \bibinfo {pages} {092003}}, \Eprint
  {http://arxiv.org/abs/1509.04411}{arXiv:1509.04411 [hep-ex]}\BibitemShut
  {NoStop}%
\bibitem [{\citenamefont {Baker} \emph {et~al.}(2006)\citenamefont {Baker},
  \citenamefont {Doyle}, \citenamefont {Geltenbort}, \citenamefont {Green},
  \citenamefont {van~der Grinten} \emph {et~al.}}]{Baker:2006ts}%
  \BibitemOpen
  \bibfield  {author} {\bibinfo {author} {\bibfnamefont {C.~A.} \bibnamefont
  {Baker}}, \bibinfo {author} {\bibfnamefont {D.~D.} \bibnamefont {Doyle}},
  \bibinfo {author} {\bibfnamefont {P.}~\bibnamefont {Geltenbort}}, \bibinfo
  {author} {\bibfnamefont {K.}~\bibnamefont {Green}}, \bibinfo {author}
  {\bibfnamefont {M.~G.~D.} \bibnamefont {van~der Grinten}},  \emph {et~al.},
  }\bibfield  {title} {\emph {\bibinfo {title} {{An Improved experimental limit
  on the electric dipole moment of the neutron}}}, }\href {\doibase
  10.1103/PhysRevLett.97.131801} {\bibfield  {journal} {\bibinfo  {journal}
  {Phys. Rev. Lett.} }\textbf {\bibinfo {volume} {97}} (\bibinfo {year} {2006})
  \bibinfo {pages} {131801}}, \Eprint
  {http://arxiv.org/abs/hep-ex/0602020}{arXiv:hep-ex/0602020}\BibitemShut
  {NoStop}%
\bibitem [{\citenamefont {Baron} \emph {et~al.}(2014)}]{Baron:2013eja}%
  \BibitemOpen
  \bibfield  {author} {\bibinfo {author} {\bibfnamefont {J.}~\bibnamefont
  {Baron}} \emph {et~al.} (\bibinfo {collaboration} {ACME Collaboration}),
  }\bibfield  {title} {\emph {\bibinfo {title} {{Order of Magnitude Smaller
  Limit on the Electric Dipole Moment of the Electron}}}, }\href {\doibase
  10.1126/science.1248213} {\bibfield  {journal} {\bibinfo  {journal} {Science}
  }\textbf {\bibinfo {volume} {343}} (\bibinfo {year} {2014}) \bibinfo {pages}
  {269}}, \Eprint {http://arxiv.org/abs/1310.7534}{arXiv:1310.7534
  [physics.atom-ph]}\BibitemShut {NoStop}%
\bibitem [{\citenamefont {Cirigliano} \emph
  {et~al.}(2016{\natexlab{b}})\citenamefont {Cirigliano}, \citenamefont
  {Dekens}, \citenamefont {de~Vries}, and \citenamefont
  {Mereghetti}}]{Cirigliano:2016njn}%
  \BibitemOpen
  \bibfield  {author} {\bibinfo {author} {\bibfnamefont {V.}~\bibnamefont
  {Cirigliano}}, \bibinfo {author} {\bibfnamefont {W.}~\bibnamefont {Dekens}},
  \bibinfo {author} {\bibfnamefont {J.}~\bibnamefont {de~Vries}},  and \bibinfo
  {author} {\bibfnamefont {E.}~\bibnamefont {Mereghetti}}, }\bibfield  {title}
  {\emph {\bibinfo {title} {{Is there room for CP violation in the top-Higgs
  sector?}}}, }\href {\doibase 10.1103/PhysRevD.94.016002} {\bibfield
  {journal} {\bibinfo  {journal} {Phys. Rev.} }\textbf {\bibinfo {volume}
  {D94}} (\bibinfo {year} {2016}{\natexlab{b}}) \bibinfo {pages} {016002}},
  \Eprint {http://arxiv.org/abs/1603.03049}{arXiv:1603.03049
  [hep-ph]}\BibitemShut {NoStop}%
\bibitem [{\citenamefont {Fuyuto} and \citenamefont
  {Ramsey-Musolf}(2017)}]{Fuyuto:2017xup}%
  \BibitemOpen
  \bibfield  {author} {\bibinfo {author} {\bibfnamefont {K.}~\bibnamefont
  {Fuyuto}} and \bibinfo {author} {\bibfnamefont {M.}~\bibnamefont
  {Ramsey-Musolf}}, }\bibfield  {title} {\emph {\bibinfo {title} {{Top Down
  Electroweak Dipole Operators}}}, }\href@noop {} { (\bibinfo {year} {2017})},
  \Eprint {http://arxiv.org/abs/1706.08548}{arXiv:1706.08548
  [hep-ph]}\BibitemShut {NoStop}%
\bibitem [{\citenamefont {Benzke} \emph {et~al.}(2011)\citenamefont {Benzke},
  \citenamefont {Lee}, \citenamefont {Neubert}, and \citenamefont
  {Paz}}]{Benzke:2010tq}%
  \BibitemOpen
  \bibfield  {author} {\bibinfo {author} {\bibfnamefont {M.}~\bibnamefont
  {Benzke}}, \bibinfo {author} {\bibfnamefont {S.~J.} \bibnamefont {Lee}},
  \bibinfo {author} {\bibfnamefont {M.}~\bibnamefont {Neubert}},  and \bibinfo
  {author} {\bibfnamefont {G.}~\bibnamefont {Paz}}, }\bibfield  {title} {\emph
  {\bibinfo {title} {{Long-Distance Dominance of the CP Asymmetry in $B\to
  X_{s,d}+\gamma$ Decays}}}, }\href {\doibase 10.1103/PhysRevLett.106.141801}
  {\bibfield  {journal} {\bibinfo  {journal} {Phys. Rev. Lett.} }\textbf
  {\bibinfo {volume} {106}} (\bibinfo {year} {2011}) \bibinfo {pages}
  {141801}}, \Eprint {http://arxiv.org/abs/1012.3167}{arXiv:1012.3167
  [hep-ph]}\BibitemShut {NoStop}%
\bibitem [{\citenamefont {Amhis} \emph {et~al.}(2017)}]{Amhis:2016xyh}%
  \BibitemOpen
  \bibfield  {author} {\bibinfo {author} {\bibfnamefont {Y.}~\bibnamefont
  {Amhis}} \emph {et~al.} (\bibinfo {collaboration} {HFLAV}), }\bibfield
  {title} {\emph {\bibinfo {title} {{Averages of $b$-hadron, $c$-hadron, and
  $\tau$-lepton properties as of summer 2016}}}, }\href {\doibase
  10.1140/epjc/s10052-017-5058-4} {\bibfield  {journal} {\bibinfo  {journal}
  {Eur. Phys. J.} }\textbf {\bibinfo {volume} {C77}} (\bibinfo {year} {2017})
  \bibinfo {pages} {895}}, \Eprint
  {http://arxiv.org/abs/1612.07233}{arXiv:1612.07233 [hep-ex]}\BibitemShut
  {NoStop}%
\bibitem [{\citenamefont {Grzadkowski} and \citenamefont
  {Misiak}(2008)}]{Grzadkowski:2008mf}%
  \BibitemOpen
  \bibfield  {author} {\bibinfo {author} {\bibfnamefont {B.}~\bibnamefont
  {Grzadkowski}} and \bibinfo {author} {\bibfnamefont {M.}~\bibnamefont
  {Misiak}}, }\bibfield  {title} {\emph {\bibinfo {title} {{Anomalous Wtb
  coupling effects in the weak radiative B-meson decay}}}, }\href {\doibase
  10.1103/PhysRevD.84.059903, 10.1103/PhysRevD.78.077501} {\bibfield  {journal}
  {\bibinfo  {journal} {Phys. Rev. D} }\textbf {\bibinfo {volume} {78}}
  (\bibinfo {year} {2008}) \bibinfo {pages} {077501}}, \Eprint
  {http://arxiv.org/abs/0802.1413}{arXiv:0802.1413 [hep-ph]}\BibitemShut
  {NoStop}%
\bibitem [{\citenamefont {Drobnak} \emph {et~al.}(2012)\citenamefont {Drobnak},
  \citenamefont {Fajfer}, and \citenamefont {Kamenik}}]{Drobnak:2011aa}%
  \BibitemOpen
  \bibfield  {author} {\bibinfo {author} {\bibfnamefont {J.}~\bibnamefont
  {Drobnak}}, \bibinfo {author} {\bibfnamefont {S.}~\bibnamefont {Fajfer}},
  and \bibinfo {author} {\bibfnamefont {J.~F.} \bibnamefont {Kamenik}},
  }\bibfield  {title} {\emph {\bibinfo {title} {{Probing anomalous tWb
  interactions with rare B decays}}}, }\href {\doibase
  10.1016/j.nuclphysb.2011.10.004} {\bibfield  {journal} {\bibinfo  {journal}
  {Nucl. Phys.} }\textbf {\bibinfo {volume} {B855}} (\bibinfo {year} {2012})
  \bibinfo {pages} {82}}, \Eprint
  {http://arxiv.org/abs/1109.2357}{arXiv:1109.2357 [hep-ph]}\BibitemShut
  {NoStop}%
\bibitem [{\citenamefont {Drobnak} \emph {et~al.}(2011)\citenamefont {Drobnak},
  \citenamefont {Fajfer}, and \citenamefont {Kamenik}}]{Drobnak:2011wj}%
  \BibitemOpen
  \bibfield  {author} {\bibinfo {author} {\bibfnamefont {J.}~\bibnamefont
  {Drobnak}}, \bibinfo {author} {\bibfnamefont {S.}~\bibnamefont {Fajfer}},
  and \bibinfo {author} {\bibfnamefont {J.~F.} \bibnamefont {Kamenik}},
  }\bibfield  {title} {\emph {\bibinfo {title} {{Interplay of $t \to b W$ Decay
  and $B_q$ Meson Mixing in Minimal Flavor Violating Models}}}, }\href
  {\doibase 10.1016/j.physletb.2011.05.052} {\bibfield  {journal} {\bibinfo
  {journal} {Phys. Lett. B} }\textbf {\bibinfo {volume} {701}} (\bibinfo {year}
  {2011}) \bibinfo {pages} {234}}, \Eprint
  {http://arxiv.org/abs/1102.4347}{arXiv:1102.4347 [hep-ph]}\BibitemShut
  {NoStop}%
\bibitem [{\citenamefont {Aaboud} \emph {et~al.}(2017)}]{Aaboud:2017yqf}%
  \BibitemOpen
  \bibfield  {author} {\bibinfo {author} {\bibfnamefont {M.}~\bibnamefont
  {Aaboud}} \emph {et~al.} (\bibinfo {collaboration} {ATLAS}), }\bibfield
  {title} {\emph {\bibinfo {title} {{Analysis of the $Wtb$ vertex from the
  measurement of triple-differential angular decay rates of single top quarks
  produced in the $t$-channel at $\sqrt{s}$ = 8 TeV with the ATLAS detector}}},
  }\href {\doibase 10.1007/JHEP12(2017)017} {\bibfield  {journal} {\bibinfo
  {journal} {JHEP} }\textbf {\bibinfo {volume} {12}} (\bibinfo {year} {2017})
  \bibinfo {pages} {017}}, \Eprint
  {http://arxiv.org/abs/1707.05393}{arXiv:1707.05393 [hep-ex]}\BibitemShut
  {NoStop}%
\bibitem [{\citenamefont {Degrande} \emph {et~al.}(2012)\citenamefont
  {Degrande}, \citenamefont {Duhr}, \citenamefont {Fuks}, \citenamefont
  {Grellscheid}, \citenamefont {Mattelaer}, and \citenamefont
  {Reiter}}]{Degrande:2011ua}%
  \BibitemOpen
  \bibfield  {author} {\bibinfo {author} {\bibfnamefont {C.}~\bibnamefont
  {Degrande}}, \bibinfo {author} {\bibfnamefont {C.}~\bibnamefont {Duhr}},
  \bibinfo {author} {\bibfnamefont {B.}~\bibnamefont {Fuks}}, \bibinfo {author}
  {\bibfnamefont {D.}~\bibnamefont {Grellscheid}}, \bibinfo {author}
  {\bibfnamefont {O.}~\bibnamefont {Mattelaer}},  and \bibinfo {author}
  {\bibfnamefont {T.}~\bibnamefont {Reiter}}, }\bibfield  {title} {\emph
  {\bibinfo {title} {{UFO - The Universal FeynRules Output}}}, }\href {\doibase
  10.1016/j.cpc.2012.01.022} {\bibfield  {journal} {\bibinfo  {journal}
  {Comput. Phys. Commun.} }\textbf {\bibinfo {volume} {183}} (\bibinfo {year}
  {2012}) \bibinfo {pages} {1201}}, \Eprint
  {http://arxiv.org/abs/1108.2040}{arXiv:1108.2040 [hep-ph]}\BibitemShut
  {NoStop}%
\bibitem [{\citenamefont {Brivio} \emph {et~al.}(2017)\citenamefont {Brivio},
  \citenamefont {Jiang}, and \citenamefont {Trott}}]{Brivio:2017btx}%
  \BibitemOpen
  \bibfield  {author} {\bibinfo {author} {\bibfnamefont {I.}~\bibnamefont
  {Brivio}}, \bibinfo {author} {\bibfnamefont {Y.}~\bibnamefont {Jiang}},  and
  \bibinfo {author} {\bibfnamefont {M.}~\bibnamefont {Trott}}, }\bibfield
  {title} {\emph {\bibinfo {title} {{The SMEFTsim package, theory and tools}}},
  }\href {\doibase 10.1007/JHEP12(2017)070} {\bibfield  {journal} {\bibinfo
  {journal} {JHEP} }\textbf {\bibinfo {volume} {12}} (\bibinfo {year} {2017})
  \bibinfo {pages} {070}}, \Eprint
  {http://arxiv.org/abs/1709.06492}{arXiv:1709.06492 [hep-ph]}\BibitemShut
  {NoStop}%
\bibitem [{\citenamefont {Alloul} \emph {et~al.}(2014)\citenamefont {Alloul},
  \citenamefont {Christensen}, \citenamefont {Degrande}, \citenamefont {Duhr},
  and \citenamefont {Fuks}}]{Alloul:2013bka}%
  \BibitemOpen
  \bibfield  {author} {\bibinfo {author} {\bibfnamefont {A.}~\bibnamefont
  {Alloul}}, \bibinfo {author} {\bibfnamefont {N.~D.} \bibnamefont
  {Christensen}}, \bibinfo {author} {\bibfnamefont {C.}~\bibnamefont
  {Degrande}}, \bibinfo {author} {\bibfnamefont {C.}~\bibnamefont {Duhr}},  and
  \bibinfo {author} {\bibfnamefont {B.}~\bibnamefont {Fuks}}, }\bibfield
  {title} {\emph {\bibinfo {title} {{FeynRules 2.0 - A complete toolbox for
  tree-level phenomenology}}}, }\href {\doibase 10.1016/j.cpc.2014.04.012}
  {\bibfield  {journal} {\bibinfo  {journal} {Comput. Phys. Commun.} }\textbf
  {\bibinfo {volume} {185}} (\bibinfo {year} {2014}) \bibinfo {pages} {2250}},
  \Eprint {http://arxiv.org/abs/1310.1921}{arXiv:1310.1921
  [hep-ph]}\BibitemShut {NoStop}%
\bibitem [{\citenamefont {Durieux} \emph {et~al.}(2015)\citenamefont {Durieux},
  \citenamefont {Maltoni}, and \citenamefont {Zhang}}]{Durieux:2014xla}%
  \BibitemOpen
  \bibfield  {author} {\bibinfo {author} {\bibfnamefont {G.}~\bibnamefont
  {Durieux}}, \bibinfo {author} {\bibfnamefont {F.}~\bibnamefont {Maltoni}},
  and \bibinfo {author} {\bibfnamefont {C.}~\bibnamefont {Zhang}}, }\bibfield
  {title} {\emph {\bibinfo {title} {{Global approach to top-quark
  flavor-changing interactions}}}, }\href {\doibase 10.1103/PhysRevD.91.074017}
  {\bibfield  {journal} {\bibinfo  {journal} {Phys. Rev.} }\textbf {\bibinfo
  {volume} {D91}} (\bibinfo {year} {2015}) \bibinfo {pages} {074017}}, \Eprint
  {http://arxiv.org/abs/1412.7166}{arXiv:1412.7166 [hep-ph]}\BibitemShut
  {NoStop}%
\end{thebibliography}%
